\documentclass[preprint,12pt]{elsarticle} 

\usepackage{lineno}
\usepackage{color}
\usepackage[usenames,dvipsnames]{xcolor}
\usepackage{fullpage}
\usepackage{amsmath}
\usepackage{amssymb}
\usepackage{mathrsfs}
\usepackage{newfloat}
\usepackage{graphicx}
\usepackage{subfig}
\usepackage{multirow}
\usepackage{stackengine}
\usepackage{caption}
\usepackage{tikz}
\usepackage{float}
\usepackage{multicol}
\usetikzlibrary{arrows,positioning,calc}
\usepackage{pstool}
\usepackage{textcomp}
\usepackage{fixltx2e}
\usepackage[hidelinks]{hyperref}
\hypersetup{colorlinks,bookmarksopen,linkcolor=blue,citecolor=blue,urlcolor=blue}
\usepackage{blindtext}
\usepackage{bbold}
\usepackage[inline]{enumitem}

\newcommand{\res}{\ensuremath{r}}
\newcommand{\map}{\ensuremath{\boldsymbol{\phi}}}
\newcommand{\argmin}{\arg\min}
\newcommand{\x}{\ensuremath{\mathbf{x}}}
\newcommand{\mode}{\ensuremath{\boldsymbol{\varphi}}}
\newcommand{\smooth}{\ensuremath{\mathbf{v}}}
\newcommand{\mufluc}{\ensuremath{\mathbf{w}}}
\newcommand{\dofs}{\ensuremath{\underline{\mathbb{a}}}}
\newcommand{\dof}{\ensuremath{a}}

\definecolor{myred}{RGB}{222,45,38}
\definecolor{myblue}{RGB}{0,115,189}
\definecolor{mygreen}{rgb}{0.19,0.61,0.21}

\newcommand{\bs}[1]{{\boldsymbol{#1}}}

\graphicspath{{figs/}}

\journal{Int. J. Solids Struct.}

\biboptions{authoryear}

\begin{document}

\begin{frontmatter}

\title{Harvesting Deformation Modes for Micromorphic Homogenization from Experiments on Mechanical Metamaterials\tnoteref{mytitlenote}}
\tnotetext[mytitlenote]{The post-print version of this article is published in \emph{Int. J. Solids Struct.}, \href{https://doi.org/10.1016/j.ijsolstr.2024.112916}{10.1016/j.ijsolstr.2024.112916}. This manuscript version is made available under the \href{https://creativecommons.org/licenses/by/4.0/}{CC-BY 4.0} license.}

\author[TUe]{S.~Maraghechi} 
\ead{S.Maraghechi@tue.nl}

\author[TUe]{O.~Roko\v{s}}
\ead{O.Rokos@tue.nl}

\author[TUe]{R.H.J.~Peerlings}
\ead{R.H.J.Peerlings@tue.nl}

\author[TUe]{M.G.D.~Geers}
\ead{M.G.D.Geers@tue.nl}

\author[TUe]{J.P.M.~Hoefnagels\corref{correspondingauthor}}
\ead{J.P.M.Hoefnagels@tue.nl}

\address[TUe]{Mechanics of Materials, Department of Mechanical Engineering, Eindhoven University of Technology, P.O.~Box~513, 5600~MB~Eindhoven, The~Netherlands}
\cortext[correspondingauthor]{Corresponding author.}

\begin{abstract}
	
A micromorphic computational homogenization framework has recently been developed to deal with materials showing long-range correlated interactions, i.e. displaying patterning modes. Typical examples of such materials are elastomeric mechanical metamaterials, in which patterning emerges from local buckling of the underlying microstructure. Because pattern transformations significantly influence the resulting effective behaviour, it is vital to distinguish them from the overall deformation. To this end, the following kinematic decomposition into three parts was introduced in the micromorphic scheme:
\begin{enumerate*}[label=(\roman*)]
	\item a smooth mean displacement field, corresponding to the slowly varying deformation at the macro-\mbox{scale},
	\item a long-range correlated fluctuation field, related to the buckling pattern at the meso-\mbox{scale}, and
	\item the remaining uncorrelated local microfluctuation field at the micro-\mbox{scale}.
\end{enumerate*}
The micromorphic framework has proven to be capable of predicting relevant mechanical behaviour, including size effects and spatial as well as temporal mixing of patterns in elastomeric metamaterials, making it a powerful tool to design metamaterials for engineering applications. The long-range correlated fluctuation fields need to be, however, provided \emph{a priori} as input parameters.
The main goal of this study is experimental identification of the decomposed kinematics in cellular metamaterials based on the three-part ansatz.
To this end, a full-field micromorphic Integrated Digital Image Correlation (IDIC) technique has been developed. The methodology is formulated for finite-size cellular elastomeric metamaterial specimens deformed in
\begin{enumerate*}[label=(\roman*)]
	\item virtually generated images and 
	\item experimental images attained during \textit{in-situ} compression of specimens with millimetre sized microstructure using optical microscopy.
\end{enumerate*}
The proposed IDIC method identifies the different kinematic fields, both before and after the microstructural buckling, and without any prior knowledge determines correctly the relevant patterning modes required by the homogenization scheme. It is further argued that patterning modes are independent of the unit cell size, the hole diameter to cell size ratio, as well as local material properties, allowing for modelling and design of (finite- and infinite-size) metamaterials and specimens with graded microstructures in terms of geometry and/or material properties.
It is shown that the proposed methodology is also applicable to cellular metamaterials and structures with different microstructural designs.

\end{abstract}


\end{frontmatter}



%
%
\section{Introduction}
\label{sec5:intro}
Mechanical metamaterials have recently gained substantial attention in the literature due to their potential for various engineering applications such as strain tunable photonic crystals \citep{Krishnan2009}, tunable acoustic properties~\citep{Zhao2015,Bilal2017,Wormser2017,qida_lin_three-dimensional_2023}, soft robotics \citep{Yang2015,Mirzaali2018a,Yu2018,jiajia_shen_active_2023}, programmable mechanical metamaterials \citep{Florijn2014,Bertoldi2017,Kolken2017,linzhi_li_magnetically_2023}, biomedical prosthetics \citep{Yavari2015,Zadpoor2017}, or functionally graded metamaterials \citep{Francesconi2019,amin_farzaneh_sequential_2022}. Various aspects of such metamaterials with 2D \citep{Bertoldi2008} and 3D \citep{Javid2016,Hedayati2019} microstructures have been investigated. A number of studies focused on the effect of the hole geometry and stacking \citep{Bertoldi2010,Overvelde2012,Overvelde2014,Hu2013}. Several studies reported in the literature made use of full-field displacement measurement techniques to investigate cellular metamaterials, among which \cite{Slann2015} used local Digital Image Correlation (DIC) to validate their numerical study on auxetic metamaterials, \cite{Xu2016} used DIC in the design of thermally expanding tunable 3D metamaterials, \cite{Easey2019} studied different designs of auxetic dome-shaped structures using stereo DIC, and \cite{Tang2015} compared DIC and finite element analysis results on highly stretchable and reconfigurable metamaterials. Apart from displacement and strain assessment, \cite{Mullin2007} made a qualitative comparison between numerically and experimentally assessed stresses in cellular elastomers. Besides the geometrical and morphological effects, also the type of material is important. \cite{Mirzaali2018} used 3D printing to manufacture and study multi-material mechanical metamaterials, \cite{Wu2014} investigated different pattern transformations as a result of swelling in cellular hydrogel membranes, whereas~\cite{Hu2018} investigated the effect of friction on buckling pattern formation in cellular structures. A large number of works also focused on experimental investigation of pantographic metamaterials, e.g., \citep{patrick_auger_poynting_2020,francois_hild_multiscale_2020,emilio_barchiesi_validation_2021}.

\begin{figure}
	\centering
	\subfloat[Undeformed microstructure]{\includegraphics[width=.3\textwidth]{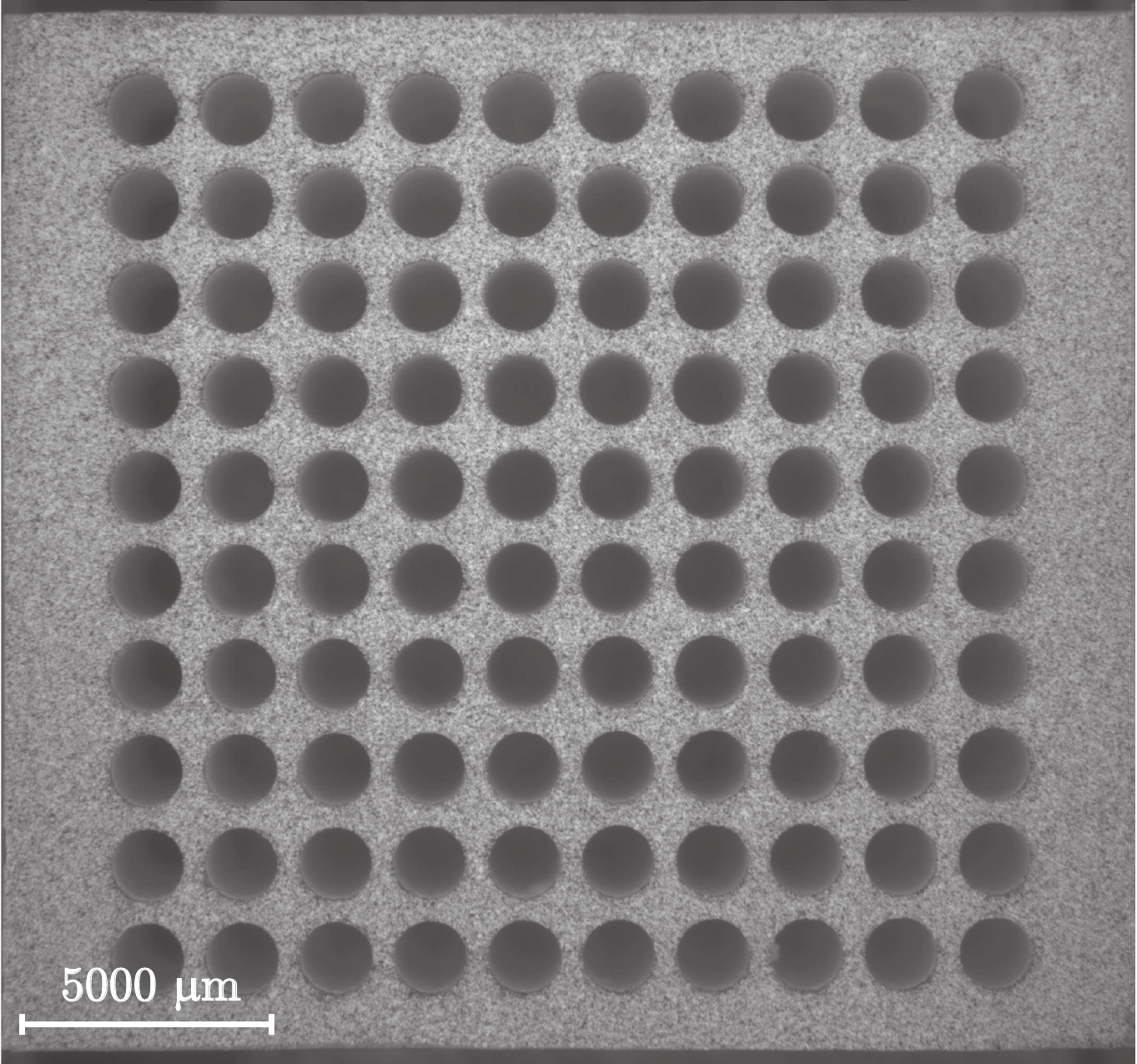}\label{fig5:undef}}\hspace{1.0em}
	\subfloat[Deformed microstructure]{\includegraphics[width=.3\textwidth]{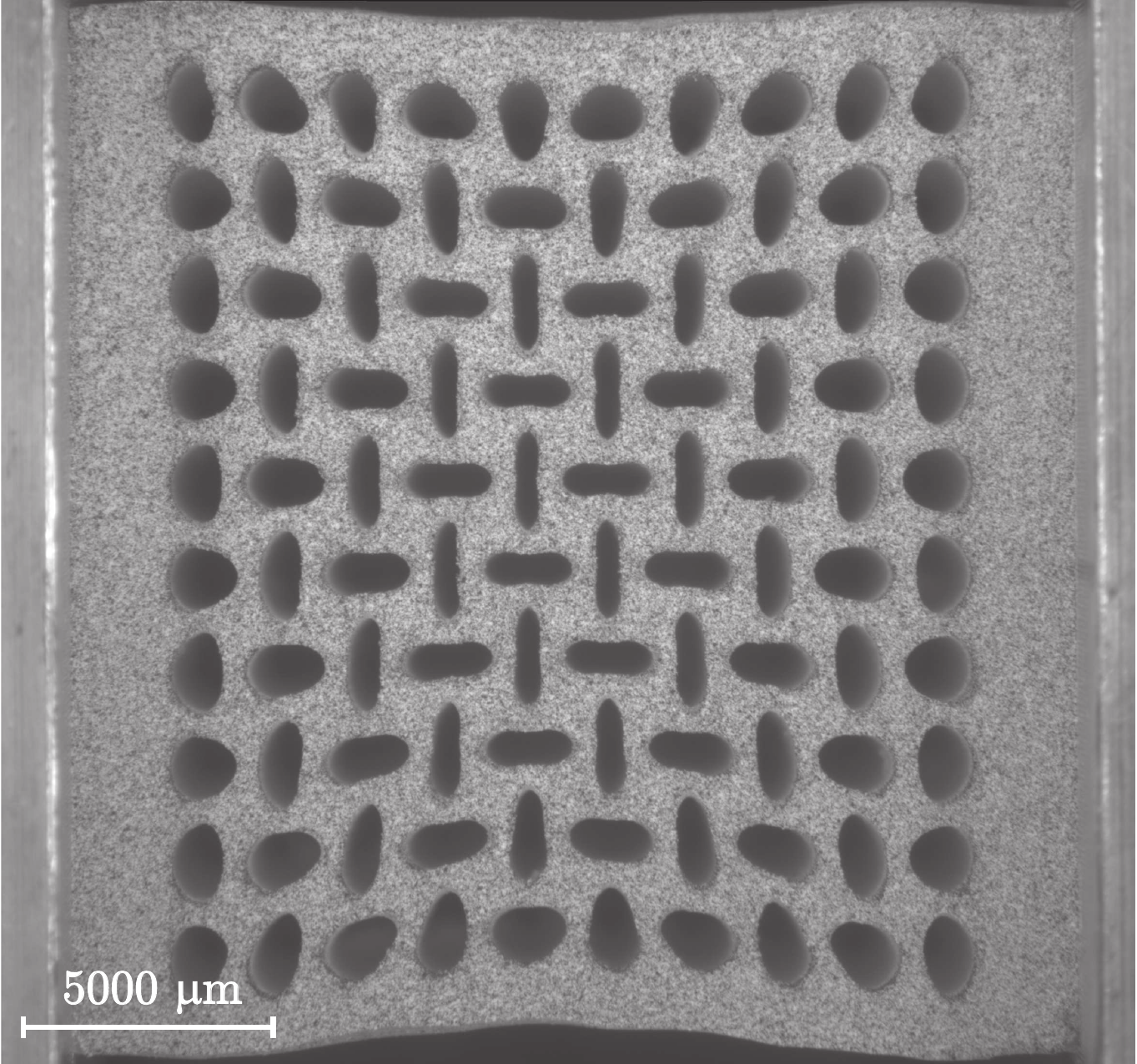}\label{fig5:def}}
	\caption{Elastomeric metamaterial specimen \protect\subref{fig5:undef} before and \protect\subref{fig5:def} after the onset of microstructural buckling. The emerging antisymmetric pattern is clearly visible, which is a result of a compression in the horizontal direction.}
	\label{fig5:pattern}
\end{figure}

The common denominator to the listed applications are exotic properties of metamaterials, typically emerging from a coordinated motion of their microstructure, i.e., from the so-called patterning. The patterning usually occurs upon reaching a critical level of compressive load, resulting in a local buckling of the underlying microstructure and a significant change in the overall mechanical properties. Fig.~\ref{fig5:undef} shows an example of a cellular elastomeric metamaterial specimen with a regular grid of circular holes, where the local buckling results in a pattern of alternating elliptical holes (Fig.~\ref{fig5:def}). Here it can be observed that the patterning is locally constrained by external boundaries, and depends on the size of the holes relative to the size of the entire specimen (i.e., on the scale ratio), studied computationally in detail by~\cite{Ameen2018}. Since such irregularities in the pattern directly influence the overall behaviour of the specimen---even on a larger scale---, it is important from the engineering perspective to be able to identify these distributions, independently of the global deformation, for a proper assessment of and design for certain mechanical properties.
	
Apart from classical homogenization approaches mostly limited to linear elastic behaviour, such as, e.g., \citep{toupin_elastic_1962,mindlin_micro-structure_1964,mindlin_second_1965}, an effective tool for predicting mechanical behaviour of materials with complex microstructures on an engineering scale and operating in a non-linear regime is first-order computational homogenization~\citep{Kouznetsova2001,Miehe2002}. To determine effective constitutive behaviour, computational homogenization solves a microscopic boundary value problem associated with each macroscopic integration point. Obtained results are used to compute effective stresses, which are equilibrated at the macroscale. Because separation of scales is assumed, such a scheme is able to properly model only macroscopically local continuum. For better approximation, extensions to second-order homogenization schemes were proposed~\citep{Kouznetsova2004,dellisola_pantographic_2019,giorgio_second-grade_2023}, which use gradient of the deformation gradient to account for non-local effects. Non-locality can also be addressed using generalized micromorphic theories, e.g., \citep{Forest2009,Hutter2017,Biswas2017,Rokos2019,Rokos2019b}, in which additional kinematic fields are introduced to account for the underlying microfluctuations. For a recent overview and further approaches see, e.g., \cite{Findeisen2020}.

Of particular interest in this contribution is the micromorphic computational homogenization scheme based on recent theoretical and numerical studies \citep{Rokos2019,Rokos2019b,van_bree_newton_2020}, which relies on a general decomposition of the displacement field according to the following ansatz
\begin{equation}\label{eq5:ansatz}
\ensuremath{\mathbf{u}}(\x) = \smooth_0(\x) + \sum_{i=1}^n v_i (\x)\mode_i(\x)+\mufluc(\x).
\end{equation}
Here, the total displacement $\ensuremath{\mathbf{u}}(\x)$, which is a vector field expressed as a function of the material coordinate in the undeformed configuration $\x$, is divided into three parts:
\begin{enumerate*}[label=(\roman*)]
	\item the mean smooth displacement field, $\smooth_0(\x)$, corresponding to the slow variations at the macro-scale;
	\item long-range correlated fluctuations at the meso-scale, represented by the vector fields $\mode_i(\x)$, $i = 1, \dots, n$, scaled by their spatial magnitudes $v_i(\x)$; and
	\item a remaining microfluctuation field, $\mufluc(\x)$, representing the uncorrelated fluctuations at the micro-scale.
\end{enumerate*}
By means of the patterning modes, $\mode_i$, the micromorhpic scheme is capable of capturing spatial non-local effects in the microstructure, present typically because of the lack of separation of scales and emerging due to effects such as boundary layers or temporal and spatial mixing of modes. Non-local effects have been studied numerically as well as experimentally within the context of metamaterial structures in, e.g., \citep{Ameen2018,Maraghechi2019b}. Using a variational formulation towards a numerical approximation, \citep[cf.][]{van_bree_newton_2020}, an effective potential energy is minimized with respect to the kinematic decomposition of Eq.~\eqref{eq5:ansatz}, resulting in a micromorphic type of effective continuum, i.e., not of the standard Cauchy--Born type. Although the standard balance of momentum equation~$\boldsymbol{\nabla}_0\cdot\mathbf{P} = \boldsymbol{0}$ governs the evolution of the~$\smooth_0(\x)$ field, where~$\mathbf{P}$ is the homogenized first Piola--Kirchhoff stress tensor energetically conjugated with the deformation gradient tensor~$\mathbf{F} = \mathbf{I} + (\boldsymbol{\nabla}_0\smooth_0(\x))^\mathsf{T}$, an additional coupled micromorphic balance equation emerges. Herein, $\boldsymbol{\nabla}_0$ denotes the gradient operator with respect to the reference configuration. This equation governs the evolution of~$v_i (\x)$, which is in the form of~$\boldsymbol{\nabla}_0\cdot\boldsymbol{\Lambda} - \Pi_i = 0$, where~$\boldsymbol{\Lambda}$ is energetically conjugated with~$\boldsymbol{\nabla}_0v_i (\x)$ while~$\Pi_i$ is conjugated with~$v_i (\x)$. Since~$v_i (\x)$ corresponds to the magnitude of the patterning fluctuation field~$\mode_i(\x)$, its kinematic interpretation is straightforward and essential boundary conditions can be identified as $v_i (\x) = 0$ along any fixed boundaries. On the contrary, $\boldsymbol{\Lambda}\cdot\boldsymbol{N} = 0$ on the free boundaries may be adopted (where~$\boldsymbol{N}$ denotes the outer normal). We refer the interested reader for further details and derivations to~\cite[][Section~3.3]{Rokos2019b}. Note that unlike second-order computational homogenization \citep[e.g.,][]{Kouznetsova2004}, which uses continuity of the gradient of the deformation gradient tensor field~$\mathbf{F}$ to account for non-locality and size effects, micromorphic computational homogenization introduces an additional independent kinematical field with a governing equation complementing the standard balance of linear momentum. The homogenization scheme can thus facilitate optimized designs of engineering systems, including deeper understanding of their behaviour and development of new metamaterials. Although the kinematic fields $\smooth_0$, $v_i$ and $\mufluc$ are an outcome of a numerical analysis, the microfluctuation patterning fields $\mode_i$ are input parameters characteristic to the geometry of the underlying microstructure that need to be provided \emph{a priori}, in analogy to material constants of an underlying constitutive model. The patterning modes can be estimated through an educated guess \cite[cf.][]{Rokos2019}, or a numerical Bloch-type analysis~\citep[cf.][]{Bertoldi2008}. It would, however, be much better to measure the patterning modes directly from experiments, as this would include the real material behaviour, all micro-fabrication flaws, and errors in the correction of spatial distortions in the geometrical images.

Therefore, the goal of this contribution is to develop a novel full-field Integrated Digital Image Correlation (IDIC) framework for quantitative experimental identification of the kinematic fields decomposed according to Eq.~\eqref{eq5:ansatz}, including the meso-scale long-range correlated fluctuation fields $\mode_i$ for the micromorphic homogenization scheme. To this end, high-resolution images captured at different stages of deformation during an experimental test of a finite-sized specimen are analysed with this micromorphic IDIC framework. IDIC is a powerful full-field displacement measurement technique based on regularization of the kinematics according to specific knowledge about the problem at hand, resulting in identification of the optimal input model parameters minimizing the difference between experimentally measured data and model predictions~\citep{Roux2006,Neggers2015,Blaysat2015}. Knowledge of the nature of the problem at hand can be integrated in the DIC scheme in the form of a parametric analytical description, which is utilized here in the form of the kinematic fields of interest. Specifically, the total displacement field is regularized based on the ansatz of Eq.~\eqref{eq5:ansatz} by parametrizing all relevant kinematic components, providing upon correlation the smooth mean field, $\smooth_0$, the long-range correlated fluctuation modes carried by the microstructure, $\mode_i$, as well as their spatial distribution magnitudes, $v_i$.

The proposed methodology is tested on virtually generated and deformed images, as well as optical microscopy images taken during an \textit{in-situ} compression test on a finite-size cellular elastomeric metamaterial with rectangular stacking of circular holes, cf.\ Fig.~\ref{fig5:pattern}. A spectral density analysis is shown to be an effective tool for estimation, parametrization, and initialization of the fluctuation mode for a specific metamaterial. It is demonstrated that the micromorphic IDIC scheme leads to a proper decomposition of the kinematic fields of interest, both before and after the emergence of the fluctuation pattern. Considering geometrical arguments, the patterning mode is shown to be independent of the considered periodic cell size, although effective mechanical behaviour of such a material is still size dependent. Moreover, it is furthermore shown that the patterning mode is only weakly dependent on the hole diameter versus unit cell size ratio, which enables the design of metamaterials with graded microstructures. 

Although a square stacking of holes is used throughout to guide and validate our developments, the proposed methodology is applicable for a general class of cellular metamaterials including hexagonal stacking of holes and chiral or re-entrant microstructures, as well as graded microstructures in terms of geometry (stacking of holes) and material properties (e.g., stiffness). The reason for focusing on the square stacking of holes herein is twofold. First, our detailed analysis requires testing inside an \textit{in-situ} micro-tensile stage underneath an optical microscope, which means that the total specimen size needs to fit within a small, rectangular space. Therefore, square stacking of holes is preferred as it results in a thin, constant-thickness boundary layer along the edge, in contrast to other microstructures that may result in a boundary layer with a varying thickness along the edge. Second, a rectangular specimen shape with square stacking of holes is easiest to manufacture with high quality and apply a constant load along the boundary edge. Finally, the remaining boundary layer effect of square-holed microstructure has been studied in detail (for precisely the same rectangular specimen shape) elsewhere~\citep{Ameen2018,Maraghechi2019b}, so this configuration is known to work well. But, as indicated earlier, it will be demonstrated that the herein proposed methodology is generally applicable also to various (graded) geometries.
%
%
\section{Micromorphic Integrated Digital Image Correlation}
\label{sec5:method}
%
%
\subsection{Integrated Digital Image Correlation}
\label{sec5:idic}
Integrated digital image correlation is a full-field displacement measurement method based on the minimization of the residual field, which is the difference of the reference image, $f$, and the deformed image, $g$, probed at, respectively, reference coordinates and the deformed positions corresponding to those coordinates based on the displacement field. The problem is solved over the whole region of interest, where the residual field is defined as
\begin{equation}\label{eq5:r}
\res(\x,\dofs) = f \left( \x \right) - g \left( \map_M(\x,\dofs) \right),
\end{equation}
with~$\map_M(\x,\dofs) = \x + \ensuremath{\mathbf{u}}\left(\x,\dofs\right)$ being the mapping function corresponding to the mechanical displacement field, $\ensuremath{\mathbf{u}}$. The displacement field, in turn, is based on the regularization according to a model with specific knowledge of the kinematics at hand, thereby reducing the number of degrees of freedom (dof) contained in $\dofs$, and improving the robustness of the solution \citep{Ruybalid2016}. This model can be a numerical model in which the dofs would be the material parameters or an analytical model describing the deformation field. The optimal values of the degrees of freedom, $\dofs^{opt}$, are found as
\begin{equation}\label{eq5:dof_opt}
\dofs^{opt} = \underset{\dofs}{\argmin} \left( \frac{1}{2} \int \res^2 \left(\x,\dofs \right) d\x \right).
\end{equation}
This minimization problem is then solved by an iterative scheme such as Gauss--Newton algorithm \citep{Neggers2016}.
%
%
\subsection{Regularization of the Kinematics Based on the Micromorphic Kinematical Ansatz}
\label{sec5:Midic}
The kinematical ansatz introduced in Eq.~\eqref{eq5:ansatz} is used to experimentally obtain the decomposed kinematic fields of cellular metamaterials. The local microfluctuations are the uncorrelated part of the kinematics and by definition unknown, therefore $\mufluc(\x)$ is not included in the micromorphic IDIC parametrization for cellular metamaterials (in fact, $\mufluc(\x)$ is directly related to the residual field~$\res(\x,\dofs)$ in Eq.~\eqref{eq5:r} via the image gradient). Accordingly, the displacement field for the IDIC problem is defined as
\begin{equation}\label{eq5:u_dic}
\ensuremath{\mathbf{u}}(\x,\dofs) = \smooth_0(\x,\dofs_{v_0}) + \sum_{i=1}^n v_i(\x,\dofs_{v_i})\mode_i(\x,\dofs_{\varphi_i}),
\end{equation}
where $\dofs = \left[ \dofs_{v_0} \ \dofs_{v_i} \ \dofs_{\varphi_i} \right]^{\mathsf{T}}$  is the column of degrees of freedom, with $\dofs_{v_0}$, $\dofs_{v_i}$ and $\dofs_{\varphi_i}$ the sets of dofs for $\smooth_0$, $v_i$ and $\mode_i$. Since each mode $\mode_i$ is periodic within an integer multiple of the unit cell size \citep{Rokos2019}, a truncated 2D Fourier series is a natural choice for its parametrization. As will be shown below, the selection of relevant sine waves can be easily done by means of a spectral density analysis of one example for any family of cellular metamaterials and loading conditions. Note that the number of modes that are activated by different loading conditions in cellular metamaterials is often only one, e.g., Fig.~\ref{fig5:pattern}, and always limited to a few \citep{Rokos2019b}. Each long-range correlated fluctuation mode $\mode_i$ is defined such that it is non-zero where the cellular microstructure exists and decays linearly to zero inside a region of surrounding (non-cellular) bulk material over a width of half the unit cell size. The boundary between bulk and cellular material is a priori selected manually by visual inspection of the reference image of the undeformed geometry; high accuracy in boundary identification is not needed. Because $\smooth_0$ and $v_i$ are slow macroscopic fields, smooth functions such as globally-supported polynomials are the best choice for their parametrization. To this end, Chebyshev polynomials of $5$th and $6$th order will be used hereafter to parametrize $\smooth_0$ and $v_1$. Since the behaviour of these polynomials outside the specimen's domain is irrelevant (because no kinematics is considered there), potentially large oscillations may be ignored in these regions.

Finally, in any \textit{in-situ} test under optical visualization, in addition to the mechanical displacement field, the images are affected by optical distortions of the optical lenses. Such distortions may introduce large errors if neglected. Therefore, the residual field of Eq.~\eqref{eq5:r} is rewritten as
\begin{equation}\label{eq5:r2}
\res(\x,\dofs) = f \left( \map_S ( \x ) \right) - g( \map_S ( \map_M(\x,\dofs) ) ),
\end{equation}
where $\map_S(\x) = \x + \mathbf{S}(\x)$ is the mapping function describing the spatial distortion \citep{Maraghechi2018}. The spatial distortion field, $\mathbf{S}(\x)$, is determined a priori in a calibration step by the method introduced in \citep{Maraghechi2019}. Hence, there are no additional degrees of freedom related to spatial distortion that need to be identified in the micromorphic IDIC problem of Eq.~\eqref{eq5:dof_opt}. 
%
%
\section{Results and Discussion}
\label{sec5:results}
To validate the micromorphic IDIC framework, it is applied to both virtual (Section~\ref{sec5:virt}) and real \textit{in-situ} micro-compression experiments (Section~\ref{sec5:real}) on cellular elastomeric metamaterial. Its general applicability to various types of metamaterial microstructures in terms of spectral analysis is first demonstrated in Section~\ref{sec5:modes} for square and hexagonal stacking of holes, as well as for a chiral metamaterial, whereas considerations towards graded microstructures are discussed in Section~\ref{sec5:graded}.
%
%
\subsection{Parametrization and Initial Guess of the Patterning Modes~$\mode_i$}
\label{sec5:modes}
\begin{figure}
	\centering
	\begin{tabular}{@{}cccc@{}}
	\subfloat[square ref.]{\includegraphics[scale=0.035]{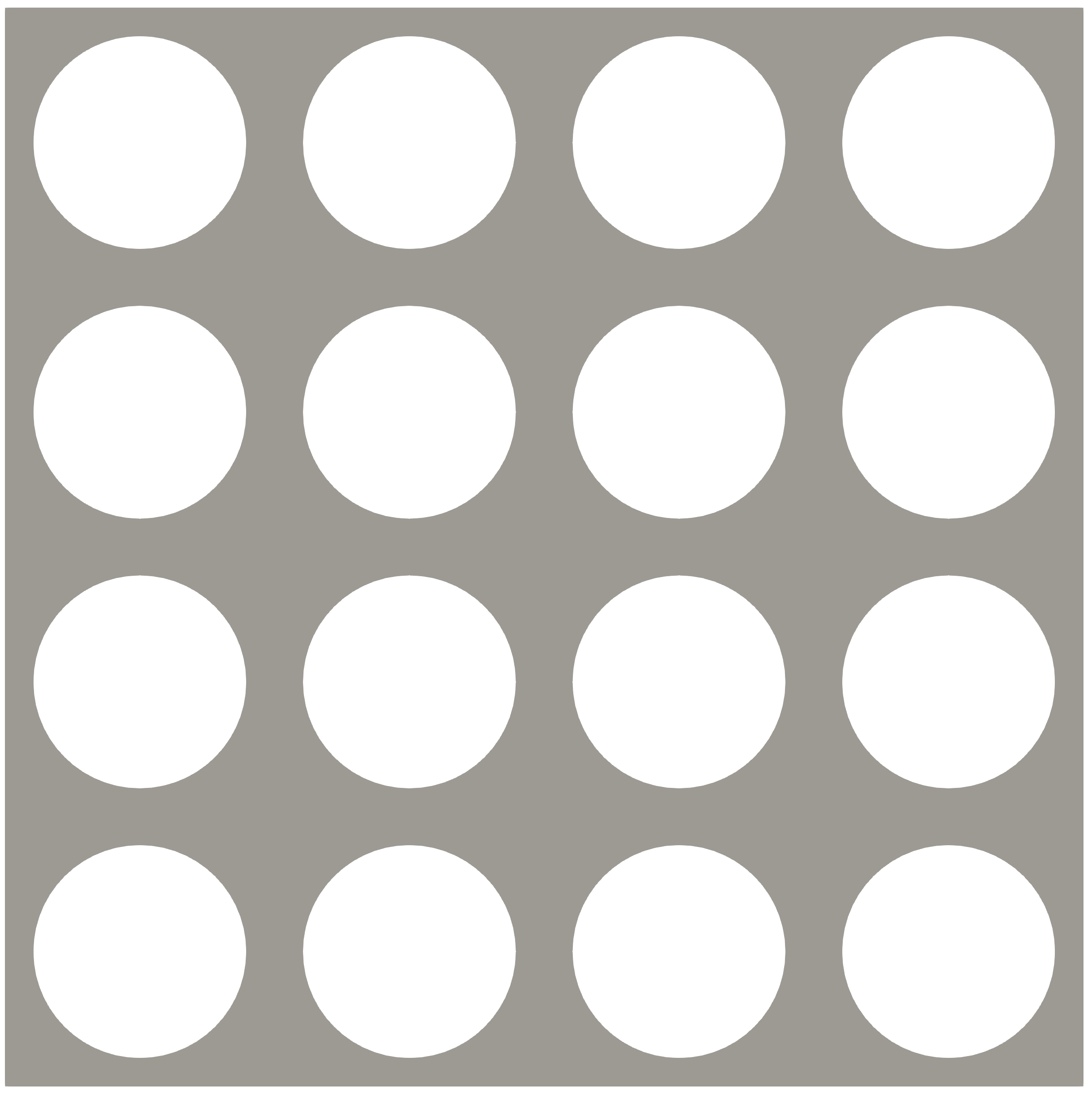}\label{fig:modesa}}
	&
	\subfloat[square def.]{\includegraphics[scale=0.035]{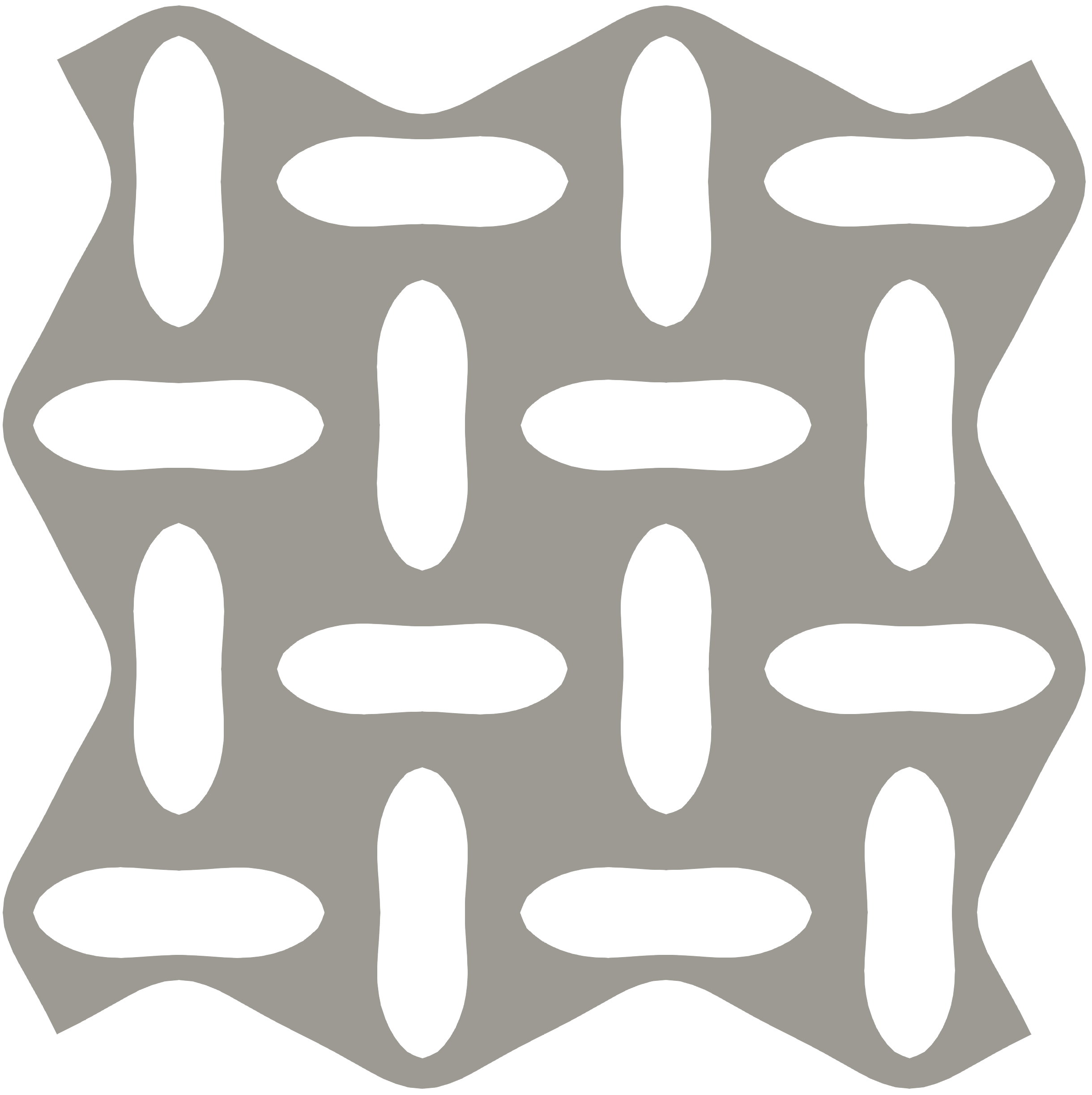}\label{fig:modesb}}
	&
	\subfloat[$S_{xx}$~\mbox{[mm$^6$]} square]{\includegraphics[scale=0.625]{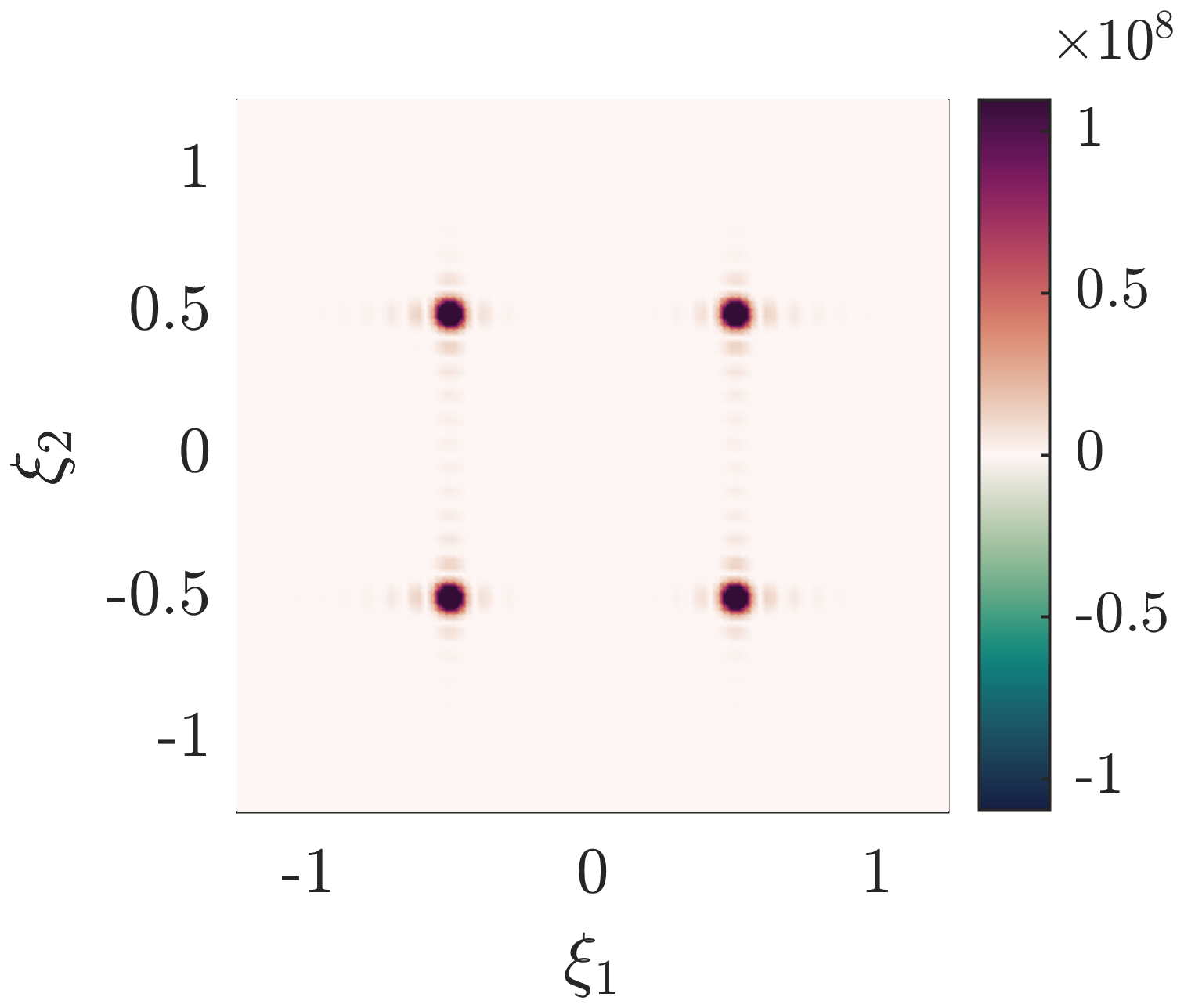}\label{fig:modesc}}
	&
	\subfloat[$S_{yy}$~\mbox{[mm$^6$]} square]{\includegraphics[scale=0.625]{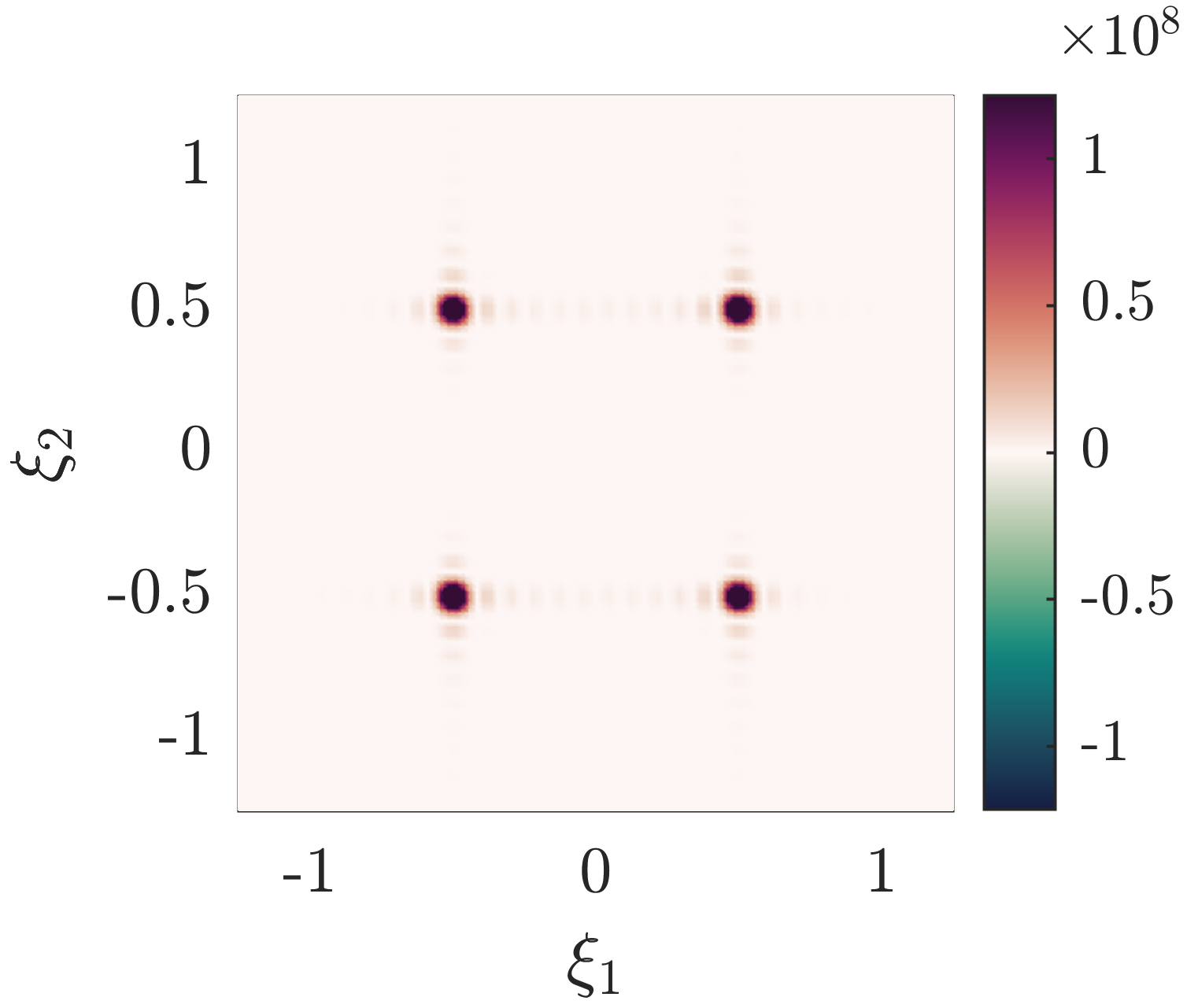}\label{fig:modesd}}
	\\
	\subfloat[hexagonal ref.]{\includegraphics[scale=0.035]{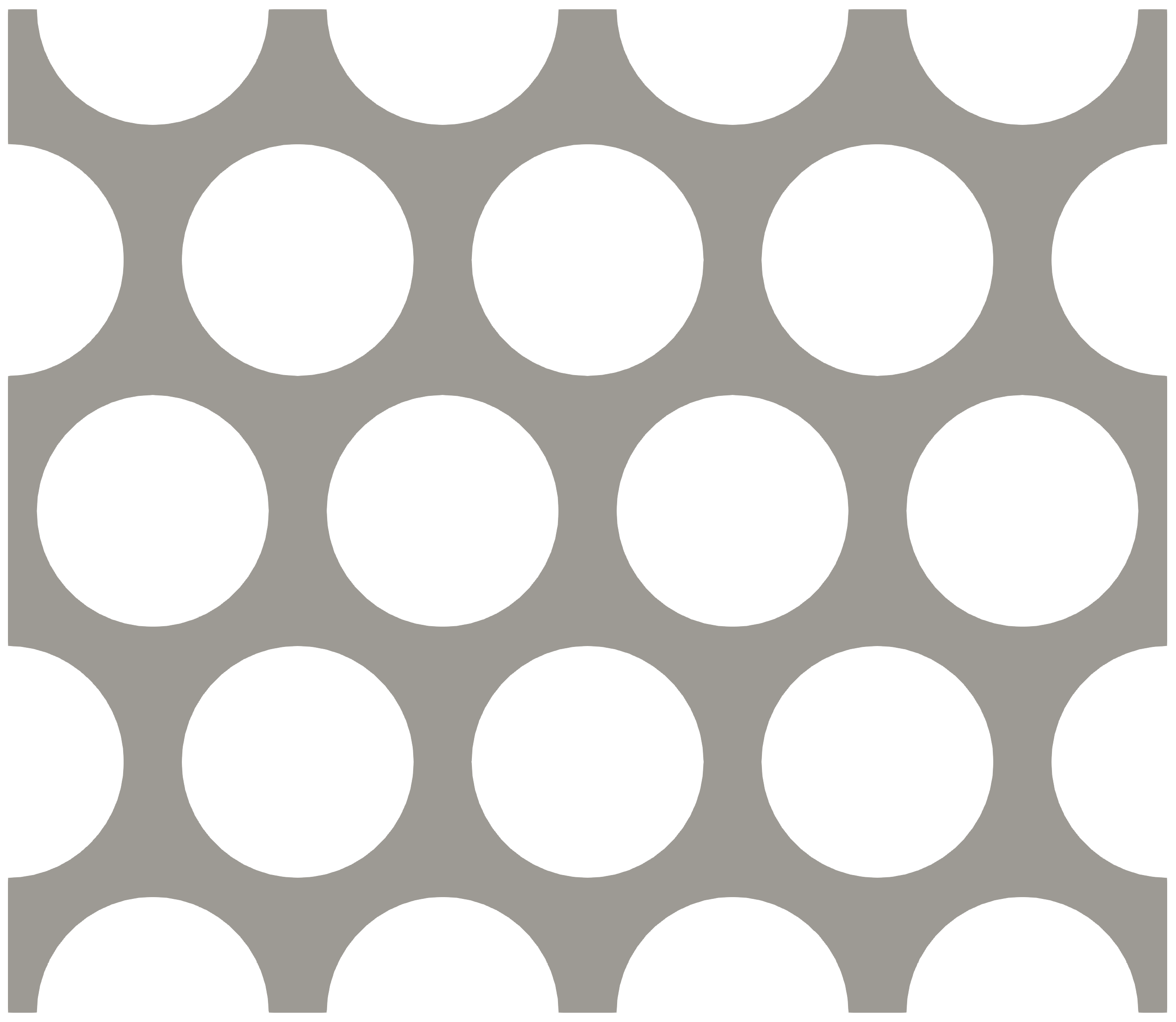}\label{fig:modese}}
	&
	\subfloat[hexagonal def.]{\includegraphics[scale=0.035]{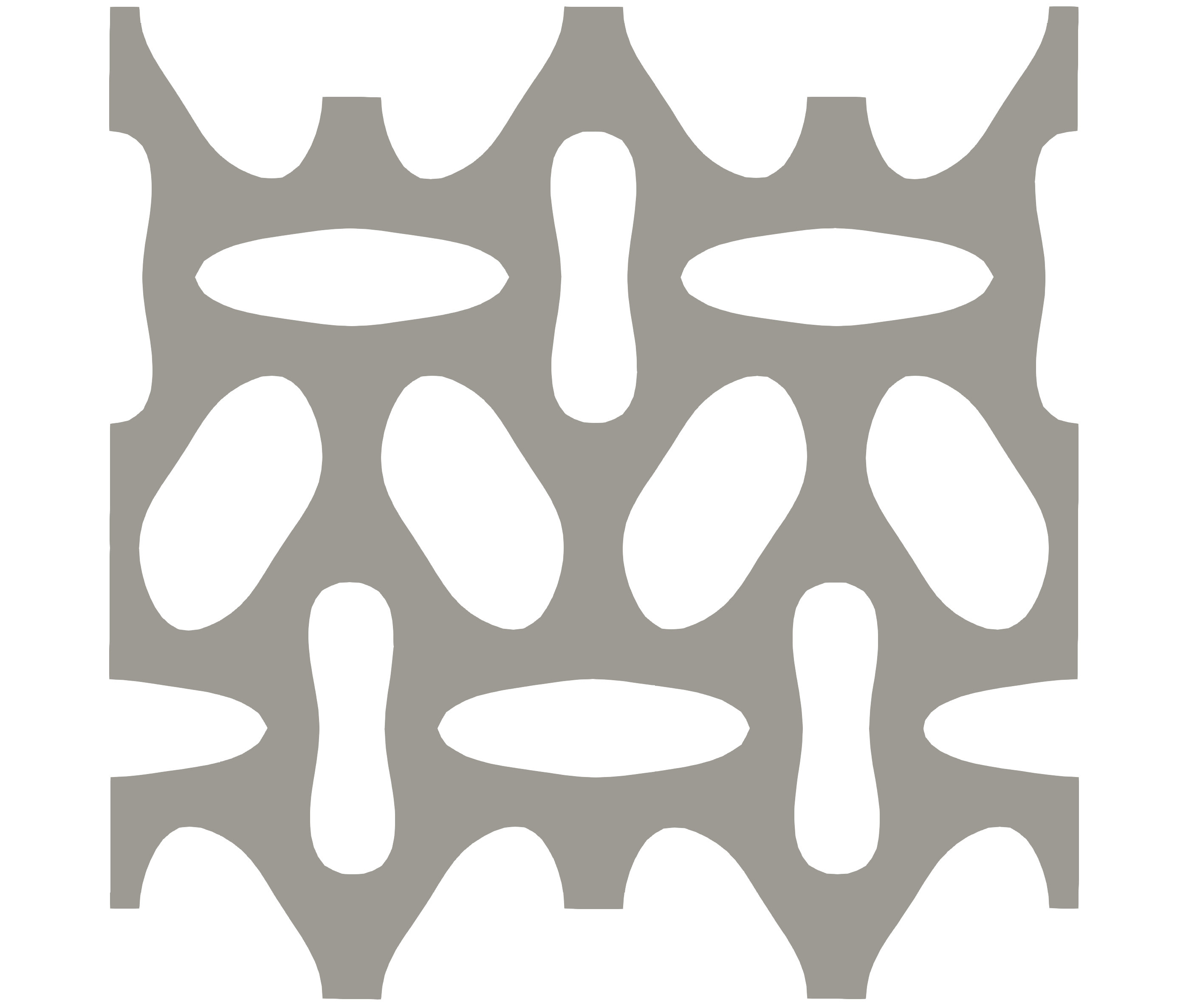}\label{fig:modesf}}
	&
	\subfloat[$S_{xx}$~\mbox{[mm$^6$]} hexagonal]{\includegraphics[scale=0.625]{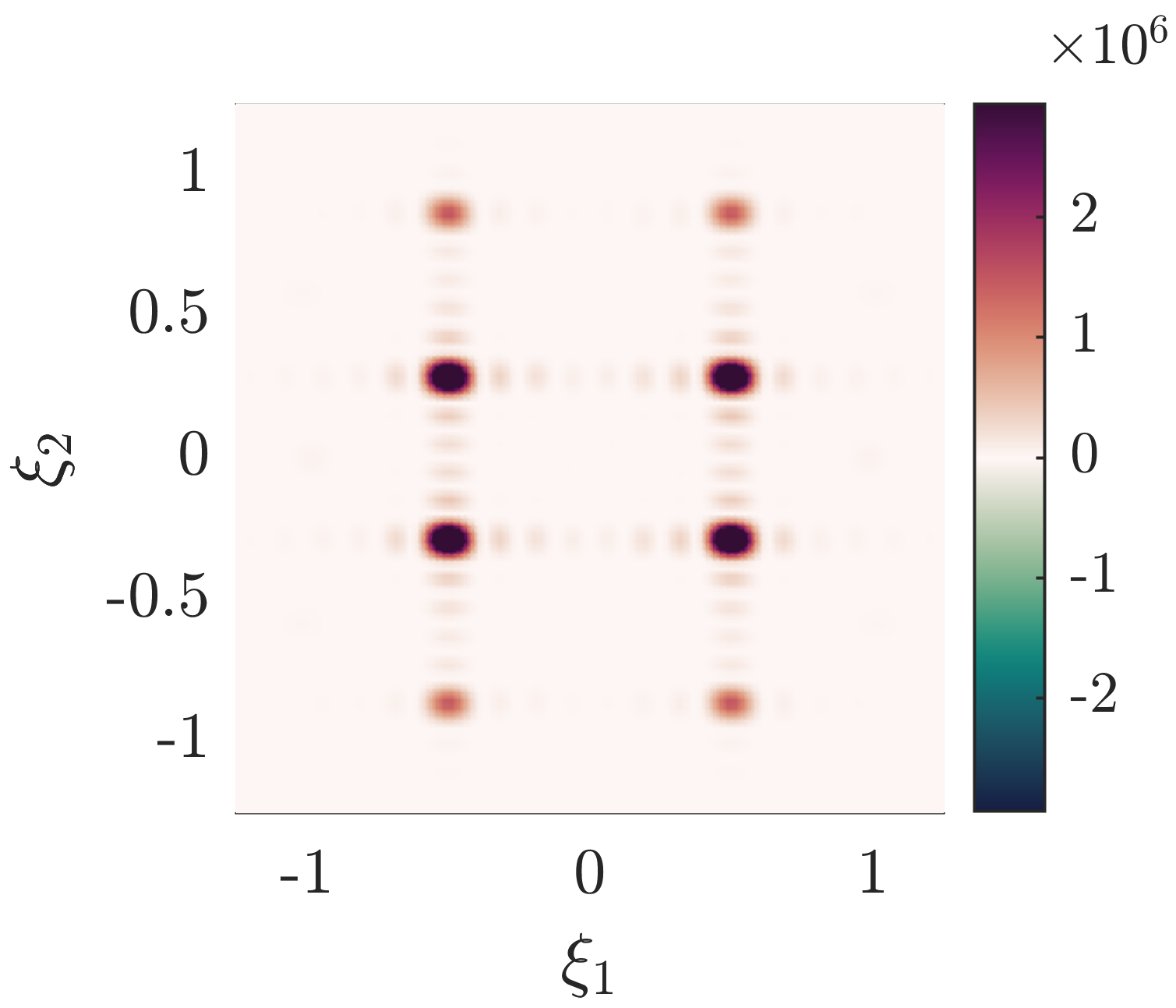}\label{fig:modesg}}
	&
	\subfloat[$S_{yy}$~\mbox{[mm$^6$]} hexagonal]{\includegraphics[scale=0.625]{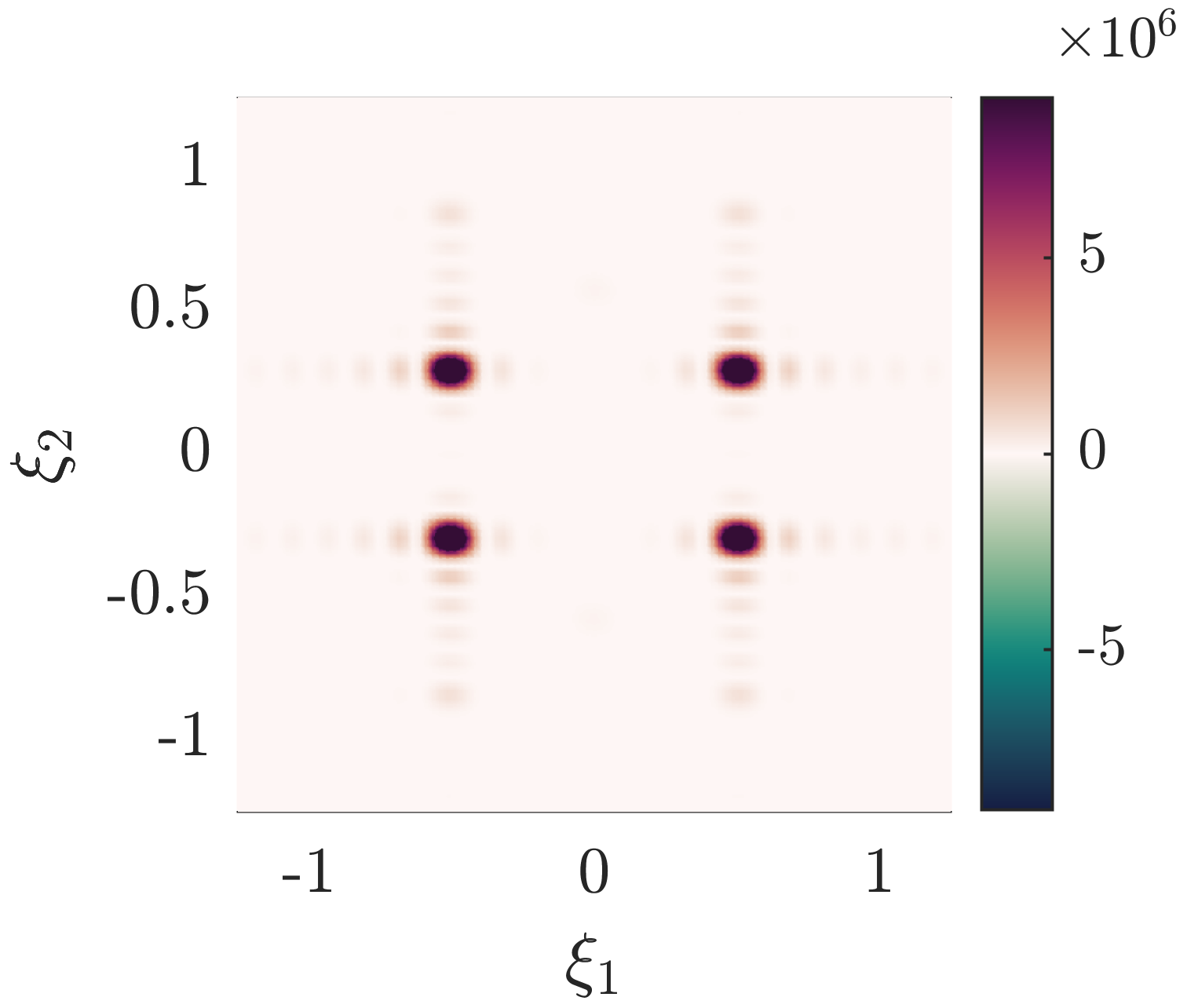}\label{fig:modesh}}
	\\
	\subfloat[chiral ref.]{\includegraphics[scale=0.15]{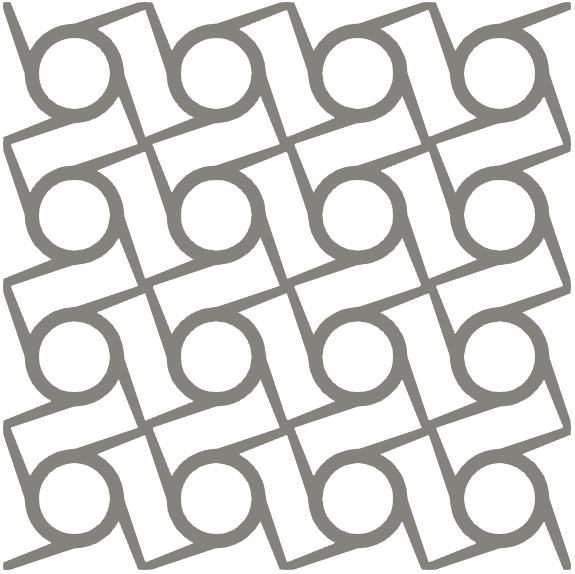}\label{fig:modesi}}
	&
	\subfloat[chiral def.]{\includegraphics[scale=0.177]{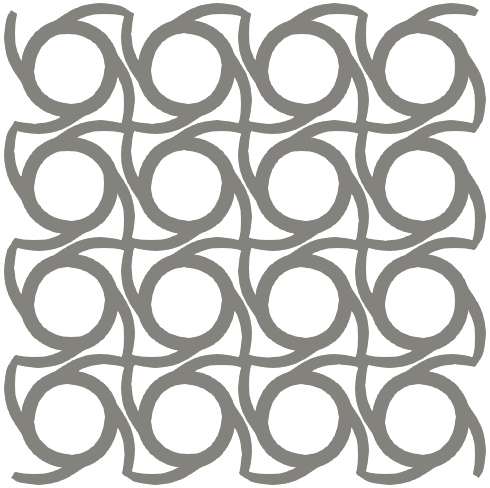}\label{fig:modesj}}
	&
	\subfloat[$S_{xx}$~\mbox{[mm$^6$]} chiral]{\includegraphics[scale=0.625]{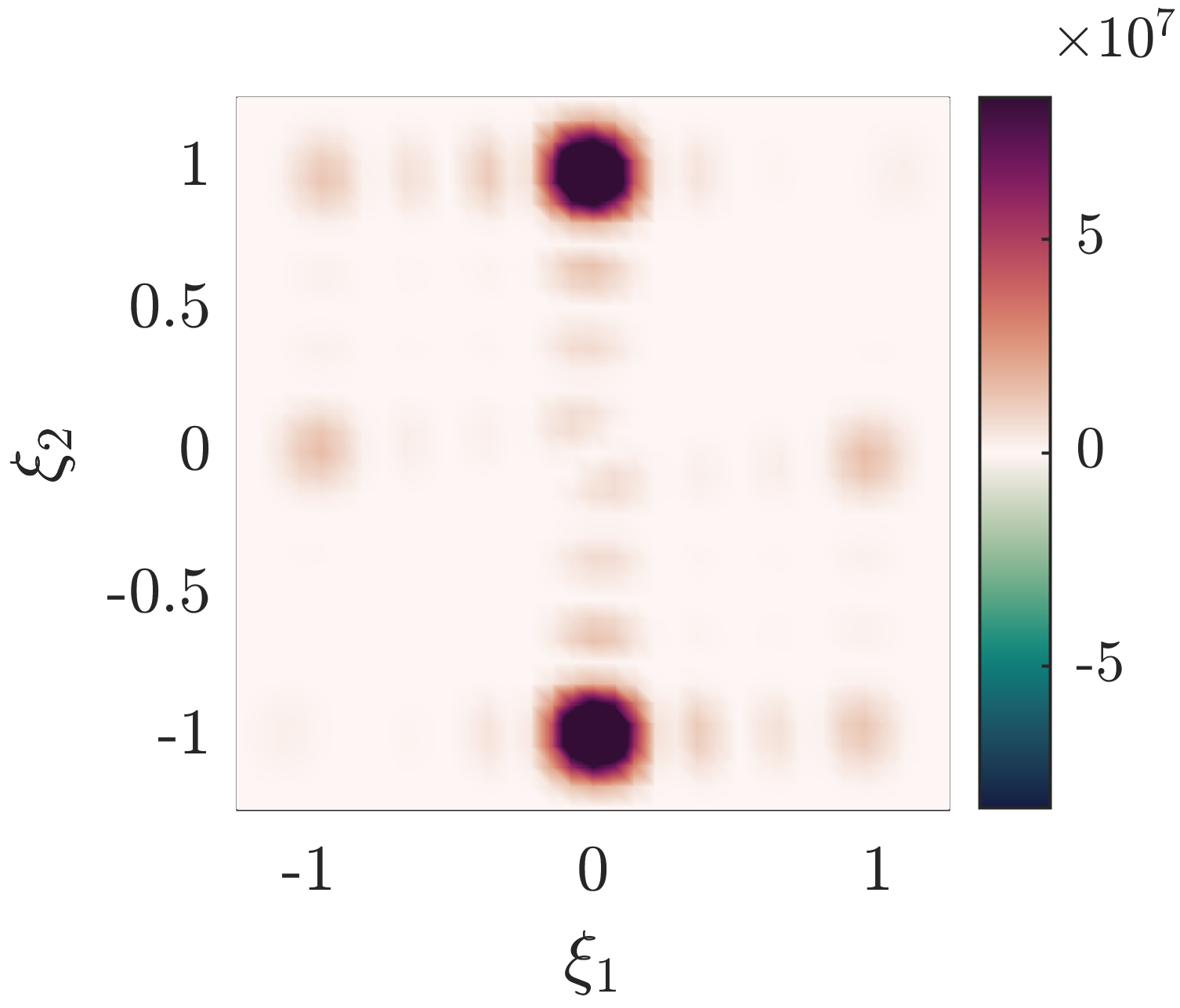}\label{fig:modesk}}
	&
	\subfloat[$S_{yy}$~\mbox{[mm$^6$]} chiral]{\includegraphics[scale=0.625]{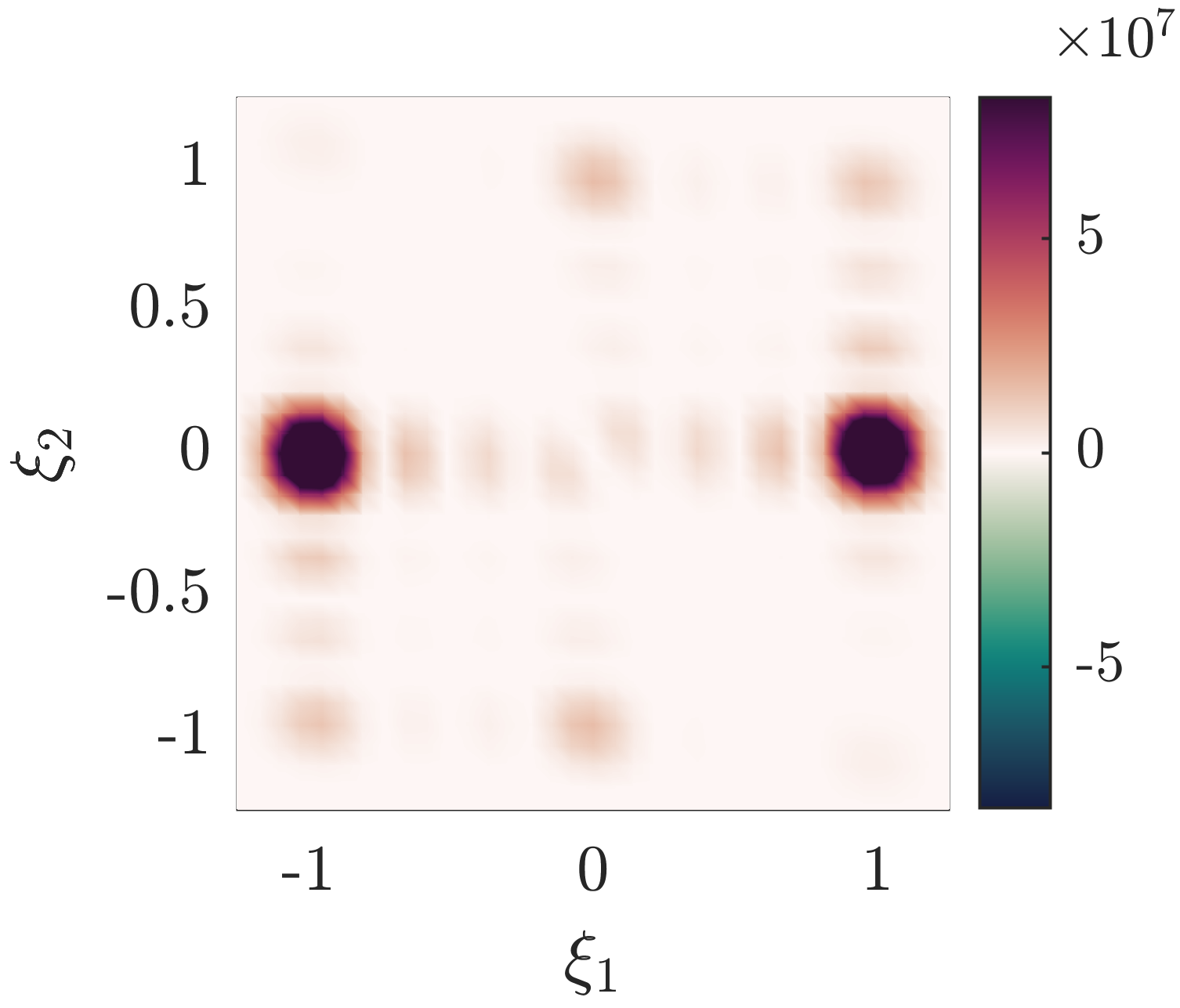}\label{fig:modesl}}
	\end{tabular}
	\caption{Reference configuration, one particular deformed configuration, and spectral densities~$S_{xx}(\bs{\xi})$ and~$S_{yy}(\bs{\xi})$, normalized with respect to the unit cell size~$l$, expressed as a function of the wave vector~$\bs{\xi} = [\xi_1,\xi_2]$, corresponding to a square stacking of holes \protect\subref{fig:modesa}--\protect\subref{fig:modesd}, hexagonal stacking of holes \protect\subref{fig:modese}--\protect\subref{fig:modesh}, and a chiral microstructure \protect\subref{fig:modesi}--\protect\subref{fig:modesl}. The most of the pattern energies localize around a few particular wave vectors, allowing in all cases for effective approximation through a truncated Fourier series.}
	\label{fig:modes}
\end{figure}

In the underlying deformed or buckled microstructure, corresponding to an infinitely large specimen or a periodic arrangement of unit cells, typically a long-range correlated pattern emerges, implying non-local behaviour. Such an order is a periodically repeating pattern which can well be represented with a few lowest-frequency terms of a truncated Fourier series. To show this, we perform a spectral density analysis~\citep{Grigoriu} on the reference displacement field, available from numerical simulations obtained for infinite periodic specimens and for which the affine part of the deformation has been subtracted. Note that in real experiments the displacement fields will also be available from regular DIC, as shown below in Section~\ref{sec5:real}. Three typical metamaterial microstructures of square and hexagonal stacking of holes and a chiral metamaterial are considered. All three geometries are depicted in the first column of Fig.~\ref{fig:modes}, their corresponding deformed states for various types of compression (uniaxial or biaxial) in the second column, whereas spectral densities in the last two columns. The spectral densities are the energy spectral densities of the displacement components in the horizontal and vertical directions, $S_{xx} = \hat{u}_x(\mathbf{\xi}) \hat{u}_x^*(\mathbf{\xi})$ and $S_{yy} = \hat{u}_y(\mathbf{\xi}) \hat{u}_y^*(\mathbf{\xi})$, where $u_x$ and $u_y$ are the $x$- and $y$-components of $\mathbf{u}(\x)$, the hat indicates the Fourier transformation, and $*$ complex conjugate. Inspection of spectral density graphs indicates that most of the energy is localized only around a few particular wave vectors, hence allowing to make a specific choice on the parametrization and initialization of the long-range correlated fluctuation field, see, e.g., \cite{Rokos2019} and Eq.~\eqref{eq5:mode} below, and this parametrization plus initialization is then applicable to a broad range of metamaterials. Note that in the spectral density graphs there are multiple secondary peaks present, corresponding to the microfluctuation~$\mufluc$, which are (almost) invisible in the images by the naked eye due to their small magnitudes.
%
%
\subsection{Virtual Experiments}
\label{sec5:virt}
\begin{figure}
	\centering
	\includegraphics[height=.3\textwidth]{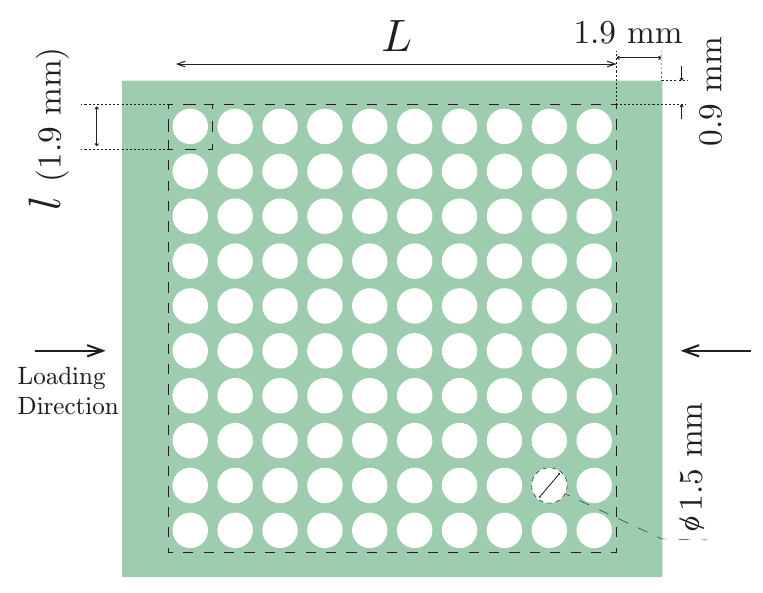}
	\caption{Analysed cellular elastomeric metamaterial specimen, depicting the total length of specimen excluding the bulk side edges ($L$: the large dashed square), the size of the unit cells ($l=1.9$~mm: small dashed square), hole diameter ($d = 1.5$~mm), the size of the bulk side edges and the loading direction.}
	\label{fig5:sample_sketch}
\end{figure}

\begin{figure}
\centering
\subfloat[Deformed virtual image]{\includegraphics[trim=100 150 100 120,clip,height=.3\textwidth]{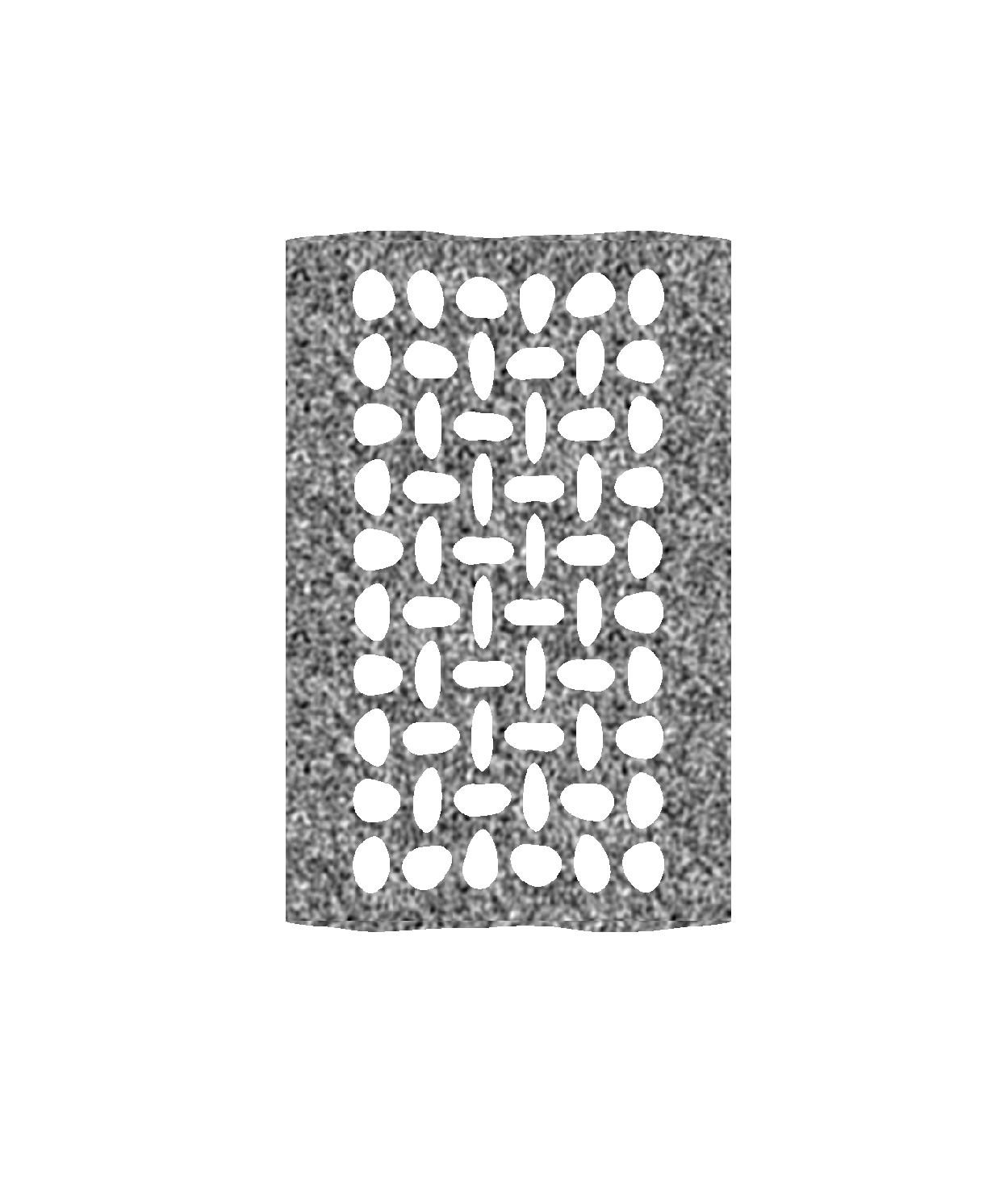}\label{fig5:def_virt}}
\subfloat[$S_{xx}$~\mbox{[mm$^6$]}]{\includegraphics[trim=0 0 0 0,clip,height=.3\textwidth]{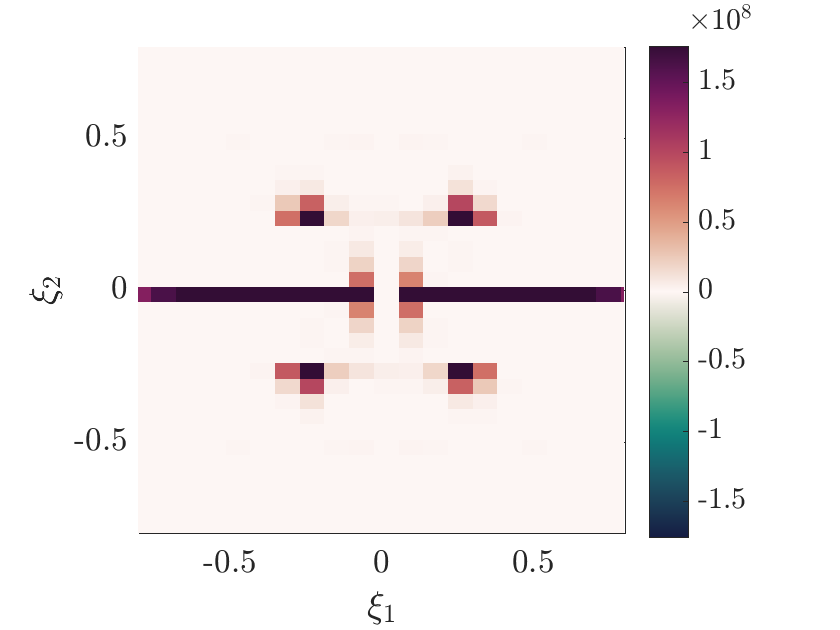}\label{fig5:sxx_virt}}
\subfloat[$S_{yy}$~\mbox{[mm$^6$]}]{\includegraphics[trim=0 0 0 0,clip,height=.3\textwidth]{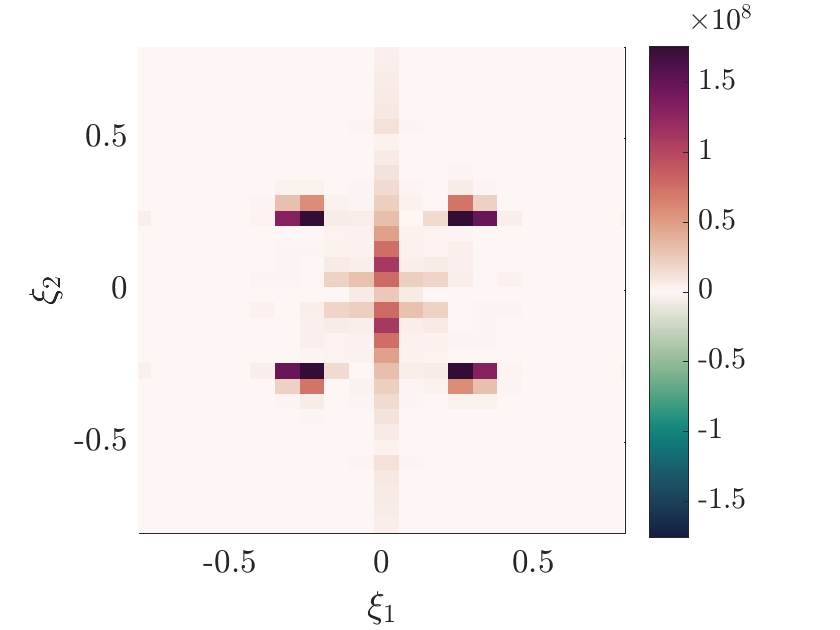}\label{fig5:syy_virt}}
\caption{\protect\subref{fig5:def_virt} Virtually generated and deformed speckle pattern, corresponding to compression in horizontal direction after the onset of pattern transformation in a cellular elastomeric metamaterial, revealing the presence of one fluctuation mode inside the specimen. Energy spectral density functions of the corresponding displacement field in \protect\subref{fig5:sxx_virt} $x$ and \protect\subref{fig5:syy_virt} $y$ direction, revealing the principal wave vectors related to the long-range correlated fluctuation mode (four peaks approximately on the diagonals). The range of the colour bars is reduced to highlight the frequencies related to the long-range correlated fluctuation mode.
}\label{fig5:spect_virt}
\end{figure}

The micromorphic IDIC scheme will be first tested on virtually generated and deformed speckle images. By this means the method is tested in a case where there is no influence of image distortions and noise in the images, and there are no imperfections in the specimen and the loading, etc. The geometry of the specimen analysed virtually and later experimentally (although for a different scale ratio) is depicted in Fig.~\ref{fig5:sample_sketch}. It has hole diameters $d = 1.5$~mm and centre to centre pitch $l = 1.9$~mm (i.e., unit cell size $l$) with edges of 1.9~mm and 0.9~mm bulk material in loading and transverse directions. The length of the specimen, excluding the bulk edges, is denoted $L$. The scale ratio of the specimen, $L/l$, is equal to the number of holes in the loading direction. Details on the specimen fabrication are discussed thoroughly in the PhD thesis of \cite{SiavashPhD}. To obtain virtual data, a plane strain finite element simulation, using quadratic isoparametric triangular elements with three-point Gauss integration rule and large-strain Total Lagrangian formulation with stability control and bifurcation, is performed for the compression of a cellular elastomeric metamaterial specimen with scale ratio of $6$, i.e., $6$ and $10$ holes in loading and transverse directions, respectively. Dirichlet boundary conditions are applied on the two vertical edges perpendicular to the loading direction to apply overall compression to the specimen, while the two remaining edges are free. Upon reaching the critical overall strain, the imposed deformation triggers an instability in the underlying microstructure leading to patterning. The material, i.e., PDMS, is modelled by a hyper-elastic Ogden material model. The material considered, although rubber-like, is of sufficiently high compressibility (i.e., the initial Poisson's ratio $< 0.5$) to avoid any numerical problems with locking of elements. The compressibility of the material has been verified numerically as well as experimentally. More details on the numerical simulations are presented in PhD thesis \citep{SiavashPhD}. The displacement fields obtained from this simulation are used to deform a speckle pattern for two different time steps, one before microstructural buckling and the other after the emergence of the patterned fluctuation field, corresponding to~$7.8\%$ and~$12.8\%$ applied nominal strain. A total formulation within the context of IDIC was employed, in which two snapshots in a deformed state (pre-bifurcation and post-bifurcation images) have been correlated with the reference undeformed image, although an updated formulation correlating incremental states is equally possible as well. To generate deformed images, pixel intensities of the reference image are mapped by displacements, interpolated at nodal or pixel positions, by inverting the iso-parametric mappings of the underlying FE approximations. This step is performed to avoid any potential additional biases due to interpolation errors. Fig.~\ref{fig5:def_virt} depicts the deformed configuration of the virtual speckle pattern after the buckling-induced pattern emerged, which exhibits the pattern transformation more pronounced in the centre and restricted at the edges, specifically the vertical edges. The employed speckle pattern is obtained as a thresholded two-dimensional scalar and stationary random Gaussian field, generated according to~\cite[][Section~5.3.1.2]{Grigoriu}. The spectral density is chosen as a radially-symmetric normal probability density function with the mean zero and standard deviation~$1/20$. No additive image white noise has been added to such generated images. The pattern is blurred using Gaussian filter with the standard deviation of $3$~pixels, resulting in the pattern shown in Fig.~\ref{fig5:def_virt}.

As the virtual images are free of spatial distortions, $\map_S(\x)=\x$ is set for the virtual experiments. Chebyshev polynomials of $5$th and $6$th order are used to parametrize $\smooth_0$ and $v_1$, respectively.
In order to attain a proper initial guess for $\mode_i(\x,\dofs_{\varphi_i})$, it is first realized that only one long-range correlated fluctuation mode is triggered in the specimen, as can be easily seen in Fig.~\ref{fig5:def_virt}, thus $n=1$ is set in Eq.~\eqref{eq5:ansatz} with only $\mode_1$ considered. It may also be clear that the triggered mode is periodic with a periodicity of two unit cells, recall Figs.~\ref{fig:modesc}--\protect\subref*{fig:modesd} where the peaks are localized at~$\xi_1 = \xi_2 = \pm 1/2$. The horizontal and vertical lines in Figs.~\ref{fig5:sxx_virt} and \ref{fig5:syy_virt} (not present in Figs.~\ref{fig:modesc}--\subref*{fig:modesd}) correspond to the discontinuities in the mean deformation due to the edges of the specimen, while the four peaks correspond to the correlated fluctuation mode, revealing two sine functions roughly in diagonal directions of the Cartesian coordinate system for the $x$ and $y$ components of $\mode_1$. The purpose of the spectral density analysis is merely the rough identification of the type of long-range correlated fluctuation modes, so only the corresponding four peaks are considered here. Inspection of $\hat{u}_x(\mathbf{\xi})$ and $\hat{u}_y(\mathbf{\xi})$, at the frequencies corresponding to the four peaks of $S_{xx}$ and $S_{yy}$, suggests that the two sine waves are of the same sign in the $x$ component, and of the opposite sign in the $y$ component of the displacement field. This inspection also suggests that the two sine waves are of the same amplitude, for both $x$ and $y$ components of the displacement in agreement with the symmetry of the problem. Based on these observations, $\mode_1$ is parametrized as the sum of two sine functions for each component, i.e.,
\begin{equation}
\begin{aligned}
\mode_1(\x,\dofs_{\varphi_1}) &= \dof_1 \left[ - \sin \frac{\pi}{l}(\dof_2 x + \dof_3 y + \dof_6) - \sin \frac{\pi}{l}(\dof_4 x + \dof_5 y + \dof_7) \right]\mathbf{e}_x \\ 
& \:\quad + \left[ \sin \frac{\pi}{l}(\dof_2 x + \dof_3 y + \dof_6) - \sin \frac{\pi}{l}(\dof_4 x + \dof_5 y + \dof_7) \right]\mathbf{e}_y, 
\end{aligned}
\label{eq5:mode}
\end{equation}
where $\dofs_{\varphi_1} = [\dof_1,\dof_2,\dof_3,\dof_4,\dof_5,\dof_6,\dof_7]^{\mathsf{T}}$ is the column of degrees of freedom describing the mode. Note that the direction of both sine waves and their wavelengths are free and independent, which compensates for inaccuracies in the measurement of the unit cell size due to processing uncertainties. This allows for an accurate identification of the shape and magnitude of the fluctuation mode in the micromorphic IDIC analysis, whereas the spectral density analysis is only used for a proper parametrization of the fluctuation mode and for an approximate initial guess of all these parameters. Note also that it is sufficient to perform the spectral density analysis only once for a family of cellular metamaterials triggering the same pattern. In cases where more than one mode is activated, the spectral analysis should be done separately on each of the subregions containing the individual modes, cf. Section~\ref{sec5:modes}.

The dofs describing $\mode_1$ are initialized as: $\dofs_{\varphi_1} = [1,1,1,1,1,0,0]^{\mathsf{T}}$. The initial values for $\dof_2$ to $\dof_5$ are set to a wavelength of $2$ unit cells and their signs and ratios are set such that the initial orientation of the two sine waves is aligned with the two diagonals of the coordinate system. Note that $\dof_1$ is only defining the ratio of the amplitude of the mode in $x$ and $y$ direction, while $v_1(\x)$ defines its absolute amplitude according to Eq.~\eqref{eq5:u_dic}. In order to assure a proper scaling of the minimization problem, $\mode_1(\x)$ is scaled such that the initial maximum value of the mode is $0.04$. Normalization of the microfluctuation field~$\boldsymbol{\varphi}_i$ is an inherent choice of the employed micromorphic computational homogenization formulation. We refer to~\cite[][Eq.~(3)]{Rokos2019b}, for more details. The scaling $\max_\mathbf{x}\| \mode_i(\x) \| = 0.04$ is chosen to avoid poor conditioning of the resulting Hessian matrix during the solution procedure using the Gauss--Newton algorithm. Moreover, since the sensitivity functions of the mode $\mode_1$ used in the Gauss--Newton optimization algorithm \citep{Neggers2016} are a function of $v_1(\x)$, the initial guess for the latter needs to be non-zero to avoid an ill-posed problem in the first iteration. To this end, coefficients of linear combination of the Chebyshev polynomials corresponding to the $v_1(\x)$ field are initiated such that a small constant non-zero spatial field results. The applied compressive global strain results in large displacement on the edges of the specimen requiring an approximate initial guess in order to ensure convergence. Thus, the smooth mean field $\smooth_0(\x)$ is initialized such that the first order term in $x$ direction approximately accounts for the applied global strain. The IDIC procedure overall identifies~$77$ dofs, out of which~$2 \cdot 21 = 42$ correspond to coefficients of linear combination~$\dofs_{v_0}$ of the Chebyshev polynomials of 5th order describing the field~$\smooth_0(\x)$, $28$ to coefficients of linear combination~$\dofs_{v_1}$ of the Chebyshev polynomials of 6th order describing the field~$v_1(\x)$, and~$7$ corresponding to the degrees of freedom~$\dofs_{\varphi_1}$ of the long-range correlated fluctuation field~$\mode_1(\x)$ specified in Eq.~\eqref{eq5:mode}. This is a negligible amount of dofs compared to local DIC which needs at least tens of thousands dofs to capture such locally-varying displacement fields.

\begin{figure}
\captionsetup[subfloat]{captionskip=-1pt,nearskip=-1pt,farskip=-1pt}
\begin{minipage}{.5\textwidth} %
\centering
Before microstructural buckling\\
\subfloat[$\smooth_0.\mathbf{e}_x$~\mbox{[$\mu$m]}]{\includegraphics[trim=0 0 0 0,clip,height =.4\textwidth]{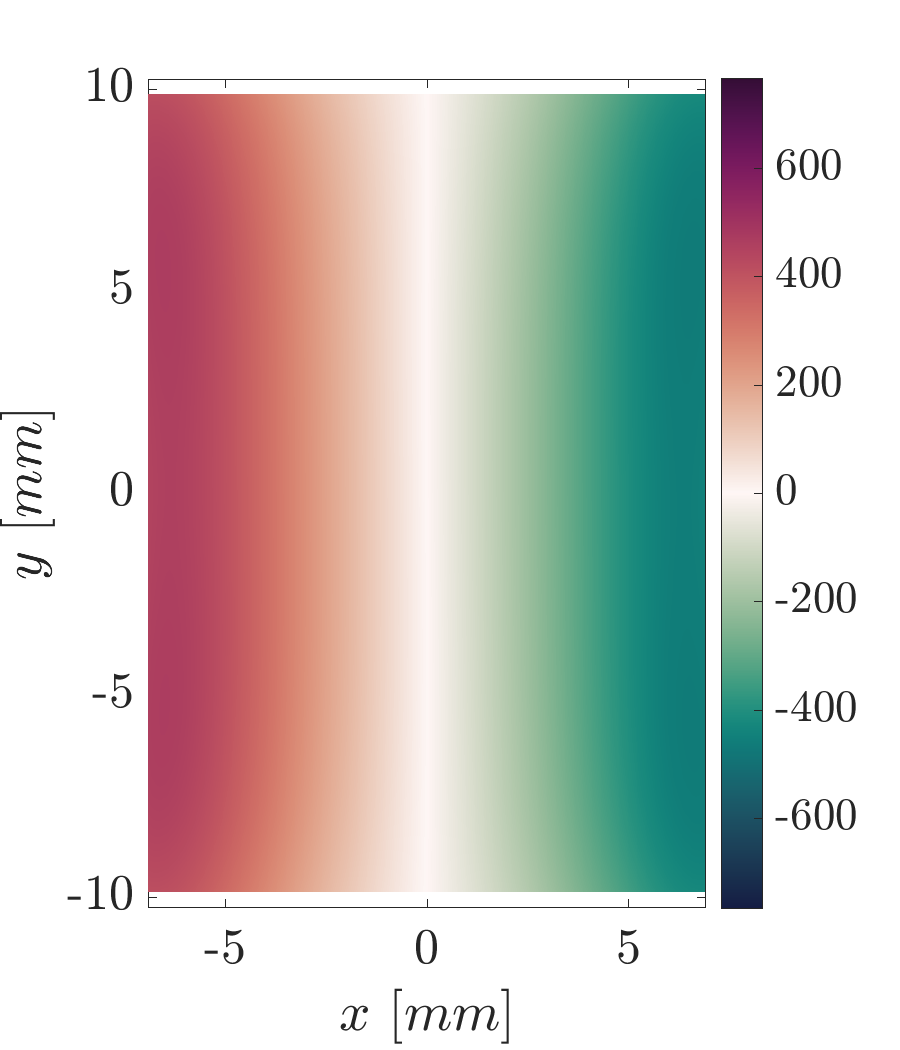}\label{fig5:virt1v0x}}\hspace{1.0em}
\subfloat[$\smooth_0.\mathbf{e}_y$~\mbox{[$\mu$m]}]{\includegraphics[trim=0 0 0 0,clip,height =.4\textwidth]{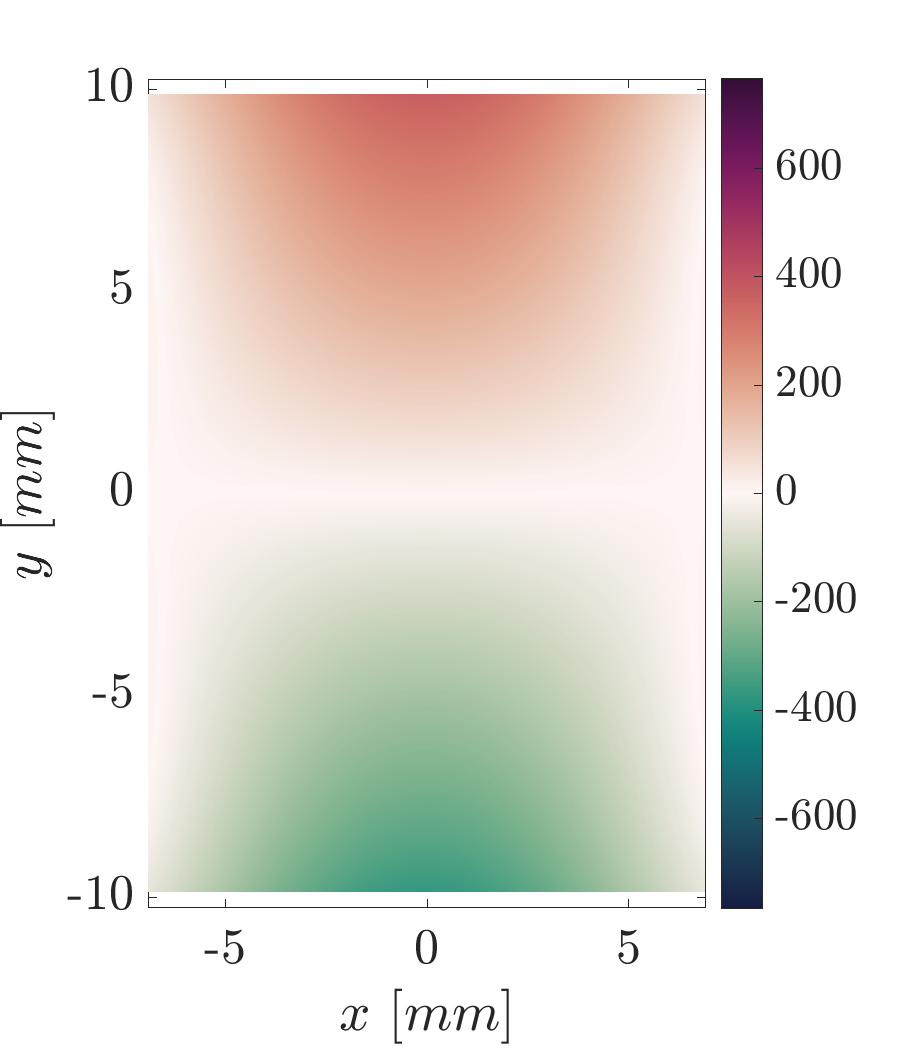}\label{fig5:virt1v0y}}\\
\subfloat[$v_1$~\mbox{[$\mu$m]}]{\includegraphics[trim=0 0 0 0,clip,height =.4\textwidth]{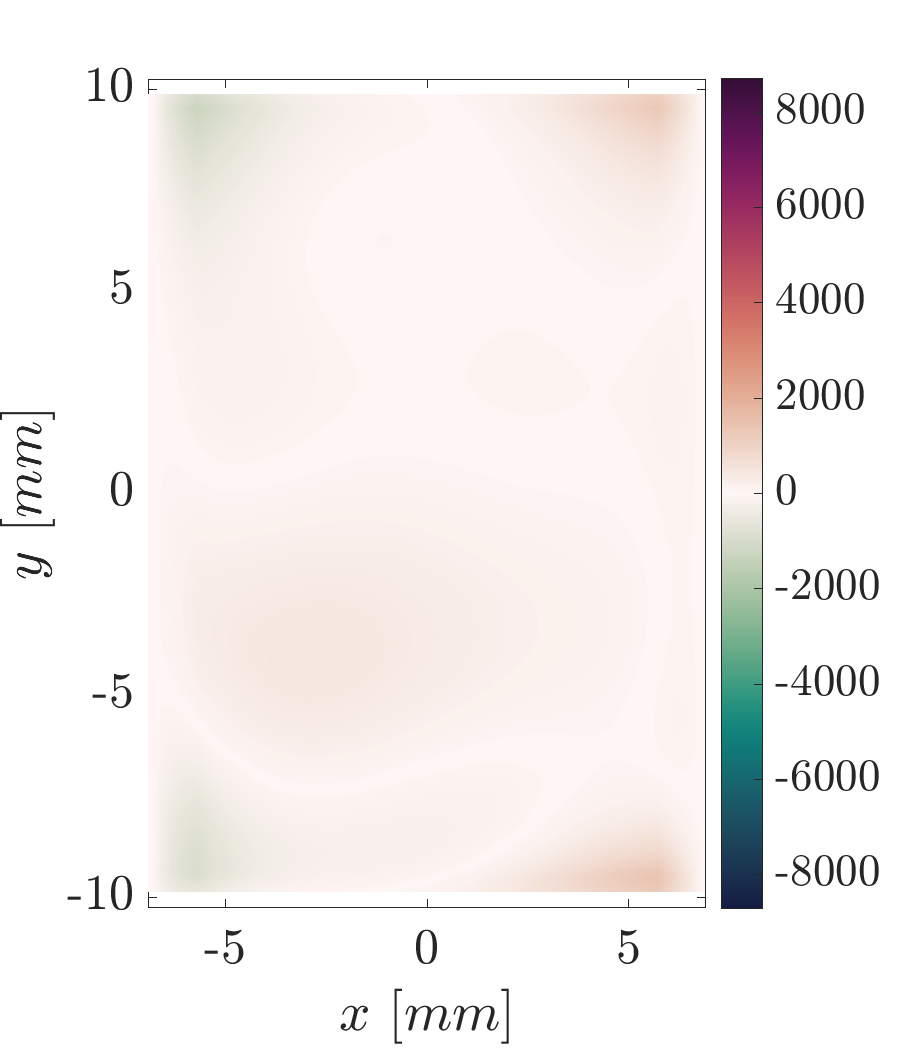}\label{fig5:virt1v1}}\hspace{1.0em}
\subfloat[$\mode_1~{[-]}$]{\includegraphics[trim=0 0 0 0,clip,height =.4\textwidth]{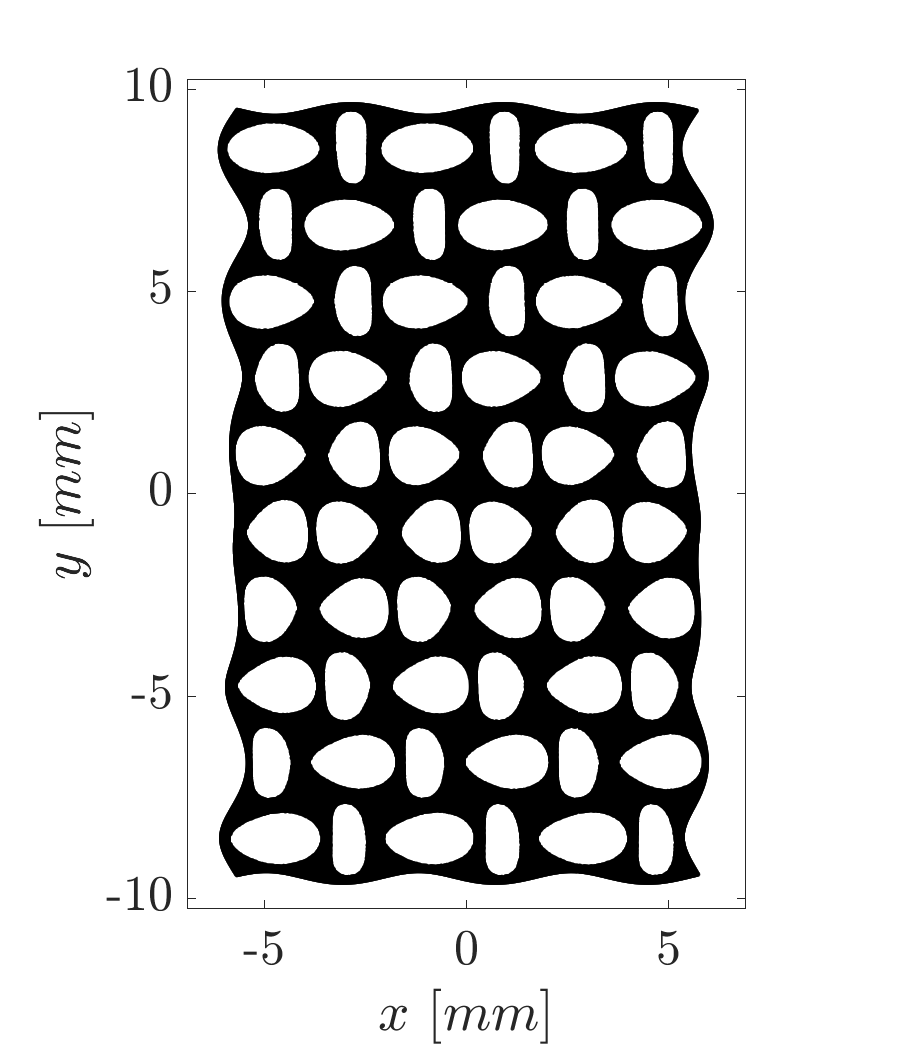}\label{fig5:virt1phi}}\\
\subfloat[$v_1\mode_1.\mathbf{e}_x$~\mbox{[$\mu$m]}]{\includegraphics[trim=0 0 0 0,clip,height =.4\textwidth]{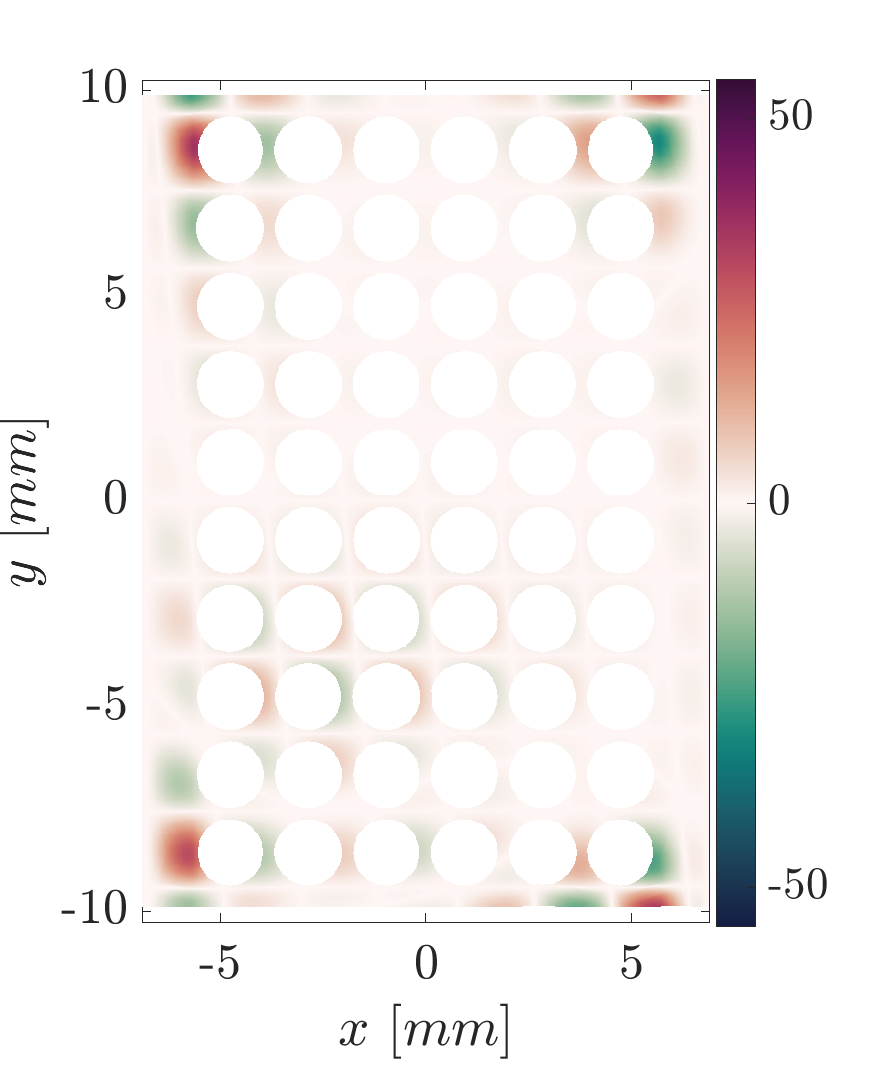}\label{fig5:virt1flucx}}\hspace{1.0em}
\subfloat[$v_1\mode_1.\mathbf{e}_y$~\mbox{[$\mu$m]}]{\includegraphics[trim=0 0 0 0,clip,height =.4\textwidth]{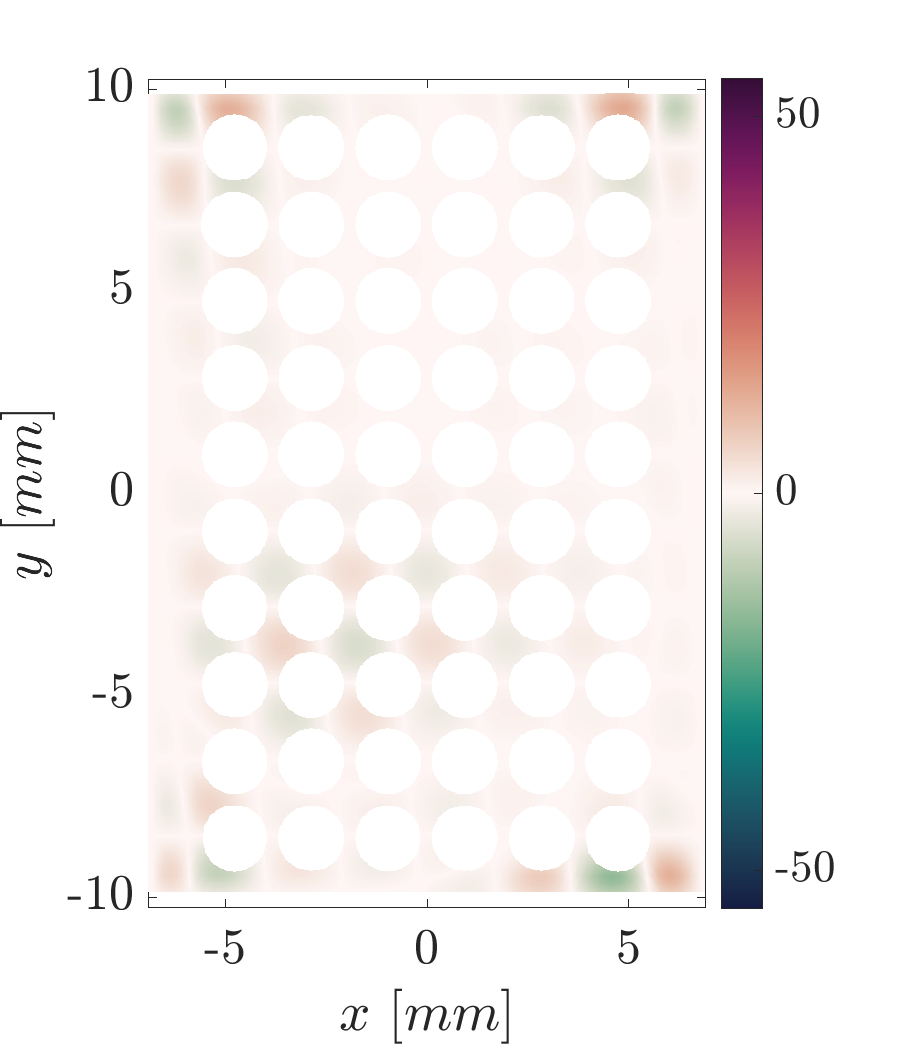}\label{fig5:virt1flucy}}\\
\subfloat[residual~{[--]}]{\includegraphics[trim=0 0 0 0,clip,height =.4\textwidth]{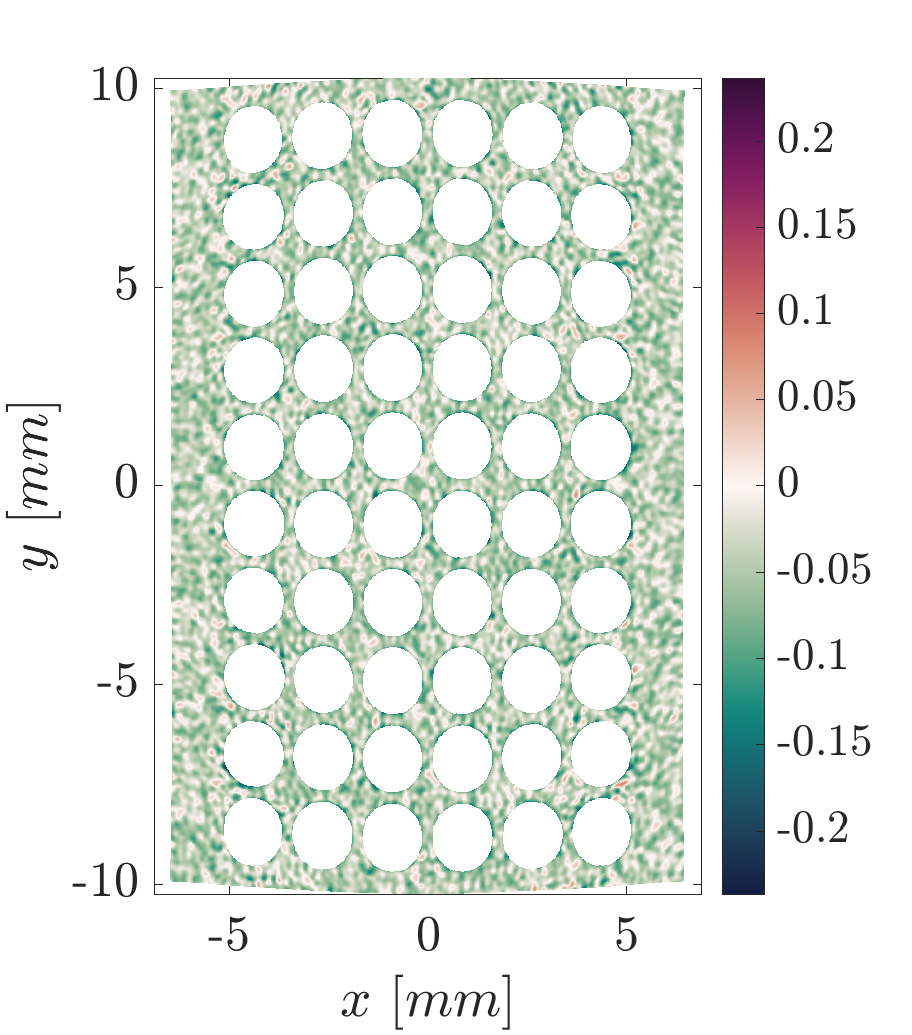}\label{fig5:virt1r}}\hspace{1.0em}
\subfloat[$\vert \vert \mufluc \vert \vert$~\mbox{[$\mu$m]}]{\includegraphics[trim=0 0 0 0,clip,height =.4\textwidth]{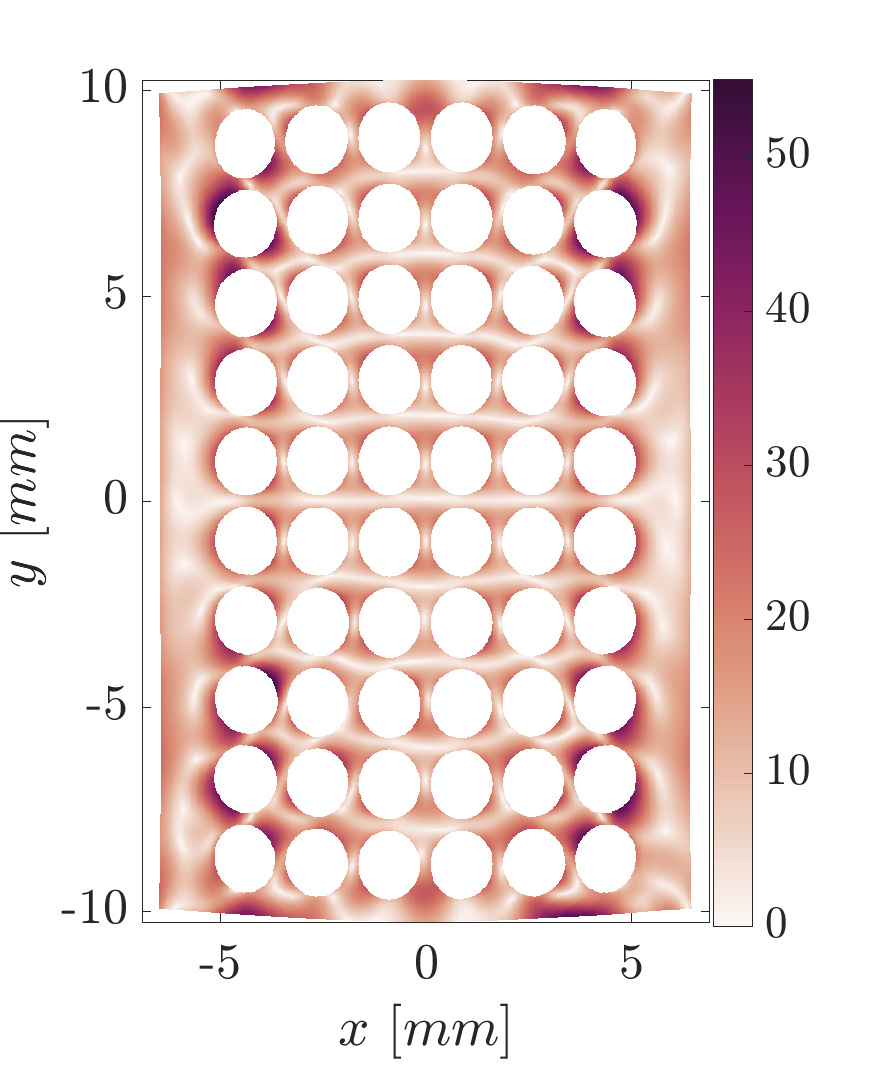}\label{fig5:virt1w}}\\
\end{minipage} %
\vline
\begin{minipage}{.5\textwidth} %
\centering
After microstructural buckling\\
\subfloat[$\smooth_0.\mathbf{e}_x$~\mbox{[$\mu$m]}]{\includegraphics[trim=0 0 0 0,clip,height=.4\textwidth]{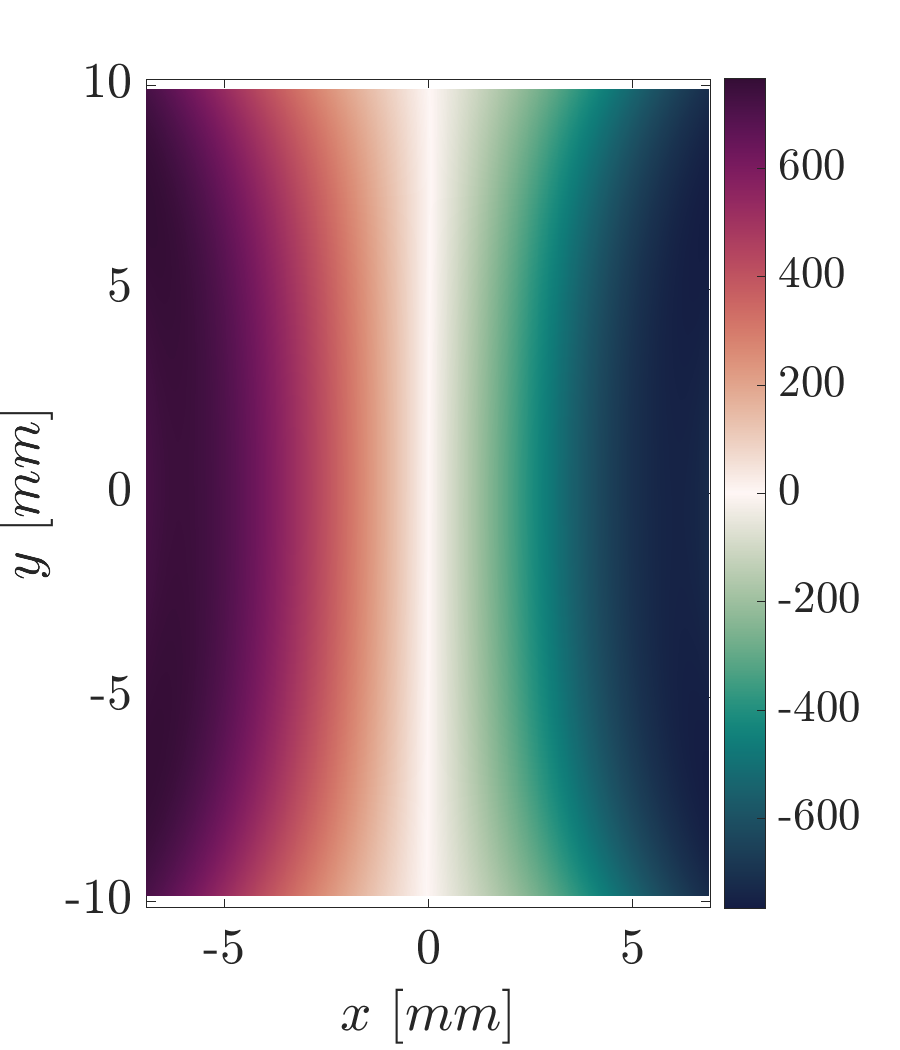}\label{fig5:virt2v0x}}\hspace{1.0em}
\subfloat[$\smooth_0.\mathbf{e}_y$~\mbox{[$\mu$m]}]{\includegraphics[trim=0 0 0 0,clip,height=.4\textwidth]{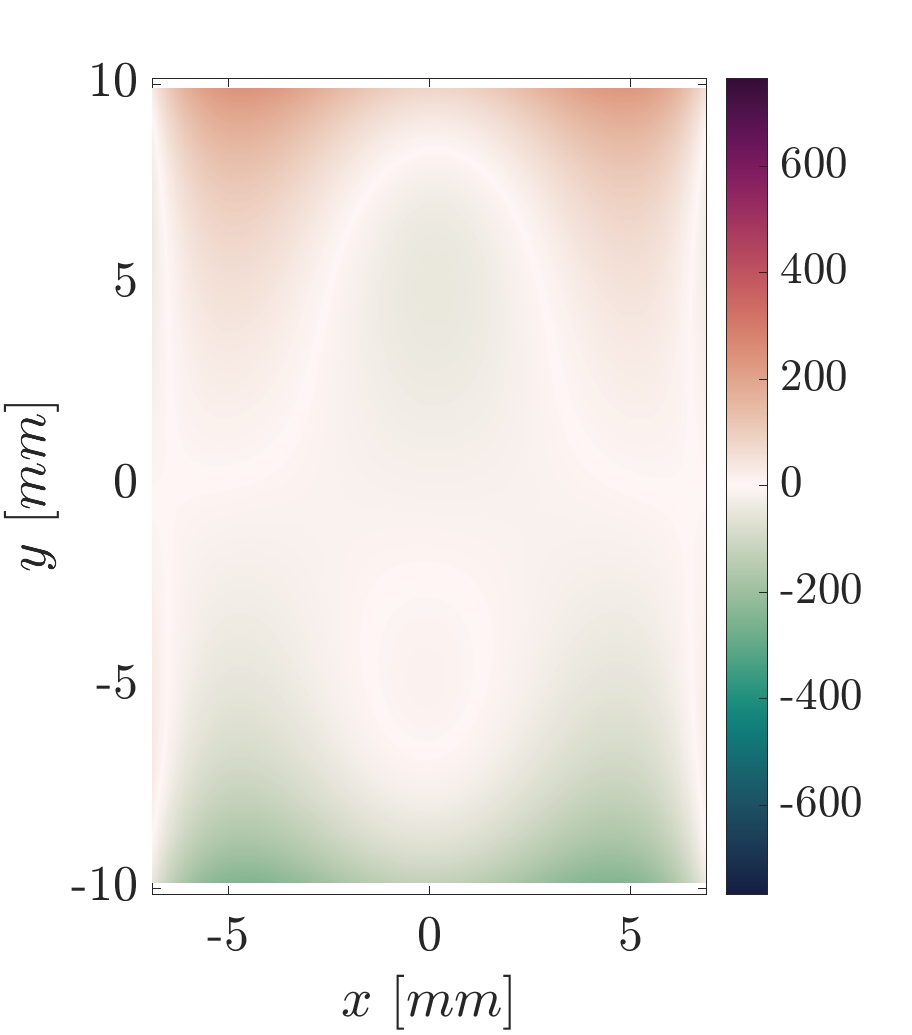}\label{fig5:virt2v0y}}\\
\subfloat[$v_1$~\mbox{[$\mu$m]}]{\includegraphics[trim=0 0 0 0,clip,height=.4\textwidth]{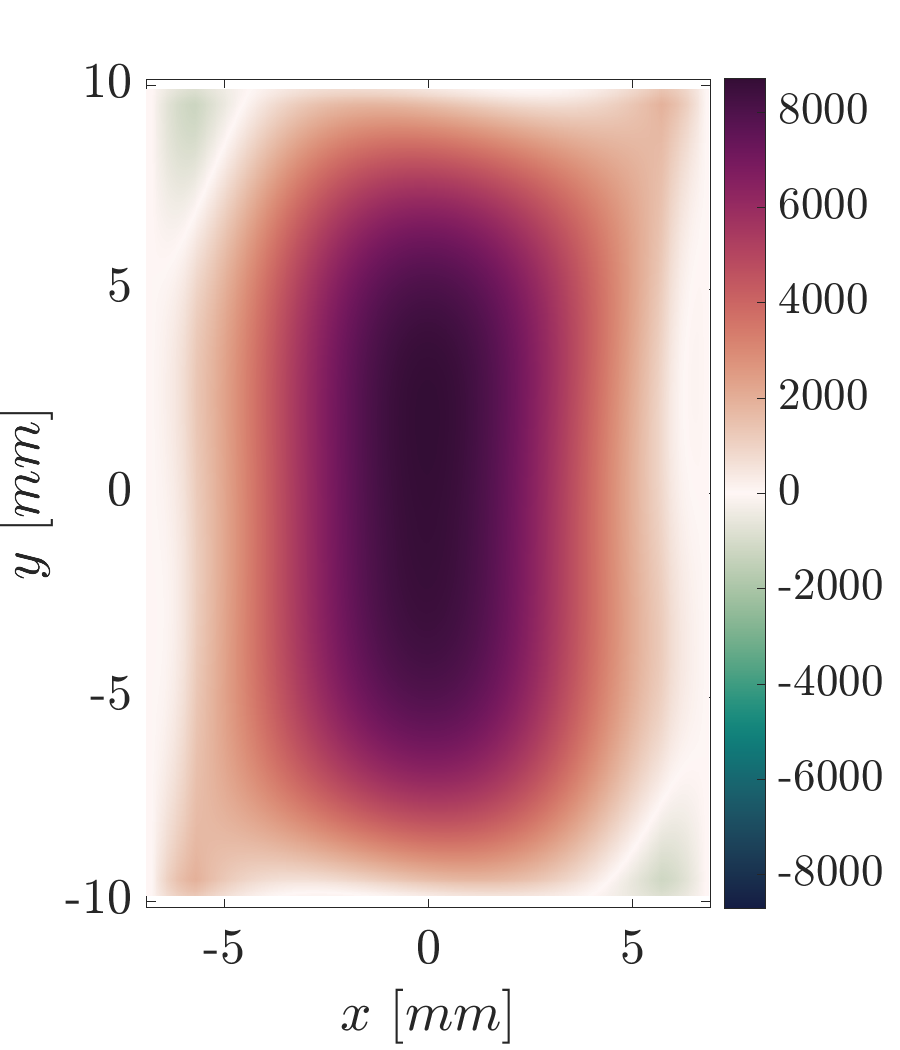}\label{fig5:virt2v1}}\hspace{1.0em}
\subfloat[$\mode_1~{[-]}$]{\includegraphics[trim=0 0 0 0,clip,height=.4\textwidth]{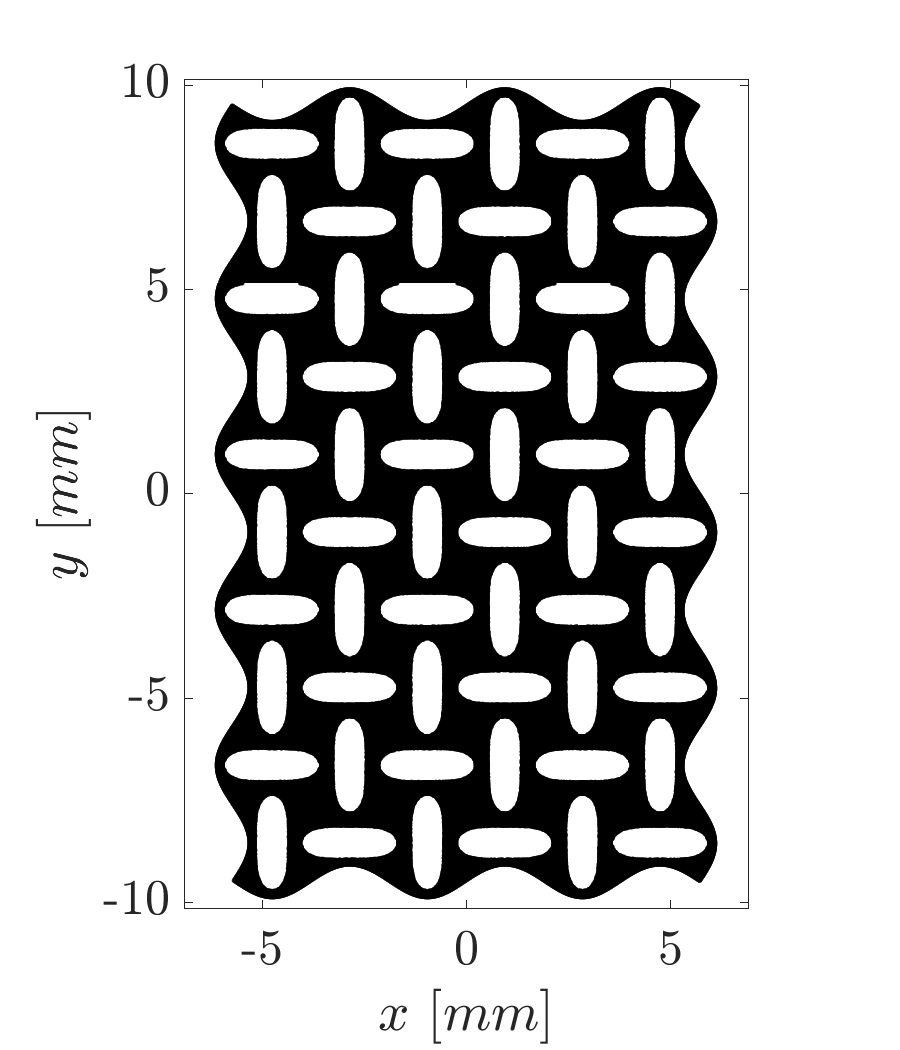}\label{fig5:virt2phi}}\\
\subfloat[$v_1\mode_1.\mathbf{e}_x$~\mbox{[$\mu$m]}]{\includegraphics[trim=0 0 0 0,clip,height=.4\textwidth]{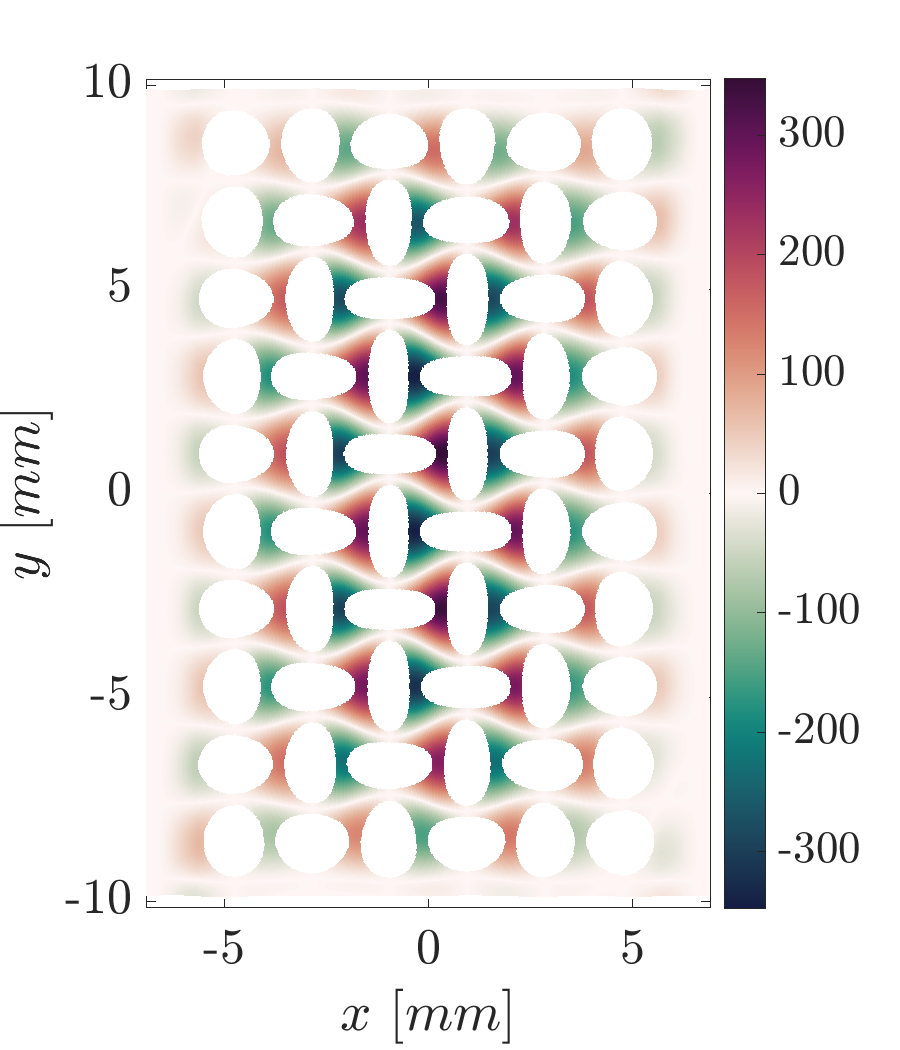}\label{fig5:virt2flucx}}\hspace{1.0em}
\subfloat[$v_1\mode_1.\mathbf{e}_y$~\mbox{[$\mu$m]}]{\includegraphics[trim=0 0 0 0,clip,height=.4\textwidth]{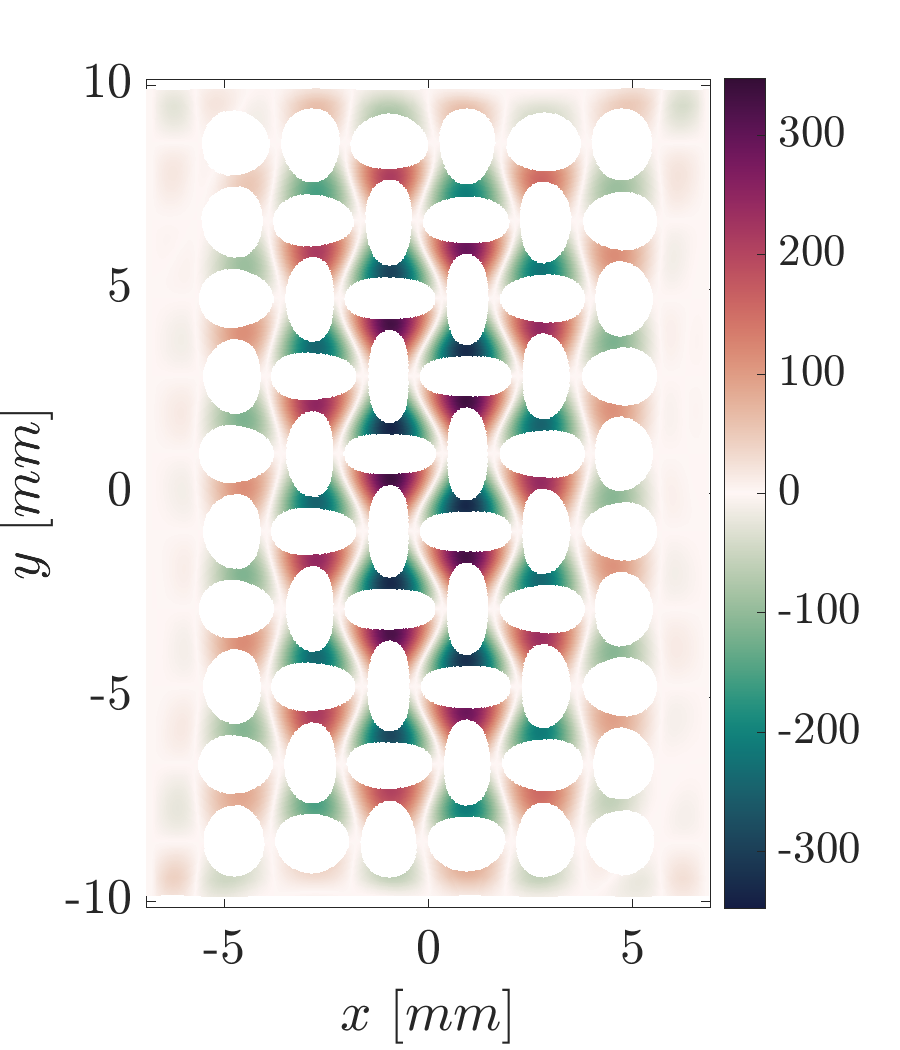}\label{fig5:virt2flucy}}\\
\subfloat[residual~{[--]}]{\includegraphics[trim=0 0 0 0,clip,height=.4\textwidth]{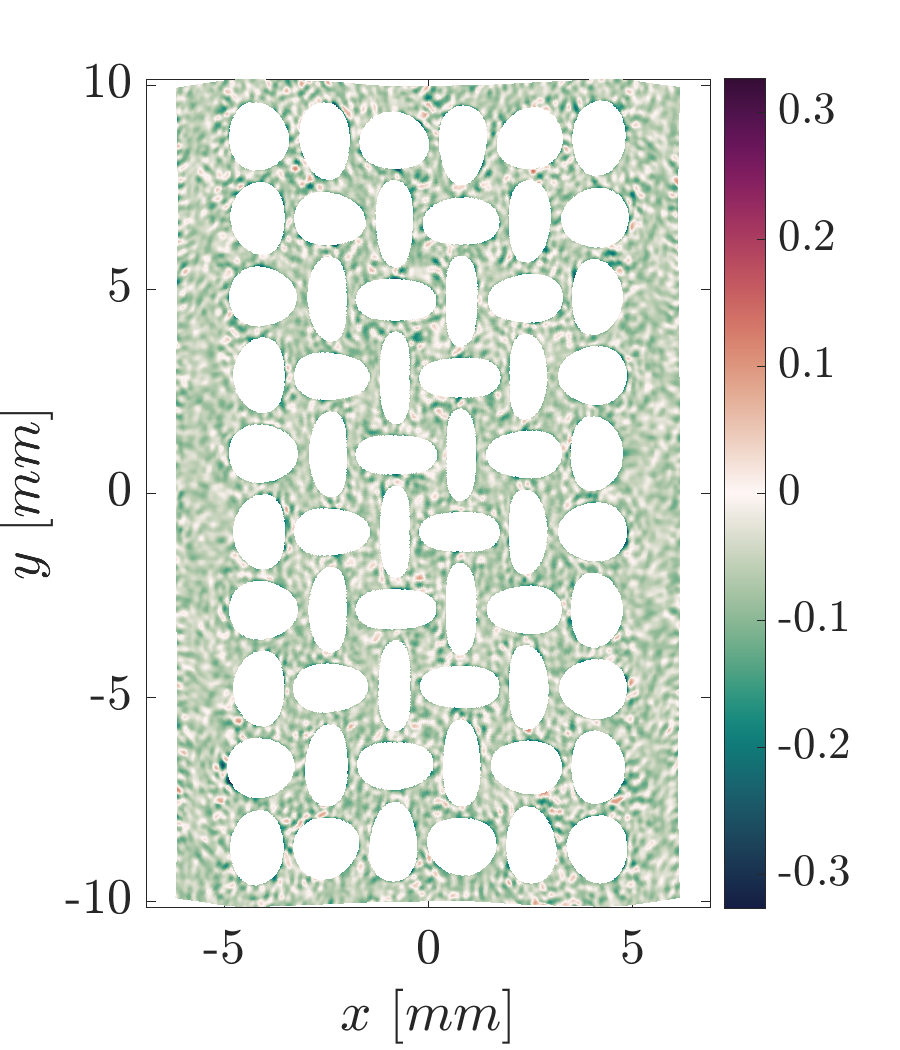}\label{fig5:virt2r}}\hspace{1.0em}
\subfloat[$\vert \vert \mufluc \vert \vert$~\mbox{[$\mu$m]}]{\includegraphics[trim=0 0 0 0,clip,height=.4\textwidth]{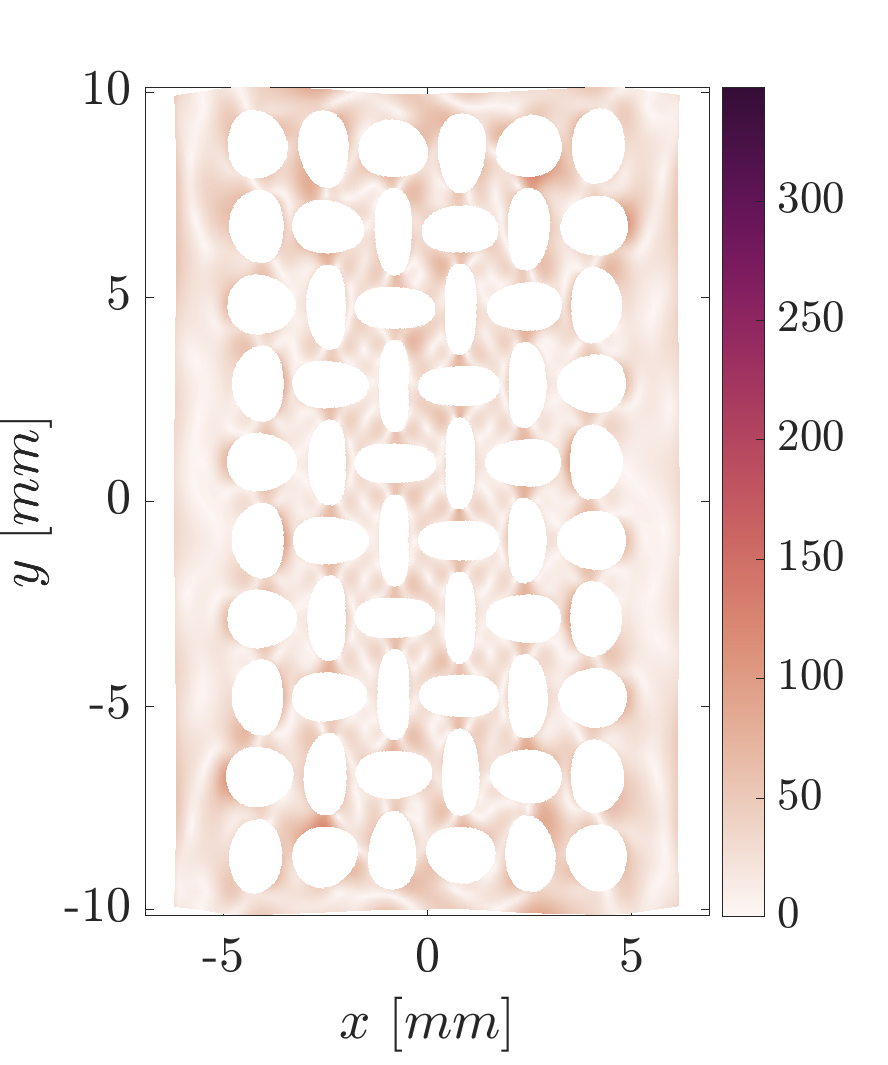}\label{fig5:virt2w}}
\end{minipage}
\caption{Results of the virtual experiments before (left) and after (right) the onset of microstructural buckling: (\protect\subref*{fig5:virt1v0x}, \protect\subref*{fig5:virt1v0y}, \protect\subref*{fig5:virt2v0x}, \protect\subref*{fig5:virt2v0y}) the smooth mean displacement fields, $\smooth_0(\x)$, in (\protect\subref*{fig5:virt1v0x}, \protect\subref*{fig5:virt2v0x}) $x$ and (\protect\subref*{fig5:virt1v0y}, \protect\subref*{fig5:virt2v0y}) $y$ directions; (\protect\subref*{fig5:virt1v1}, \protect\subref*{fig5:virt2v1}) the pattern amplitudes, $v_1(\x)$, (\protect\subref*{fig5:virt1phi}, \protect\subref*{fig5:virt2phi}) the deformed configuration of identified patterns/modes, $\mode_1(\x)$, scaled to assure the visibility of the deformed shape, and (\protect\subref*{fig5:virt1flucx}, \protect\subref*{fig5:virt1flucy}, \protect\subref*{fig5:virt2flucx}, \protect\subref*{fig5:virt2flucy}) the resulting fields, (\protect\subref*{fig5:virt1flucx}, \protect\subref*{fig5:virt2flucx}) in $x$ and (\protect\subref*{fig5:virt1flucy}, \protect\subref*{fig5:virt2flucy}) $y$ directions, corresponding to the long-range correlated fluctuations. In (\protect\subref*{fig5:virt1phi}, \protect\subref*{fig5:virt2phi}) $\mode_1(\x)$ are scaled to assure the visibility of the deformed shape of the modes. The long-range correlated fluctuation fields, $v_1(\x)\mode_1(\x)$, are plotted in the deformed configuration corresponding to their contribution only. The residual fields in normalized gray scales are in~(\protect\subref*{fig5:virt1r}, \protect\subref*{fig5:virt2r}), and (\protect\subref*{fig5:virt1w}, \protect\subref*{fig5:virt2w}) show the Euclidean norm of the microfluctuation field, $\mufluc(\x)$, plotted in deformed configuration based on the total displacement.}
\label{fig5:virt}
\end{figure}	

Fig.~\ref{fig5:virt} shows the results of the virtual experiments before (left) and after (right) the emergence of the buckling pattern. The micromorphic IDIC results in an accurate identification of the $\smooth_0(\x)$, $v_1(\x)$ and $\mode_1(\x)$, considering the residual fields that are small everywhere both before and after buckling occurs, see Figs.~\ref{fig5:virt1r} and \ref{fig5:virt2r}. For sensible initial guesses the convergence is consistent, but the procedure may be trapped in local minima during minimization for poor initial guesses. For each case, the mean smooth displacement field, $\smooth_0(\x)$, in $x$ and $y$ directions, and $v_1(\x)$ are depicted in the first two rows, i.e., Figs.~\ref{fig5:virt1v0x}--\subref*{fig5:virt1v1} before and Figs.~\ref{fig5:virt2v0x}--\subref*{fig5:virt2v1} after buckling occurs. The $\smooth_0$ fields show compression in $x$ direction both before and after buckling (a linear profile). Before pattern transformation, lateral expansion is visible in the $y$ direction, while afterwards the auxetic effect, i.e., lateral contraction, is clearly visible near the specimen's centre. As expected, the $v_1$ field is close to zero before buckling occurs, although its absolute value increases slightly in the corners of the region with cellular microstructure (the area of the specimen excluding the bulk edges), which is due to the fact that the fluctuation mode, $\mode_1$, partially captures the deformation that occurs in the corner holes where the edge effect is considerable. After the buckling occurs, the $v_1$ field is close to zero on the edges and increases towards the centre, as expected. Again, $v_1$ is not exactly zero at the edges and the corners of the metamaterial, since the fluctuation mode partially describes the deformation in the corners where the edge effect is considerable. Note that the large values in $v_1(\x)$ are due to the scaling of the corresponding mode $\mode_1$ (maximum value of 0.04).

The identified long-range correlated mode, $\mode_1(\x)$, is depicted in Figs.~\ref{fig5:virt1phi} and \ref{fig5:virt2phi} before and after buckling, where their amplitudes are scaled to assure the visibility of the deformed shape of the modes. The small $v_1$ prior to buckling entails a low sensitivity of the mode parameters. As a result, $\mode_1$ is significantly perturbed with respect to the initial guess, see Fig.~\ref{fig5:virt1phi} where the periodicity does not match twice the unit cell size any more. After buckling, however, the mode becomes representative, and it resembles the initial guess more closely and matches the mode observed in the simulations (Fig.~\ref{fig5:virt2phi}). The long-range correlated fluctuation fields, $v_1\mode_1(\x)$ in $x$ and $y$ directions, are depicted in their deformed configurations prior to buckling in Figs.~\ref{fig5:virt1flucx} and \ref{fig5:virt1flucy}, and beyond buckling in Figs.~\ref{fig5:virt2flucx} and \ref{fig5:virt2flucy}. Prior to buckling, $v_1\mode_1(\x)$ is close to zero everywhere apart from the four corners, where the identified mode, although significantly perturbed with respect to the initial guess, partially captures the edge effect described above. Post buckling, $\mode_1$ is identified properly, resulting in small modifications to the wavelengths and orientations of the sine waves compared to the initial guess based on spectral density analysis.

The residual fields in normalized grey scales before and after buckling are shown in Figs.~\ref{fig5:virt1r} and \ref{fig5:virt2r}, in the deformed configuration based on the total evaluated displacement $\mathbf{u}(\x)$. The residual fields are consistently small everywhere over the region of interest. However, the microfluctuation fields, discarded in the micromorphic IDIC scheme which seeks to extract the correlated fields only, are by definition reflected in the residual fields. In order to evaluate the performance of the method, the microfluctuation field $\mufluc(\x)$ is recovered as the difference of the reference displacement field (here available from the simulation) and the displacement field $\mathbf{u}(\x)$ according to Eq.~\eqref{eq5:u_dic}. The Euclidean norm of the microfluctuation field, $\vert\vert \mufluc \vert\vert$, is depicted in Figs.~\ref{fig5:virt1w} and \ref{fig5:virt2w} before and after the buckling, respectively. Note that the micromorphic IDIC scheme is constrained to identify displacement fluctuations with spatial frequencies bound to the highest spatial frequencies in the long-range correlated fluctuation modes, $\mode_i$. Thus, the microfluctuation field, $\mufluc$, contains all parts of the field with spatial frequencies higher than the characteristic frequency of $\mode_1$, as well visible far from the edges of the specimen in the post buckling case, see Fig.~\ref{fig5:virt2w}. As expected, the microfluctuation field, $\mufluc$, is of a significantly smaller amplitude than the correlated fluctuation field beyond buckling (compare Fig.~\ref{fig5:virt2w} with Figs.~\ref{fig5:virt2flucx} and \ref{fig5:virt2flucy}). The opposite is true prior to buckling, due to the almost negligible contribution of the long-range correlated fluctuation field (compare Fig.~\ref{fig5:virt1w} with Figs.~\ref{fig5:virt1flucx} and \ref{fig5:virt1flucy}). Note the difference in scale bars for the figures of the pre- and post-bifurcation state.

In order to assess the robustness of the method, the correlations of the virtual experiments, after the onset of buckling, are repeated with different initial guess values for the fluctuation mode. The values of $\dofs_{\varphi_i}$ are perturbed by a random value up to~$10\%$ of the nominal value stated above, resulting in random changes in the wavelengths and orientations of the sine waves as well as the ratio of the $x$ and $y$ components of $\mode_1$. The perturbed correlations are repeated 100 times, and the error in each case is evaluated as: $\epsilon = 100\times\frac{\vert\vert \mathbf{u} \vert\vert - \vert\vert \mathbf{u}_{pert} \vert\vert}{\vert\vert \mathbf{u} \vert\vert}$, where $\mathbf{u}$ is the total displacement field evaluated with no perturbation of initial guess, and $\mathbf{u}_{pert}$ is the perturbed total displacement field. The error averaged over all the repetitions is~$0.86\%$, entailing only two cases with an erroneous assessment of the total displacement field, i.e., $\epsilon > 1.5\%$, confirming the~$98\%$ robustness of the method to initial guess. The average error $\epsilon$, ignoring the two outlier cases, is~$0.45\%$, confirming also the high accuracy of the methodology.
%
%
\subsection{Real Experiments}
\label{sec5:real}
\begin{figure}
\centering
\subfloat[Deformed Image]{\includegraphics[trim=0 0 0 0,clip,height=.275\textwidth]{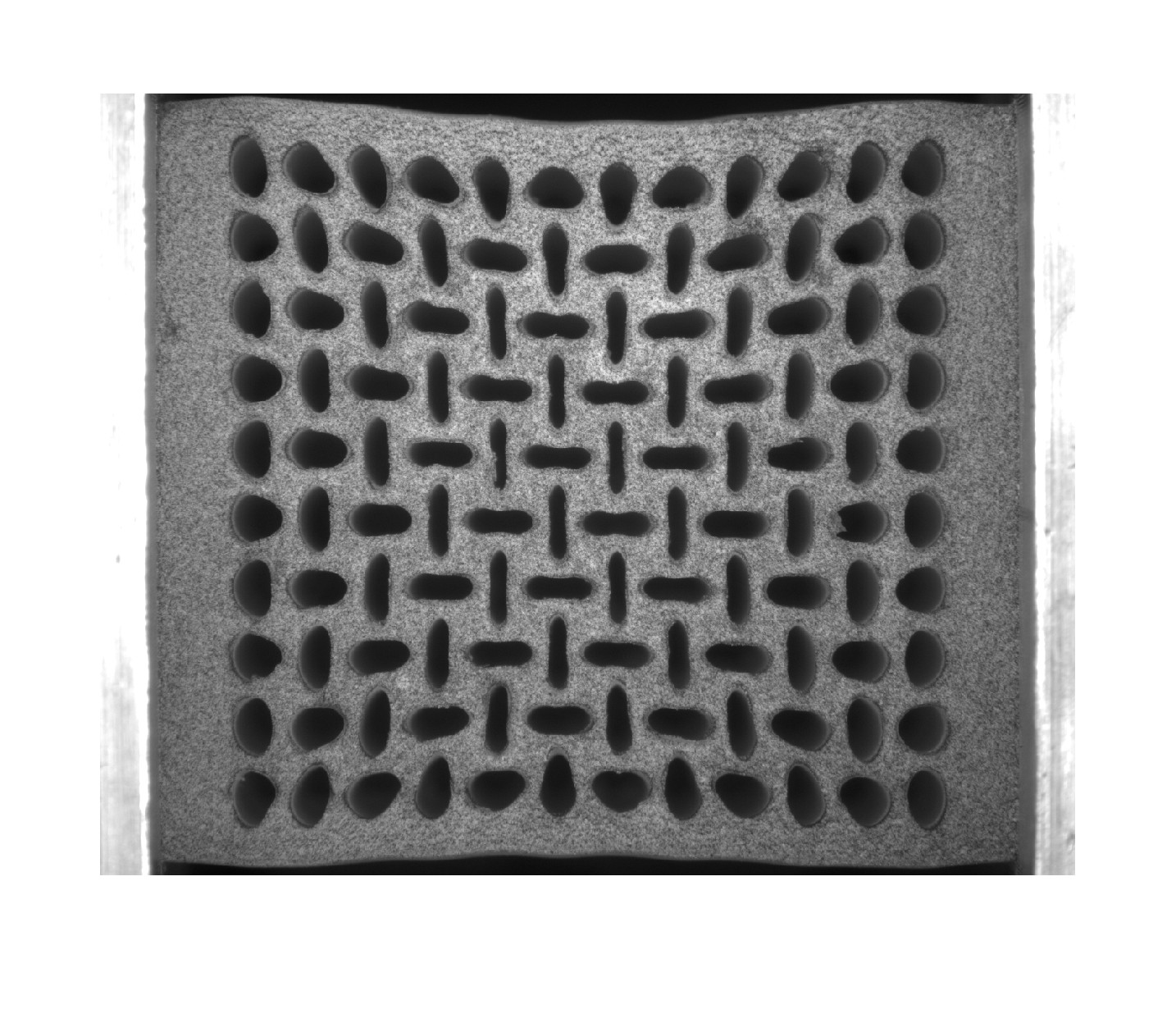}\label{fig5:def_real}}
\subfloat[$S_{xx}$~\mbox{[mm$^6$]}]{\includegraphics[trim=0 0 0 0,clip,height=.275\textwidth]{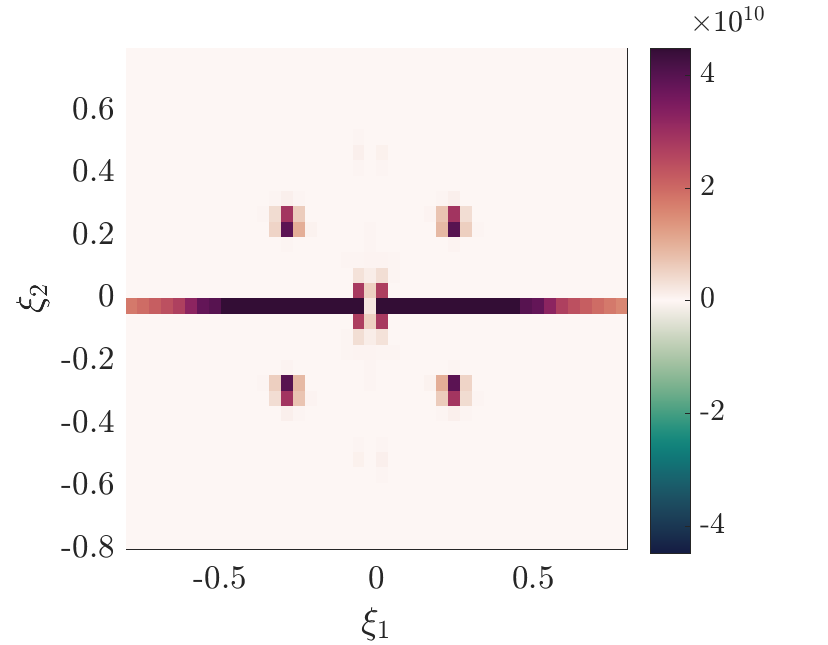}\label{fig5:sxx_real}}
\subfloat[$S_{yy}$~\mbox{[mm$^6$]}]{\includegraphics[trim=0 0 0 0,clip,height=.275\textwidth]{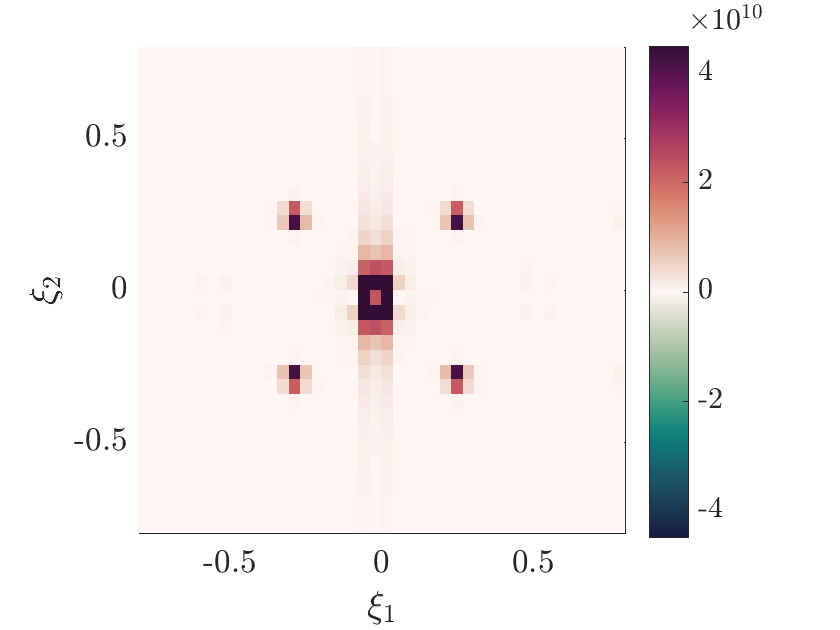}\label{fig5:syy_real}}
\caption{\protect\subref{fig5:def_real} Image of the deformed configuration of a cellular elastomeric metamaterial with millimetre-sized circular holes in a rectangular stacking, corresponding to a compressive load step after the onset of pattern transformation, acquired during an \textit{in-situ} compression test. Energy spectral density functions of the displacement field, assessed by local DIC, in \protect\subref{fig5:sxx_real} $x$ and \protect\subref{fig5:syy_real} $y$ direction, revealing the principal frequencies related to the long-range correlated fluctuation mode (four peaks approximately on the diagonals). The range of the colour bars is reduced to highlight the frequencies related to the long-range correlated fluctuation mode.}\label{fig5:spect_real}	
\end{figure}

The results of an \textit{in-situ} compression test on a cellular elastomeric metamaterial specimen, manufactured using a customized mould as presented in PhD thesis of~\cite{SiavashPhD}, are analysed with the micromorphic IDIC method. The real experiment includes a range of additional complexities compared to the virtual experiment, including the image distortions and image capturing noise as well as the imperfections in the specimens and loading conditions of the test. The specimen is $22.5$~mm thick and has 12 and 10 holes in loading and transverse directions, corresponding to a scale ratio 12. The specimen is made of Sylgard 184, with a 10:1 mixing ratio. A two-layer speckle pattern is deposited on the top surface (white powder sprayed to make a fine-grained background and India ink sprayed using an air brush, see Fig.~\ref{fig5:def_real}), resulting in a speckle size between 30 and $80~\mu$m (3 to 8 pixels). An \textit{in-situ} compression test is performed with a Kammrath \& Weiss micro tensile/compression stage with a $50$~N load cell, while a Zeiss V20 stereo microscope is used to acquire images at each load step at $7.6\times$ magnification and $2751\times2207$ pixels. The stereo microscope and the objective was used for proper imaging of the specimen providing high-enough resolution for the fine speckle pattern applied on the specimen with an adequate level of detail. More details on the specimen processing and the \textit{in-situ} test can be found in PhD thesis of~\cite{SiavashPhD}. Two images obtained during the \textit{in-situ} test, before and after the emergence of the buckling pattern, corresponding to~$8.2\%$ and~$12.7\%$ applied nominal strain on the metamaterial, are correlated to the undeformed reference configuration. The deformed configuration after emergence of the buckling pattern is shown in Fig.~\ref{fig5:def_real}.

A spatial distortion calibration step \cite{Maraghechi2018,Maraghechi2019}, is performed to measure the spatial distortions of the imaging system at the desired magnification. The resulting distortion field, which is as large as~$10\%$ of the actual displacements, is used to construct the distortion mapping function $\mathbf{S}(\x)$.

Chebyshev polynomials of 5th and 6th order are used to parametrize $\smooth_0$ and $v_1$.
The same procedure as in the case of virtual experiments is performed for estimating the reduced regularization of $\mode_1(\x)$ and the initialization of the associated parameters. To this end, first, the post buckling kinematics is determined using local DIC, in order to attain $S_{xx}$ and $S_{yy}$, shown in Figs.~\ref{fig5:sxx_real} and \ref{fig5:syy_real}. Similar to the virtual experiments, vertical and horizontal lines relate to discontinuities in the smooth mean deformations due to the specimen's edges, while the two sine waves in approximately diagonal directions for each displacement component correspond to a single long-range fluctuation mode. Based on these observations, the fluctuation mode is assigned the same way as for the virtual experiments, i.e., Eq.~\eqref{eq5:mode} with $\dofs_{\varphi_1} = [1,1,1,1,1,0,0]^{\mathsf{T}}$ as initial guess. In order to assure a proper scaling of the minimization problem, $\mode_1(\x)$ is scaled such that the initial maximum value of the mode is $0.04$. In order to avoid an ill-posed problem in the first iteration, $v_1(\x)$ is again initiated with a small constant value in space. The applied compressive global strain results in large displacement on the edges of the specimen, which requires an approximate initial guess in order to ensure convergence. Thus, the smooth mean field $\smooth_0(\x)$ is initialized such that the first order term in $x$ direction approximately accounts for the applied global strain. Similarly to the virtual experiment of Section~\ref{sec5:virt}, the IDIC procedure identifies in total~$77$ dofs~$\dofs$.

\begin{figure}
\captionsetup[subfloat]{captionskip=-1pt,nearskip=-1pt,farskip=-1pt}
\begin{minipage}{.5\textwidth}
\centering
Before microstructural buckling\\
\subfloat[$\smooth_0.\mathbf{e}_x$~\mbox{[$\mu$m]}]{\includegraphics[trim=0 0 0 0,clip,width=.46\textwidth]{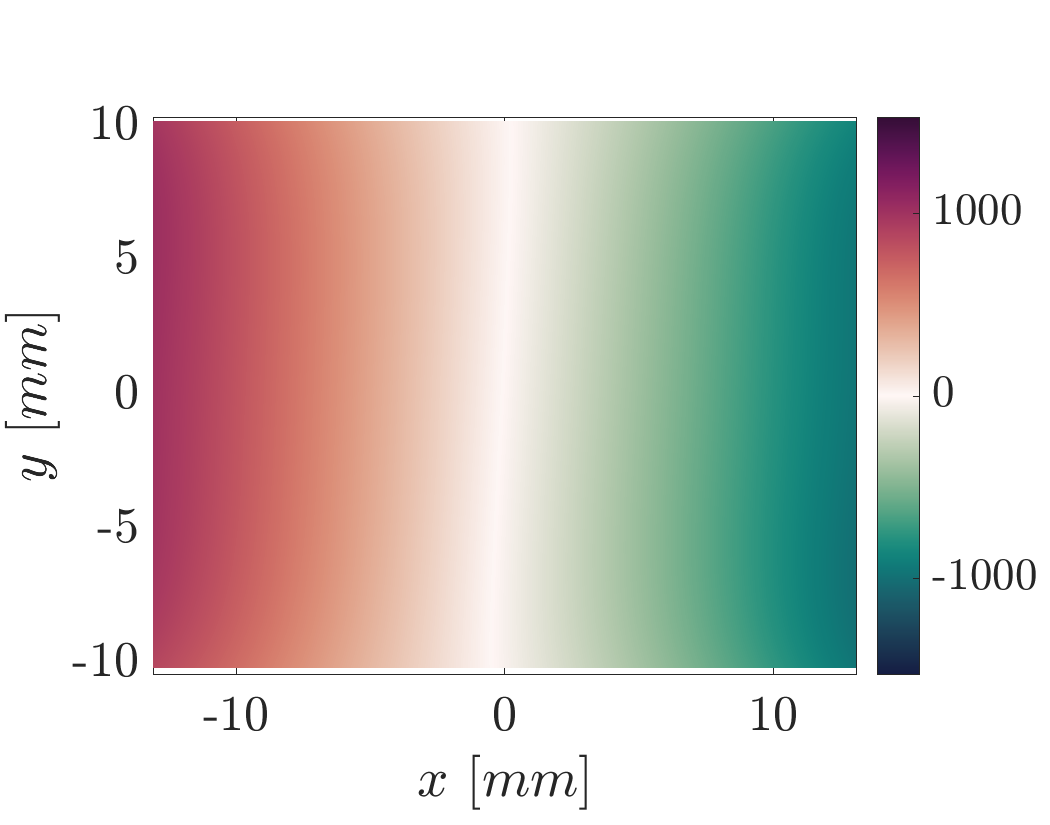}\label{fig5:real1v0x}}
\subfloat[$\smooth_0.\mathbf{e}_y$~\mbox{[$\mu$m]}]{\includegraphics[trim=0 0 0 0,clip,width=.46\textwidth]{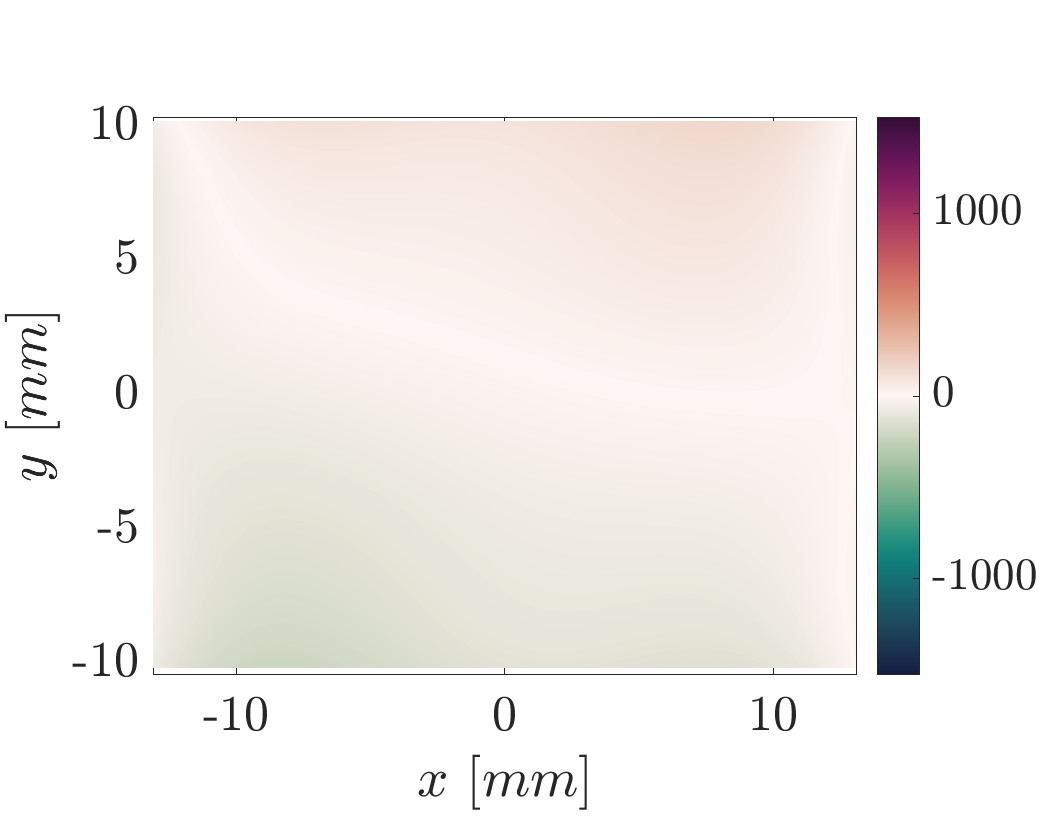}\label{fig5:real1v0y}}\\
\subfloat[$v_1$~\mbox{[$\mu$m]}]{\includegraphics[trim=0 0 0 0,clip,width=.46\textwidth]{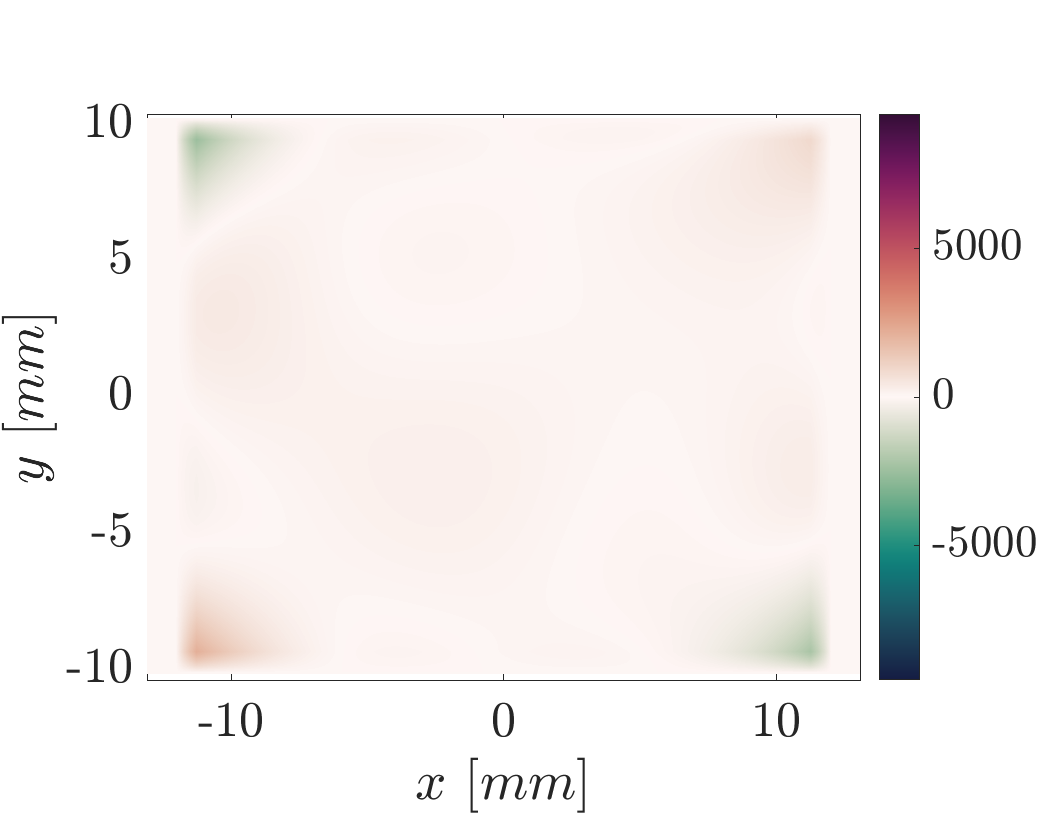}\label{fig5:real1v1}}
\subfloat[$\mode_1~{[-]}$]{\includegraphics[trim=0 0 0 0,clip,width=.46\textwidth]{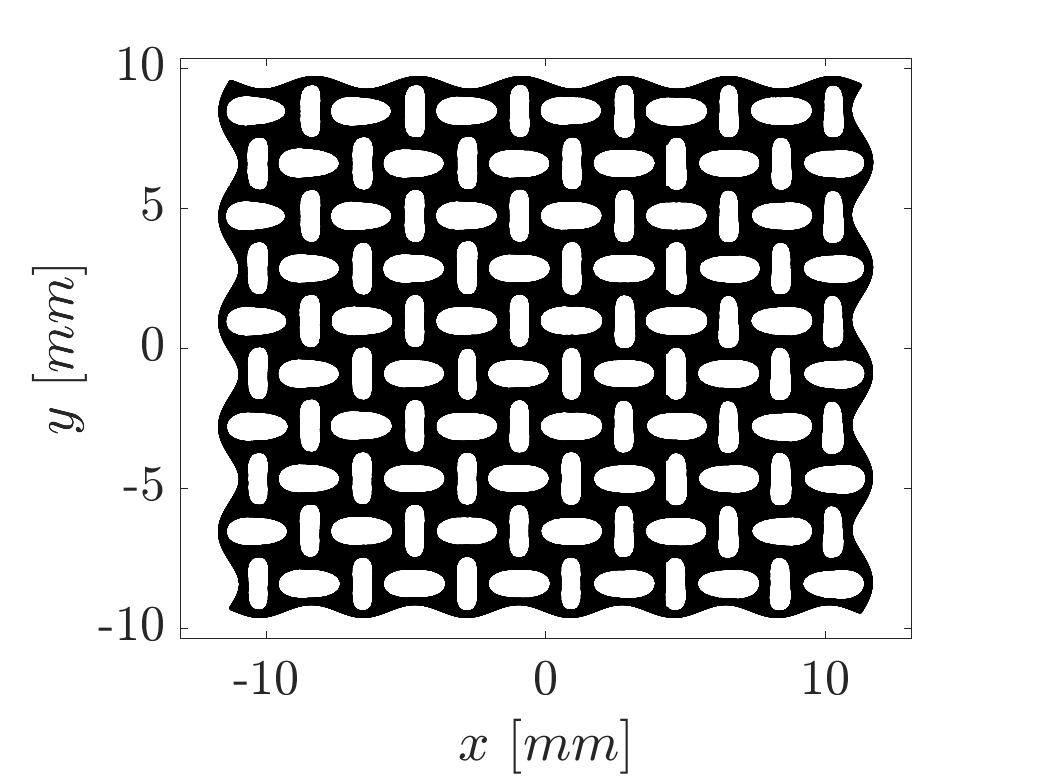}\label{fig5:real1phi}}\\
\subfloat[$v_1\mode_1.\mathbf{e}_x$~\mbox{[$\mu$m]}]{\includegraphics[trim=0 0 0 0,clip,width=.46\textwidth]{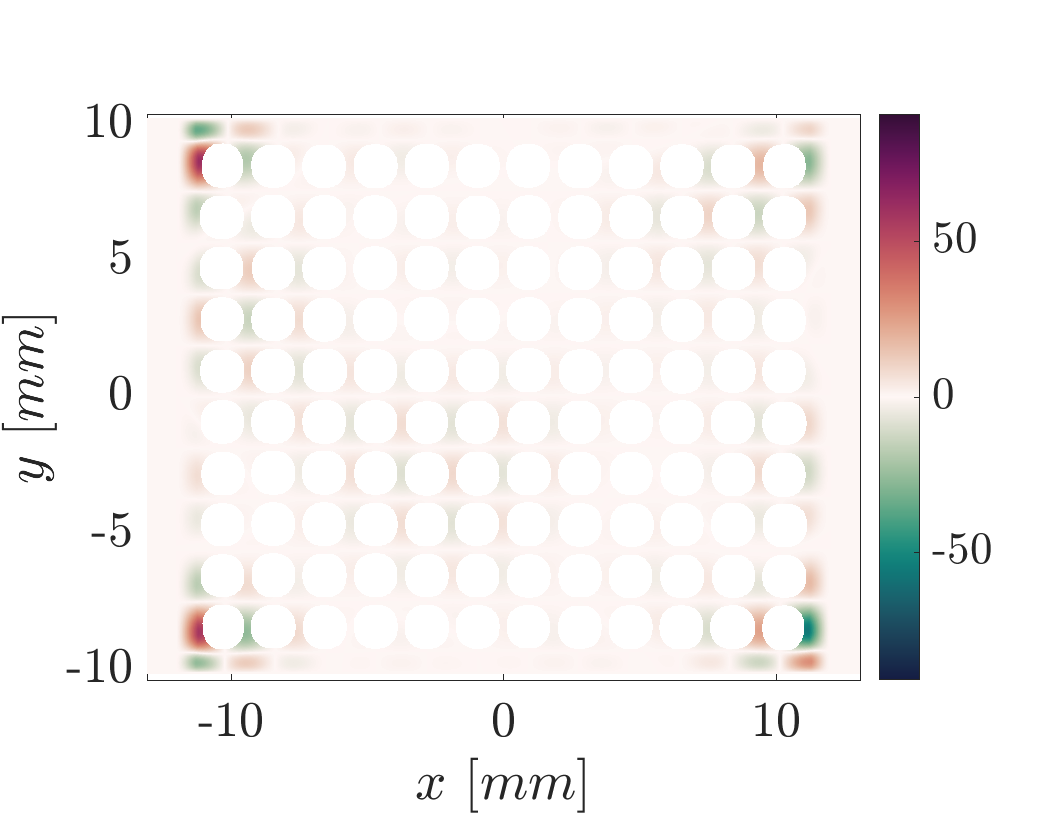}\label{fig5:real1flucx}}
\subfloat[$v_1\mode_1.\mathbf{e}_y$~\mbox{[$\mu$m]}]{\includegraphics[trim=0 0 0 0,clip,width=.46\textwidth]{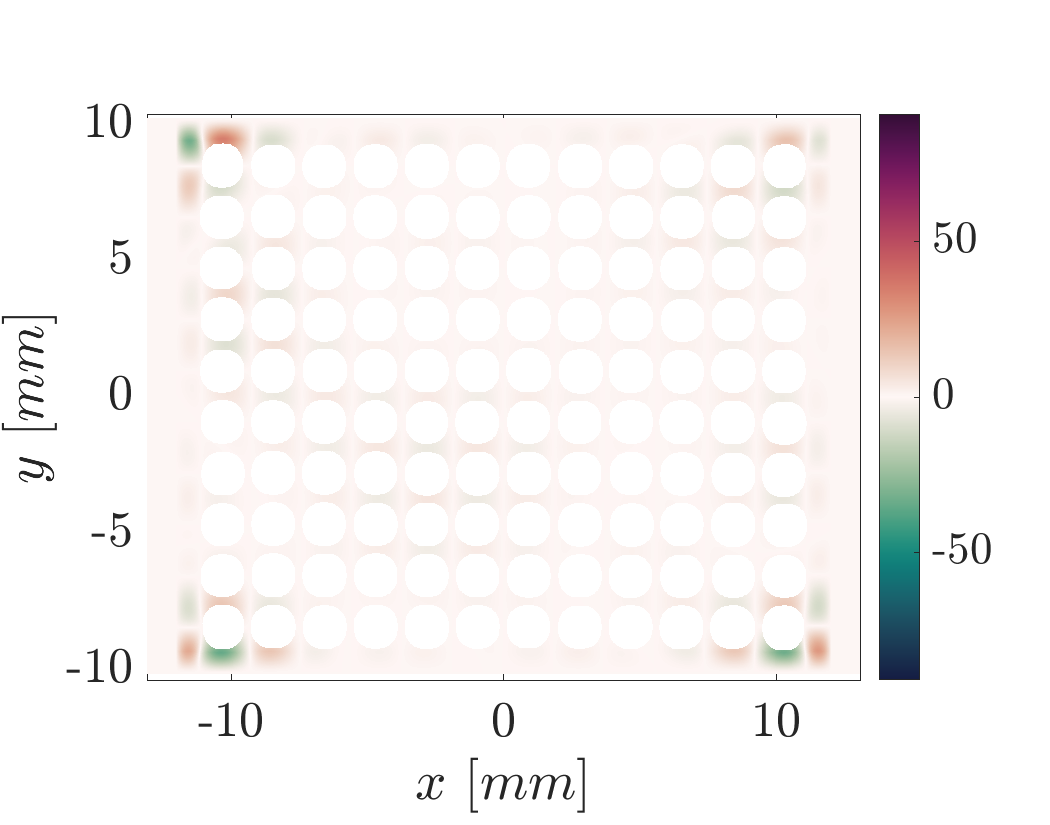}\label{fig5:real1flucy}}\\
\subfloat[residual~{[--]}]{\includegraphics[trim=0 0 0 0,clip,width=.46\textwidth]{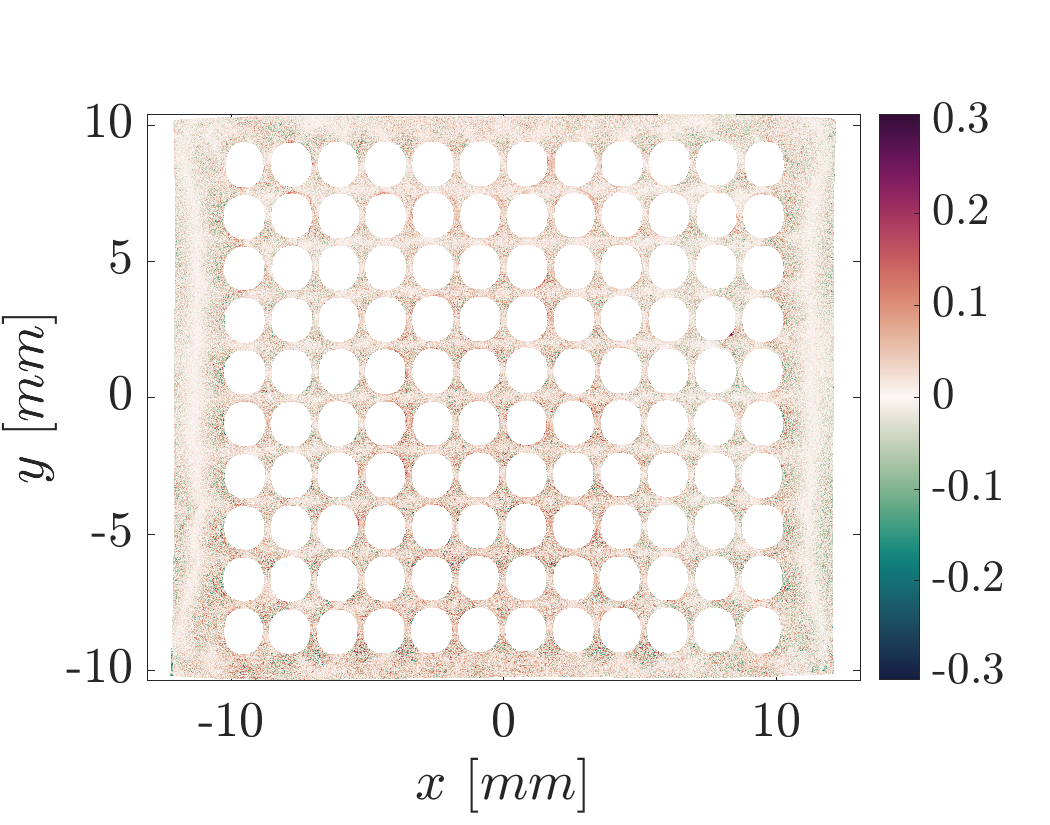}\label{fig5:real1r}}
\subfloat[$\vert \vert \mufluc \vert \vert$~\mbox{[$\mu$m]}]{\includegraphics[trim=0 0 0 0,clip,width=.46\textwidth]{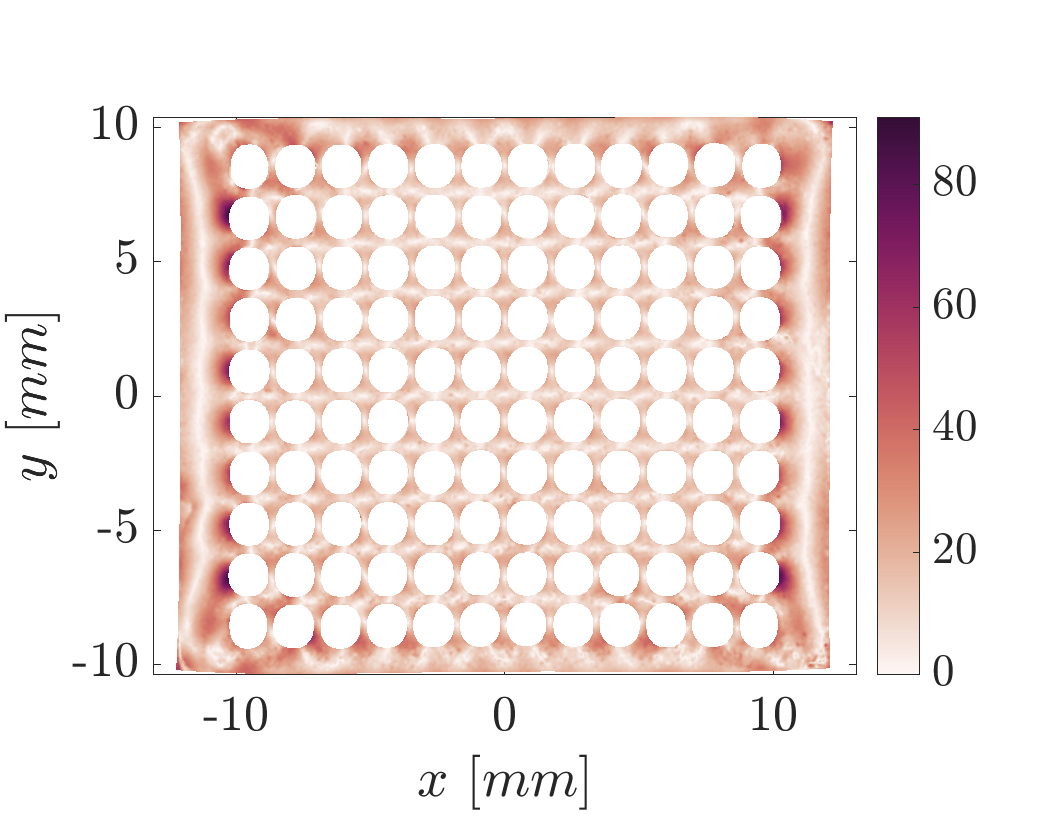}\label{fig5:real1w}}
\end{minipage} %
\vline
\begin{minipage}{.5\textwidth} %
\centering
After microstructural buckling\\
\subfloat[$\smooth_0.\mathbf{e}_x$~\mbox{[$\mu$m]}]{\includegraphics[trim=0 0 0 0,clip,width=.46\textwidth]{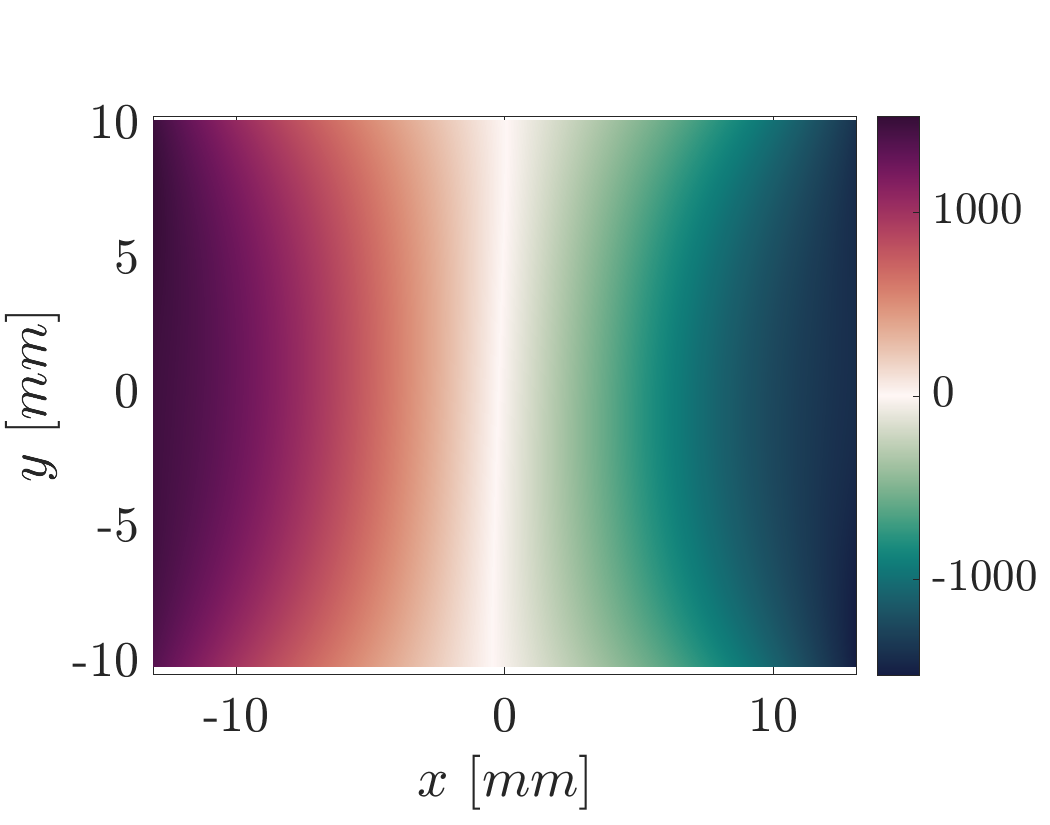}\label{fig5:real2v0x}}
\subfloat[$\smooth_0.\mathbf{e}_y$~\mbox{[$\mu$m]}]{\includegraphics[trim=0 0 0 0,clip,width=.46\textwidth]{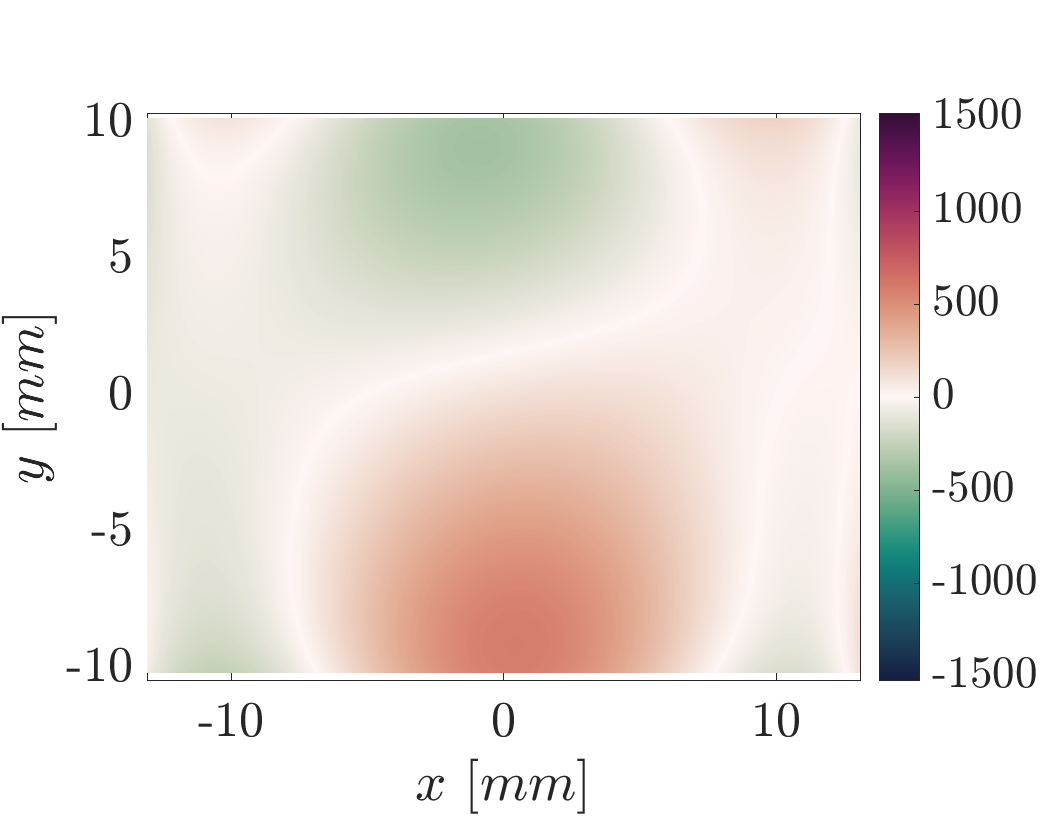}\label{fig5:real2v0y}}\\
\subfloat[$v_1$~\mbox{[$\mu$m]}]{\includegraphics[trim=0 0 0 0,clip,width=.46\textwidth]{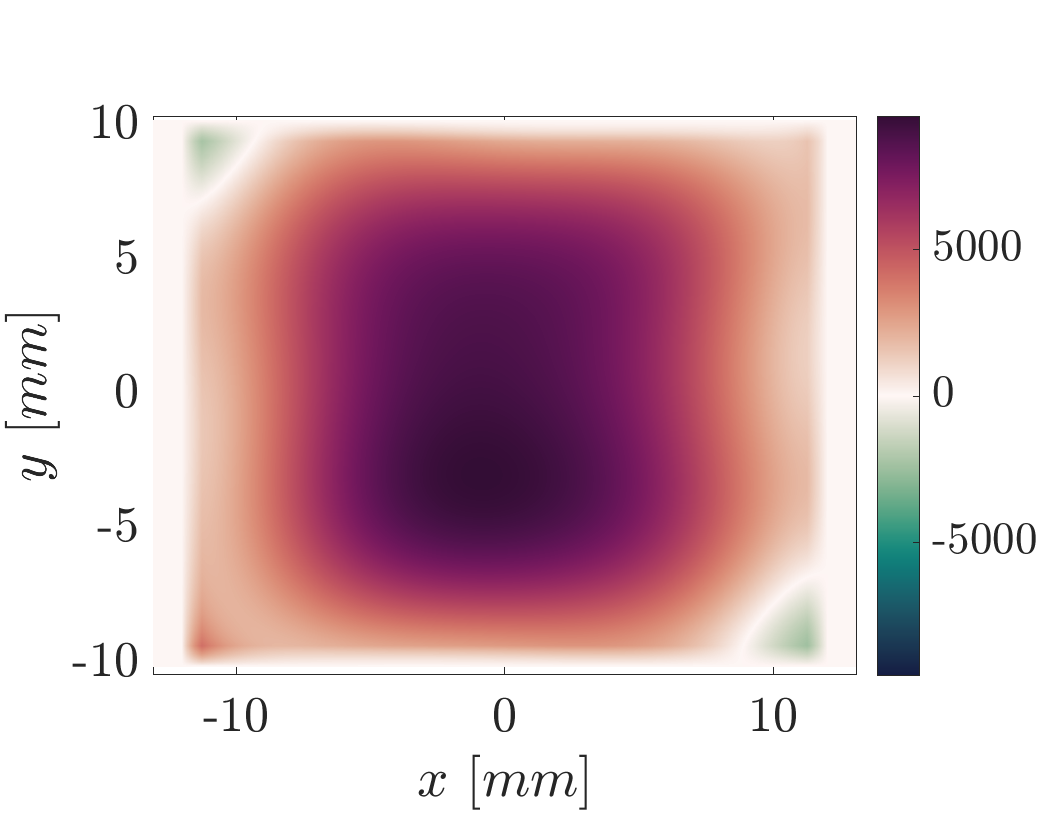}\label{fig5:real2v1}}
\subfloat[$\mode_1~{[-]}$]{\includegraphics[trim=0 0 0 0,clip,width=.46\textwidth]{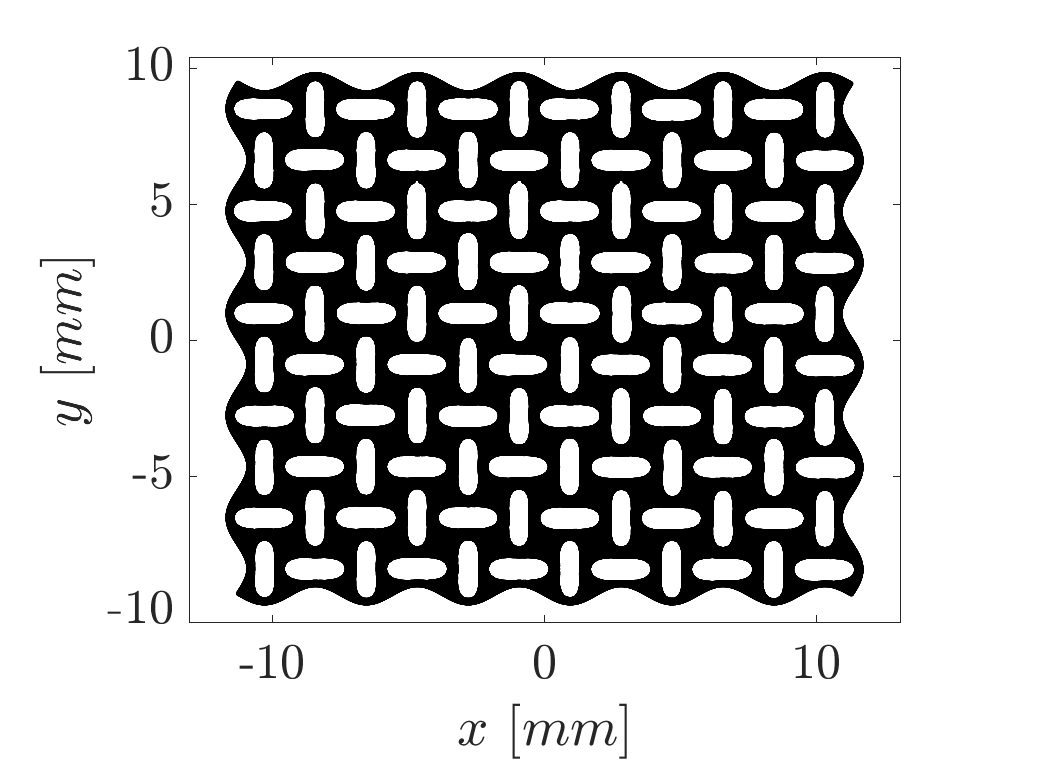}\label{fig5:real2phi}}\\
\subfloat[$v_1\mode_1.\mathbf{e}_x$~\mbox{[$\mu$m]}]{\includegraphics[trim=0 0 0 0,clip,width=.46\textwidth]{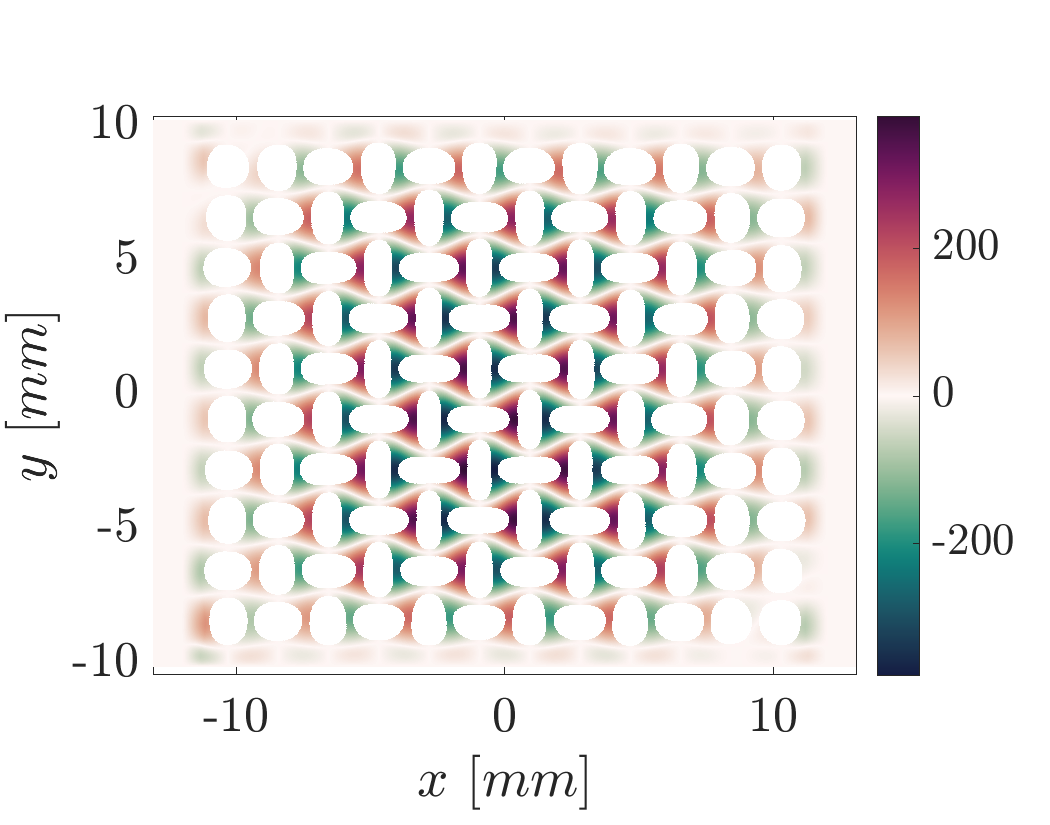}\label{fig5:real2flucx}}
\subfloat[$v_1\mode_1.\mathbf{e}_y$~\mbox{[$\mu$m]}]{\includegraphics[trim=0 0 0 0,clip,width=.46\textwidth]{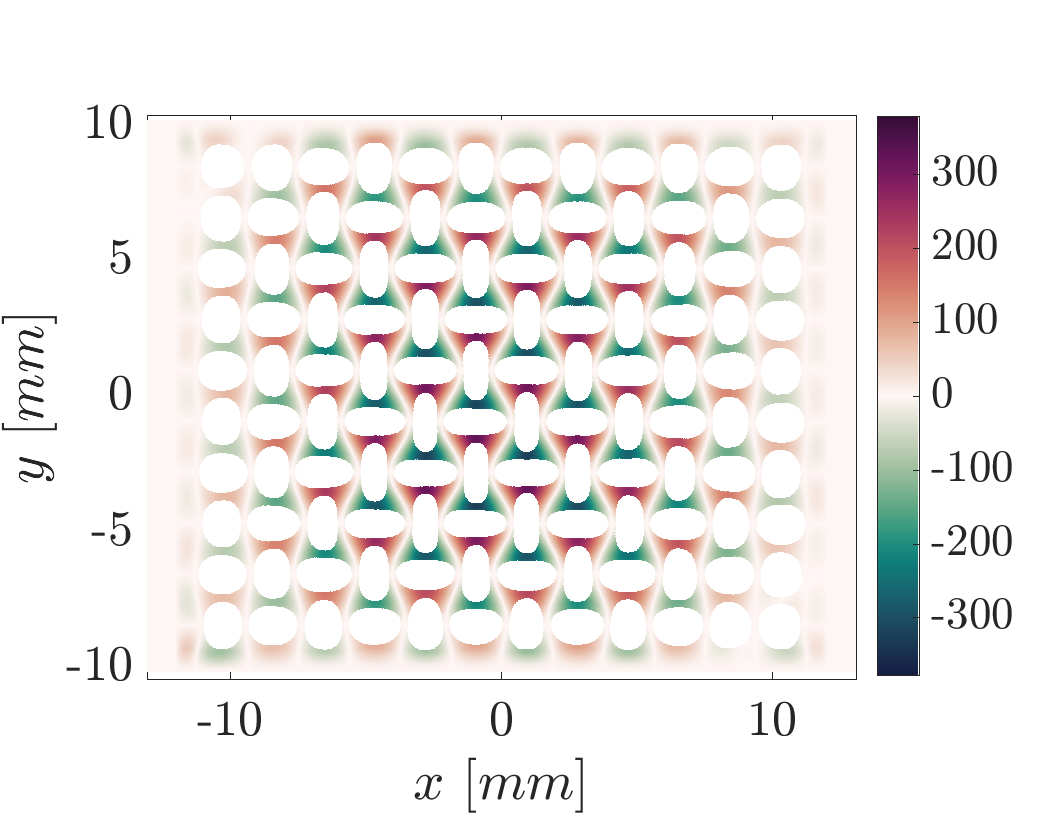}\label{fig5:real2flucy}}\\
\subfloat[residual~{[--]}]{\includegraphics[trim=0 0 0 0,clip,width=.46\textwidth]{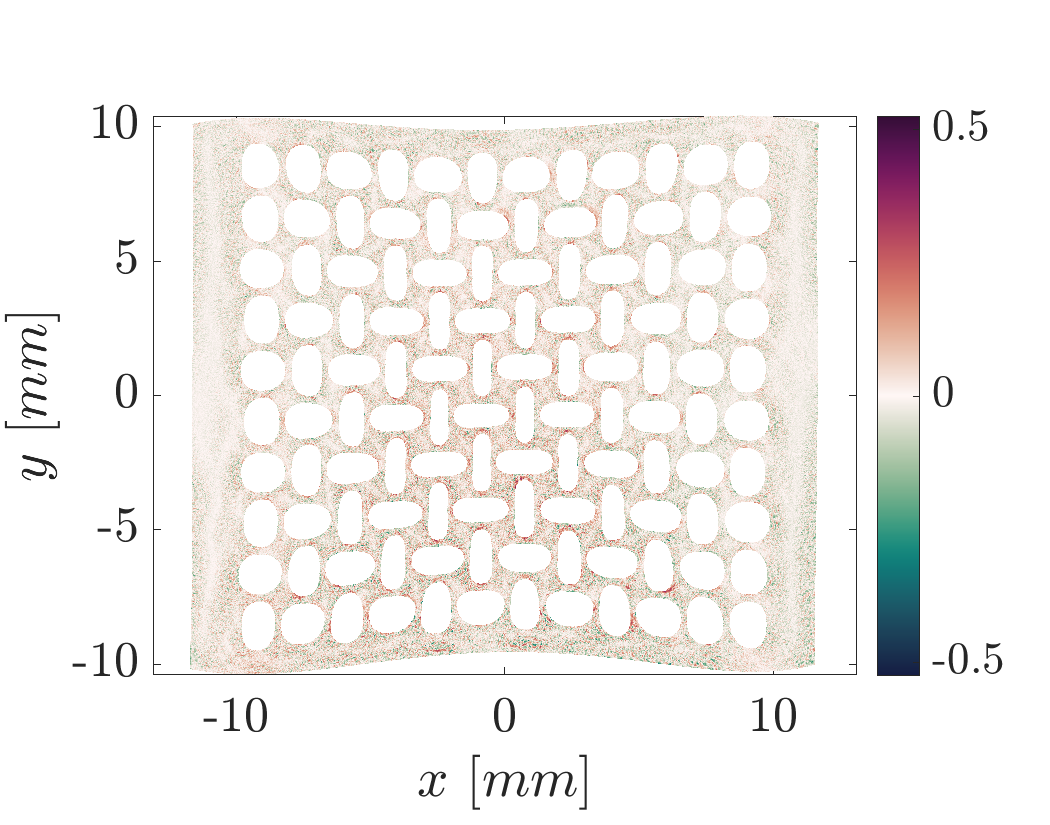}\label{fig5:real2r}}
\subfloat[$\vert \vert \mufluc \vert \vert$~\mbox{[$\mu$m]}]{\includegraphics[trim=0 0 0 0,clip,width=.46\textwidth]{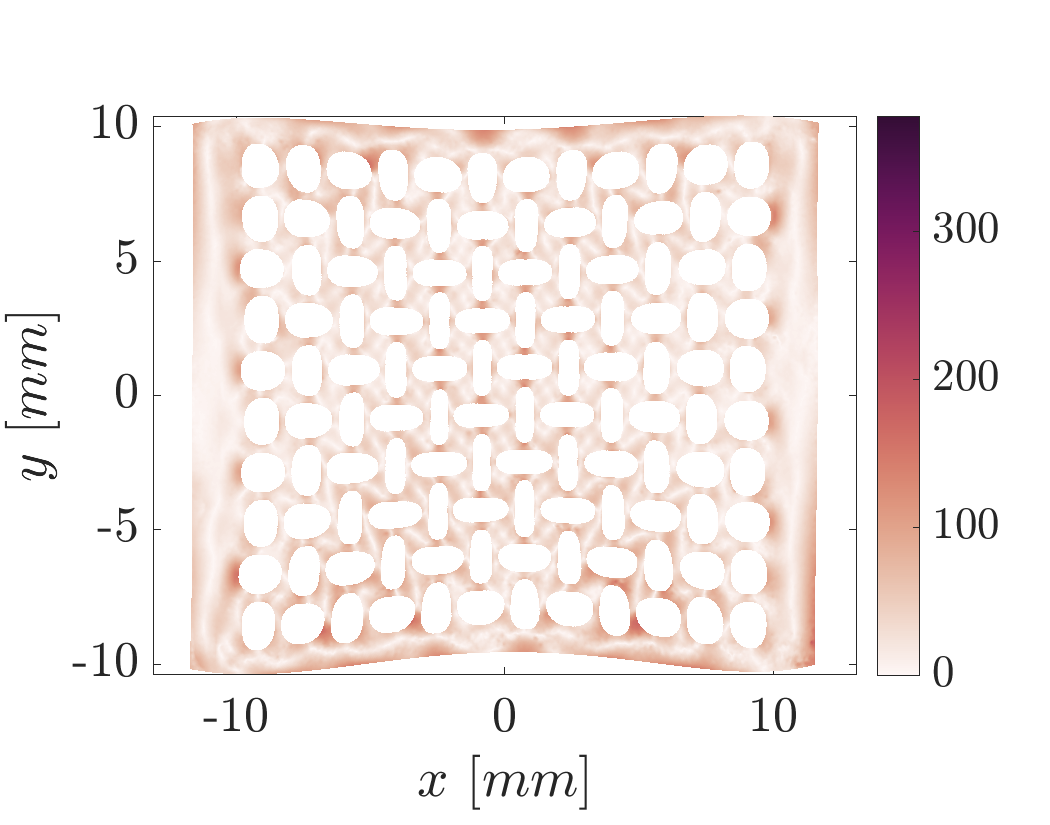}\label{fig5:real2w}}
\end{minipage}	
\vspace{-.5\baselineskip}	
\caption{Results of the real experiments, before (left) and after (right) the onset of microstructural buckling:(\protect\subref*{fig5:real1v0x}, \protect\subref*{fig5:real1v0y}, \protect\subref*{fig5:real2v0x}, \protect\subref*{fig5:real2v0y}) the smooth mean displacement fields, $\smooth_0(\x)$, in (\protect\subref*{fig5:real1v0x}, \protect\subref*{fig5:real2v0x}) $x$ and (\protect\subref*{fig5:real1v0y}, \protect\subref*{fig5:real2v0y}) $y$ directions; (\protect\subref*{fig5:real1v1}, \protect\subref*{fig5:real2v1}) the amplitudes, $v_1(\x)$, (\protect\subref*{fig5:real1phi}, \protect\subref*{fig5:real2phi}) the deformed configuration of identified modes, $\mode_1(\x)$, scaled to assure the visibility of the deformed shape, and (\protect\subref*{fig5:real1flucx}, \protect\subref*{fig5:real1flucy}, \protect\subref*{fig5:real2flucx}, \protect\subref*{fig5:real2flucy}) the resulting fields, (\protect\subref*{fig5:real1flucx}, \protect\subref*{fig5:real2flucx}) in $x$ and (\protect\subref*{fig5:real1flucy}, \protect\subref*{fig5:real2flucy}) $y$ directions, corresponding to the long-range correlated fluctuations. The long-range correlated fluctuation fields, $v_1(\x)\mode_1(\x)$, are plotted in their corresponding deformed configurations. (\protect\subref*{fig5:real1r}, \protect\subref*{fig5:real2r}) are the residual fields in normalized gray scales; and (\protect\subref*{fig5:real1w}, \protect\subref*{fig5:real2w}) the Euclidean norm of the microfluctuation field, $\mufluc(\x)$, plotted in deformed configuration based on the evaluated displacement, $\mathbf{u}(\x)$.}
\label{fig5:real}
\end{figure}

For these real \emph{in-situ} compressive experiments, the micromorphic IDIC also results in a proper and accurate identification of $\smooth_0(\x)$, $v_1(\x)$ and $\mode_1(\x)$, whereby the residual fields are small everywhere both before and after buckling, see Figs.~\ref{fig5:real1r} and \ref{fig5:real2r}. Convergence is consistent. Fig.~\ref{fig5:real} shows the results of applying micromorphic IDIC on real experiments before (left) and after (right) the emergence of the buckling pattern. For each case, the mean smooth displacement field, $\smooth_0(\x)$, in $x$ and $y$ directions, and $v_1(\x)$ are depicted in the first two rows, i.e., Figs.~\ref{fig5:real1v0x}--\subref*{fig5:real1v1} before and Figs.~\ref{fig5:real2v0x}--\subref*{fig5:real2v1} after the onset of buckling. The $\smooth_0$ field shows compression in $x$ direction both before and after buckling. Similar results as in the virtual experiments are observed in $\smooth_0$ and $v_1$. Note that the large values in $v_1(\x)$ are due to the scaling of the corresponding mode $\mode_1$ (maximum value of 0.04). The auxetic effect is more pronounced in the real experiments (Figs.~\ref{fig5:real2v0y},\subref*{fig5:real2r}) compared to the virtual experiments (Figs.~\ref{fig5:virt2v0y},\subref*{fig5:virt2r}), due to the larger specimen size, i.e., the larger scale ratio. In a specimen with a larger scale ratio the buckling pattern in the centre is less affected by the two stiffening lateral edges and, thus, it is more pronounced and results in larger auxetic effect.

The long-range correlated mode, $\mode_1(\x)$, as well as the long-range correlated fluctuation fields, $v_1\mode_1(\x)$, in $x$ and $y$ directions, are depicted in their deformed configurations in Figs.~\ref{fig5:real1phi}--\subref*{fig5:real1flucy} and Figs.~\ref{fig5:real2phi}--\subref*{fig5:real2flucy}, before and after buckling, respectively. In Figs.~\ref{fig5:real1phi} and \ref{fig5:real2phi}, $\mode_1(\x)$ is scaled to assure the visibility of the deformed shape of the modes. As expected, small values for $v_1$ prior to buckling entail a low sensitivity to the fluctuation mode, i.e., a slightly different mode is identified at this stage. Indeed, the long-range correlated fluctuation mode is as good as absent before the emergence of the buckling pattern, where it is not yet representative. The edge effect prior to buckling, which is most pronounced at the corner unit cells, is partially captured by the identified mode. This explains the non-zero values of $v_1$ and thus $v_1(\x)\mode_1(\x)$ in these areas.

The residual fields in normalized grey scales are shown in Figs.~\ref{fig5:real1r} and \ref{fig5:real2r}, in the deformed configuration based on the evaluated displacement $\mathbf{u}(\x)$ from Eq.~\eqref{eq5:u_dic}, before and after buckling. In order to extract the microfluctuation field $\mufluc(\x)$, the displacement fields obtained by local DIC on the same images are used as the reference. Note that the local DIC procedure is not a part of the methodology introduced here, and is used to assess the microfluctuation field (with errors from local DIC) in order to evaluate the performance of the method. Although local DIC gives the displacement field with minimum kinematical constraints, yet with significant statistical and systematic errors, it requires tens of thousands of dofs to fit the displacement field compared to the~$77$ dofs for the micromorphic IDIC framework. Figs.~\ref{fig5:real1w} and \ref{fig5:real2w} show the Euclidean norm of the microfluctuation field, $\vert\vert \mufluc \vert\vert$, before and after buckling. The microfluctuation field $\mufluc$, is of a higher spatial frequency than the characteristic frequency of $\mode_1$, which is more pronounced far from edges of the specimen in the post-buckling regime (Fig.~\ref{fig5:real2w}). Yet, it can be observed prior to buckling as well (Fig.~\ref{fig5:real1w}). Beyond buckling, the microfluctuation field $\mufluc$ has a significantly smaller amplitude than the correlated fluctuation field. This is not the case prior to buckling, considering the almost zero contribution of the long-range correlated fluctuation field. The residual field is larger in certain areas, which corresponds directly to higher values in the microfluctuation field, both before and after buckling.

All the identified fields, i.e., $\smooth_0(\x)$, $v_1(\x)$ and $\mode_1(\x)$, satisfy the expected mechanical behaviour, and the correlation residual, reflected also in the evaluated microfluctuation field $\mufluc(\x)$, confirms the good performance of the micromorphic IDIC method.
%
%
\subsection{Graded Microstructures and Size Independence of the Modes}
\label{sec5:graded}
\begin{figure}[htbp]
	\centering
	\subfloat[$d/\ell = 0.95$]{\includegraphics[scale=0.035]{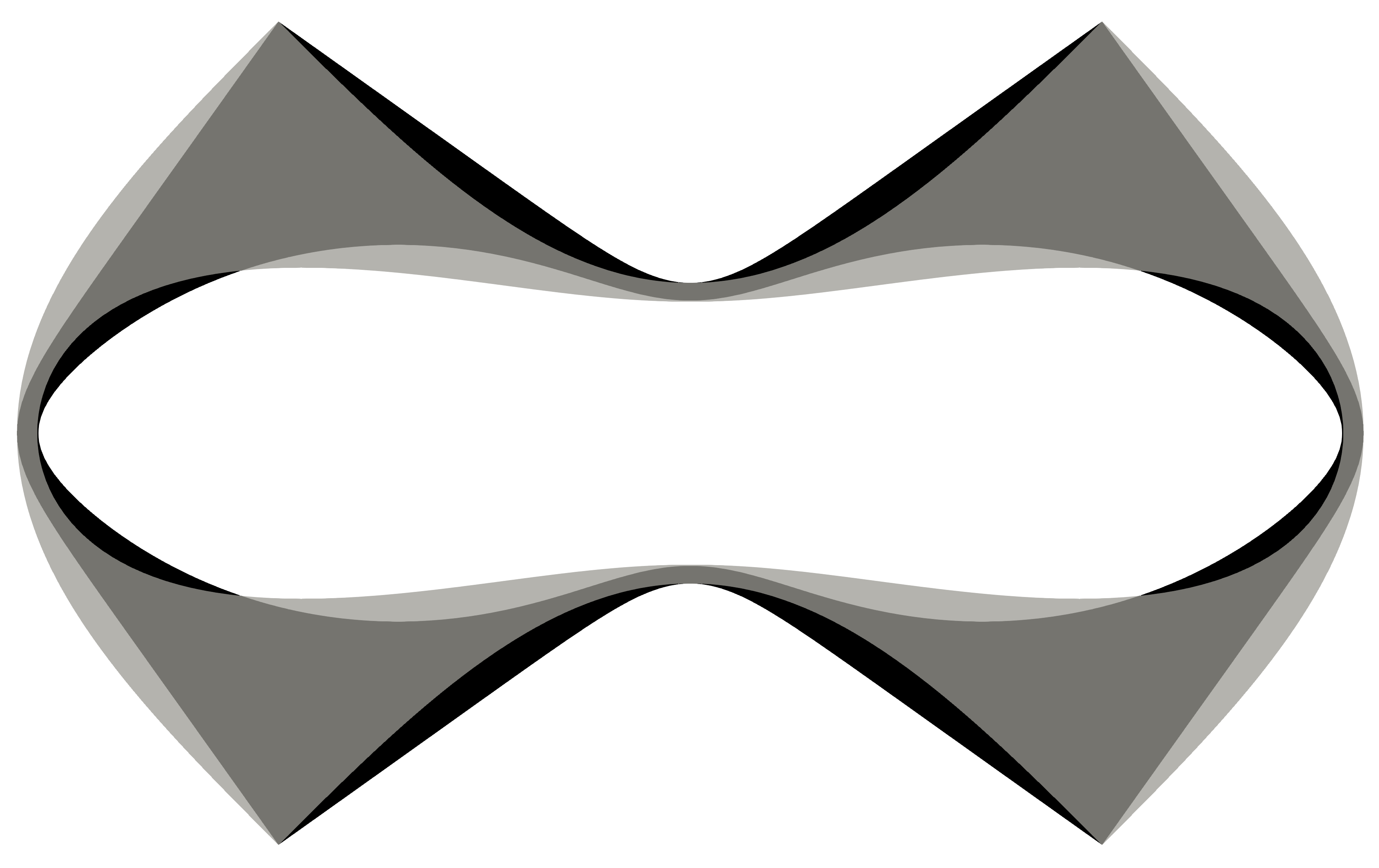}\label{fig7:gradeda}}
	\hspace{1em}
	\subfloat[$d/\ell = 0.79$ (reference)]{\includegraphics[scale=0.035]{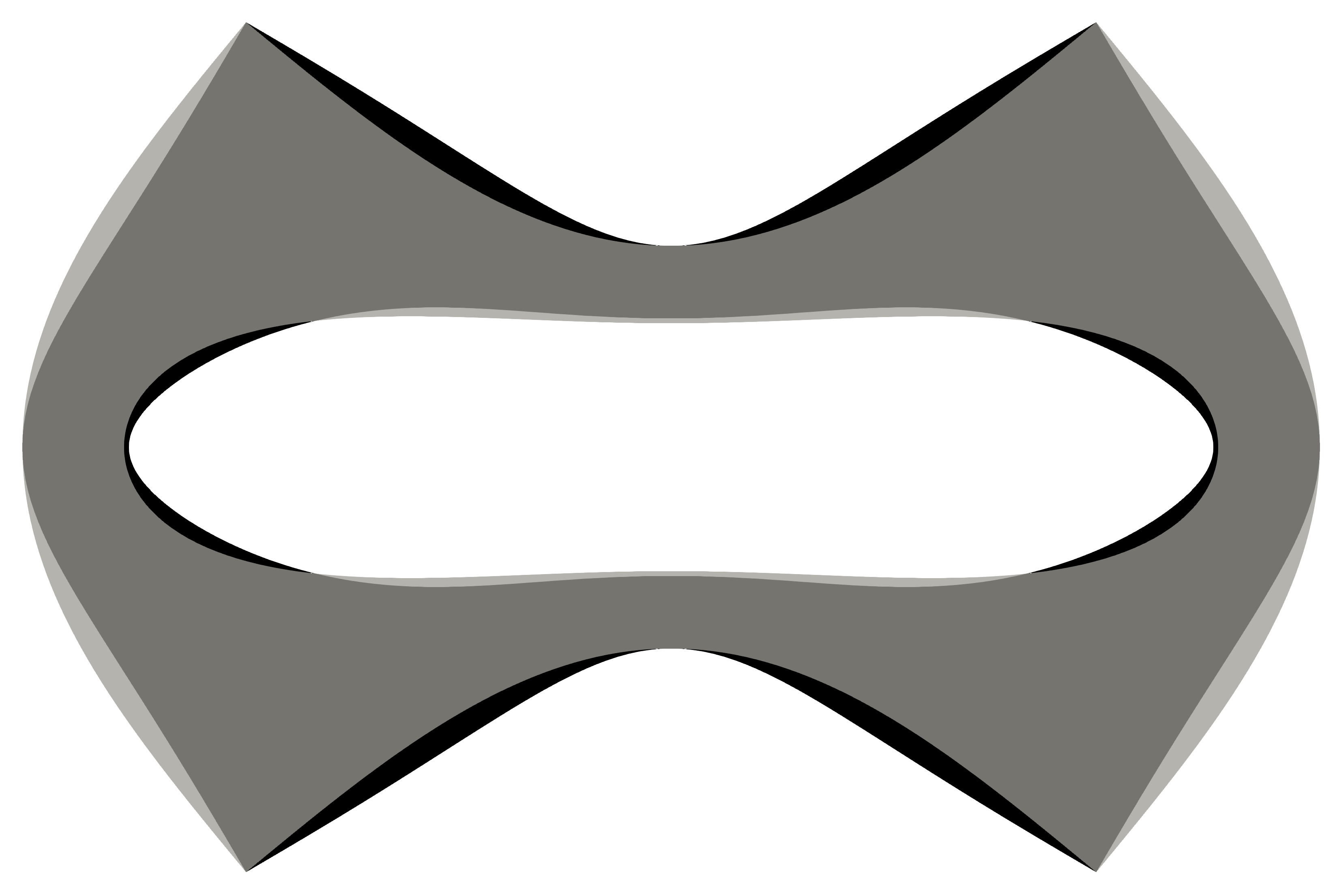}\label{fig7:gradedb}}
	\hspace{1em}	
	\subfloat[$d/\ell = 0.25$]{\includegraphics[scale=0.035]{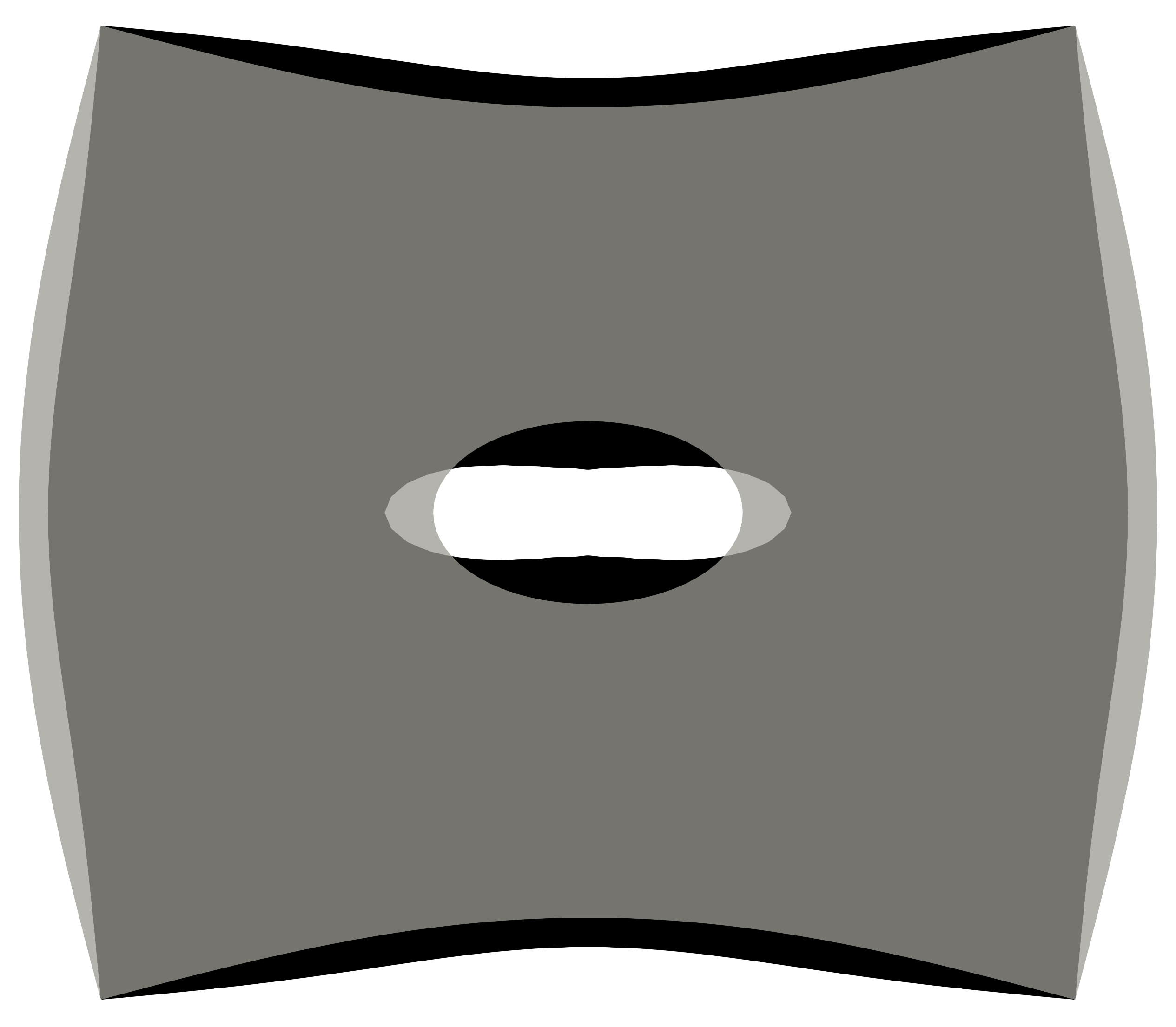}\label{fig7:gradedc}}
	\caption{Patterning modes for an infinite microstructure with a square stacking of holes for a unit cell size~$l = 1.9$~mm and a varying hole diameter~$d \in \{ 1.81, 1.50,  0.48 \}$~mm. Numerically computed modes are shown in black, whereas experimentally obtained modes according to Eq.~\eqref{eq5:mode} in Section~\ref{sec5:real} are shown in light grey. Large hole diameter to cell size ratio~$d/l = 0.95$ is shown in~\protect\subref{fig7:gradeda}, $d/l = 0.79$ corresponding to the reference geometry analysed in detail in Section~\ref{sec5:results} is shown in~\protect\subref{fig7:gradedb}, while small hole diameter $d/l = 0.25$ is shown in~\protect\subref{fig7:gradedc}.}
	\label{fig7:graded}
\end{figure}

As long as the constitutive law of the base elastomeric material is size independent (e.g., the hyperelastic Ogden material used in Section~\ref{sec5:virt}, usually used for constitutive modelling of metamaterials), the modes~$\mode_i$ are independent of the unit cell size~$l$. To show that the modes are independent also of~$d/l$ ratio (i.e., hole diameter to cell size ratio), an infinitely large specimen with a square stacking of holes is considered, cf. Figs.~\ref{fig:modesa}--\protect\subref*{fig:modesb}. In such a case the buckling analysis carried out on a~$2 \times 2$ periodic cell provides the reference (exact) mode---exact within the accuracy of the adopted numerical model---, against which the approximation of~$\mode_1$ given by Eq.~\eqref{eq5:mode} can be compared. Results are obtained for a larger and smaller $d/l$ ratio of, respectively, 0.95 and 0.25, compared to the reference value of 0.79 identified by the micromorphic IDIC procedure, see Fig.~\ref{fig7:graded}. The deformed configuration of the reference mode is shown in black colour, whereas the deformed shape of the experimentally identified mode is shown in light grey. Note that only one quarter of the~$2 \times 2$ periodic cell is shown due to various symmetries~\citep[see e.g.][]{Bertoldi2008}. Fig.~\ref{fig7:gradeda} suggests that the overall shape of the mode is captured sufficiently accurately for~$d/l = 0.95$ (recall also Fig.~\ref{fig5:virt}), although local discrepancies can be observed in higher-frequency components. Such components are not visible in Figs.~\ref{fig5:sxx_virt}--\protect\subref*{fig5:syy_virt} and~\ref{fig5:sxx_real}--\protect\subref*{fig5:syy_real}, since their contributions to spectral densities are very small. More interestingly, comparable errors are observed also in Fig.~\ref{fig7:gradedb} for~$d/l = 0.79$, meaning that identified mode coefficients~$\dofs_{\varphi_1}$ do not change substantially as a function of small variations of~$d/l$. For a very different ratio, $d/l = 0.25$, however, larger discrepancies between the two modes are observed in Fig.~\ref{fig7:gradedc}, suggesting that the experimental mode should be identified again. Although this insight, i.e., that the modes are independent for moderate variations in the hole diameter to cell size ratio, has been demonstrated here for a square stacking of holes, it is a general feature of cellular metamaterials.

This brings us to the realization that the introduced micromorphic IDIC technique, combined with the previously developed micromorphic computational homogenization scheme, provides a general tool for the design of engineering systems at two scales.
\begin{enumerate*}[label=(\roman*)]
	\item Size effects and microstructural changes are communicated through the patterning modes to the macroscale, where such information can be utilized for the design of engineering systems.
	\item Microstructures with graded properties smoothly varying in space (e.g., $d(\x)$) or material property (e.g., Young's modulus, $E(\x)$, or Poission's ratio, $\nu(\x)$), can be optimally adjusted and understood at the micro- or meso-scale level. For more details on the modelling and application of functionally graded materials the reader is referred to the extensive literature on this topic, see, e.g., \citep{Anthoine2010,Helou2019,Francesconi2019,amin_farzaneh_sequential_2022,joshua_morris_optimization_2023}.
\end{enumerate*}
It is important to realize that considerable design freedom is available even for geometrically simple microstructures, such as the herein studied square stacking of holes. Adopting different types of microstructural morphologies (e.g., hexagonal stacking of holes, chiral or re-entrant microstructures) expands the design space further.
%
%
\section{Summary and Conclusions}
\label{sec5:conclusion}
Cellular elastomeric metamaterials generally exhibit a pattern transformations based on microstructural buckling at a critical level of compressive load. These pattern transformations are correlated over long ranges in the specimen and result in considerable changes in the overall mechanical response after their emergence. A micromorphic Integrated Digital Image Correlation (micromorphic IDIC) method, was here developed to experimentally identify and quantify the decomposed kinematics for cellular metamaterials. The decomposition follows a recently introduced micromorphic computational homogenization scheme based on a kinematical ansatz consisting of three parts:
\begin{enumerate*}[label=(\roman*)]
\item mean smooth displacements, corresponding to the slow-scale material level;
\item long-range correlated fluctuation fields, representing the buckling patterns; and
\item a microfluctuation field, which captures the remaining uncorrelated part of the fluctuations. 
\end{enumerate*}
This kinematical ansatz was integrated within the digital image correlation scheme, enabling a direct identification of the smooth displacement, and the long-range correlated fluctuation pattern along with its spatial distribution field. On the basis of spectral analysis, a 2D Fourier series approximation of the long-range fluctuation mode with a proper parametrization was performed. The identified patterning modes serve as input parameters for the micromorphic computational homogenization scheme, which is capable of providing mechanical predictions of engineering systems including size effects as well as temporal and spatial mixing of patterning modes.

The methodology is validated on both virtual and real experiments for a specific case of elastomeric metamaterials with rectangular stacking of millimetre-sized circular holes. A standard spectral density analysis is used to attain a regularization with a reduced number of degrees of freedom and to initialize the associated parameters of the long-range fluctuation mode for the considered metamaterial. The method successfully decomposes the displacement field of cellular elastomers under compressive load, both before and after local microstructural buckling leading to long-range correlated pattern transformations. The performance of the method is assessed by identifying the microfluctuation fields through a comparison with local DIC data. In all cases, the microfluctuation field is of higher spatial frequency than the long-range fluctuations and of smaller amplitude after the emergence of the buckling pattern, thereby validating the kinematical ansatz as well as the micromorphic IDIC identification routine. An initial guess robustness study, performed by random perturbations up to~$10\%$ in fluctuation mode parameters, resulted in~$98\%$ robustness of the correlations and negligible errors ($0.45\%$ on average) in the results, confirming the robustness and the accuracy of the proposed method.

The micromorphic IDIC methodology may be considered novel and unique in the following aspects:
\begin{enumerate}[label=(\roman*),topsep=0pt,itemsep=-1ex,partopsep=1ex,parsep=1ex]
\item The methodology aims at and succeeds in identifying long-range correlated fluctuation fields, which is not existing yet in commercial DIC packages.
\item The decomposition of the displacement field is performed in a single minimization step, identifying the long-range correlated fluctuation mode, its amplitude in space, and the mean smooth displacement field;
\item The methodology can be readily extended to any cellular metamaterial by making proper initialization choices of the fluctuation modes;
\item It is easy to implement based on minimal modifications to conventional global DIC formulations.
\end{enumerate}

The proposed micromorphic IDIC technique, in combination with the previously developed micromorphic computational homogenization scheme, thus provides an integrated expe\-rimen\-tal-\-compu\-tati\-onal tool for optimizing the design of metamaterial-based engineering systems, including materials with various metamaterial microstructures and/or with graded properties in space, such as the hole size, or material properties, such as the Young's modulus or Poisson's ratio.

%
%
\section*{Acknowledgements}
The research leading to these results has received funding from the European Research Council under the European Union's Seventh Framework Programme (FP7/2007-2013)/ERC grant agreement \textnumero~[339392].

%
%
\bibliography{mybibfile.bib}
\end{document}